\documentclass{aa}  

\usepackage{graphicx}
\usepackage{txfonts}
\usepackage{lscape}
\usepackage{subfigure}
\usepackage{amssymb,natbib,supertabular}
\usepackage{defs}
\usepackage{appendix}
\usepackage{color}
\usepackage{array}
\usepackage[utf8]{inputenc}
\DeclareUnicodeCharacter{2212}{-}
\usepackage[breaklinks, colorlinks, citecolor=blue]{hyperref}

\usepackage{orcidlink} 
\newcommand{\orcid}[1]{\orcidlink{#1}}

\newcommand{\pz}{\phantom{0}}

\makeatletter
\renewcommand*\aa@pageof{, page \thepage{} of \pageref*{LastPage}}
\makeatother
\usepackage{lastpage}

\begin{document}

   \title{An ALMA spectroscopic survey of the \textit{Planck} high-redshift object PLCK\,G073.4$−$57.5 confirms two protoclusters}

   \subtitle{}

   \author{Ryley Hill\orcid{0009-0008-8718-0644}
        \inst{1}
          \and
          Maria del Carmen Polletta\orcid{0000-0001-7411-5386}\inst{2,3}
          \and
          Matthieu B{\'e}thermin\orcid{0000-0002-3915-2015}\inst{4,5}
          \and
          Herv{\'e} Dole\inst{6}
          \and
          R{\"u}diger Kneissl\orcid{0000-0002-5580-006X}\inst{7,8}
          \and
          Douglas Scott\orcid{0000-0002-6878-9840}\inst{1}
          }
          \offprints{R. Hill\\ \email{ryleyhill@phas.ubc.ca}}

   \institute{Department of Physics and Astronomy, University of British Columbia, 6225 Agricultural Road, Vancouver, V6T 1Z1, Canada\label{1} 
   \and 
   INAF, Istituto di Astrofisica Spaziale e Fisica Cosmica Milano,  Via A. Corti 12, I-20133 Milano, Italy\label{2} 
   \and 
   Department of Astronomy \& Astrophysics, University of California at San Diego, 9500 Gilman Drive, La Jolla, CA 92093, USA\label{3}
   \and
   Laboratoire d’Astrophysique de Marseille, CNRS, LAM, Aix Marseille Universit{\'e}, 38 Rue Fr{\'e}d{\'e}ric Joliot Curie, 13013 Marseille, France\label{4}
   \and
   Observatoire Astronomique de Strasbourg, CNRS, Universit{\'e} de Strasbourg, 11 Rue de l'Universit{\\'e}, 67000 Strasbourg, France\label{5}
   \and
   Institut d’Astrophysique Spatiale, CNRS, Universit{\'e} Paris-Saclay,  B{\^a}timent 121, 91405 Orsay, France\label{6}
   \and 
   ESO Vitacura, European Southern Observatory, Alonso de C{\'o}rdova 3107, Santiago, Chile\label{7} 
   \and 
   Joint ALMA Observatory, Alonso de Cordova 3107, Santiago, Chile\label{8} 
             }

   \date{Received ...; accepted ...}
 
\abstract
{{\it Planck\/}'s High-Frequency Instrument observed the whole sky between 350\,$\mu$m and 3\,mm, discovering thousands of unresolved peaks in the cosmic infrared background. The nature of these peaks is still poorly understood -- while some are strong gravitational lenses, the majority are overdensities of star-forming galaxies but with almost no redshift constraints. PLCK\,G073.4$-$57.5 (G073) is one of these {\it Planck\/}-selected peaks. ALMA observations of G073 suggest the presence of two structures (one around redshift 1.5 and one around redshift 2) aligned along the line of sight, but this analysis lacked robust spectroscopic confirmation. Characterizing the full redshift distribution of the galaxies within G073 is needed in order to better understand this representative example of {\it Planck\/}-selected objects, and connect them to the emergence of galaxy clusters. We used ALMA Band~4 spectral scans to search for \COthree, \COfour, and C{\sc i}(1--0) line emission, targeting eight red {\it Herschel\/}-SPIRE sources in the field, as well as four bright SCUBA-2 sources. We find 15 emission lines in 13 galaxies, and using existing photometry information, we determined the spectroscopic redshift of all 13 galaxies. Eleven of these galaxies are SPIRE-selected and lie in two structures at $\langle z \rangle\,{=}\,1.53$ and $\langle z \rangle\,{=}\,2.31$ (both with a standard deviation in redshift of 0.02), while the two SCUBA-2-selected galaxies are at $z\,{=}\,2.61$. Using multi-wavelength photometry, we constrained stellar masses and star formation rates, and using the CO and C{\sc i} emission lines we constrained gas masses. Our protocluster galaxies exhibit typical depletion timescales ($M_{\rm gas}$/SFR) for field galaxies at the same redshifts but higher gas-to-stellar mass ratios, potentially driven by emission line selection effects. The two structures confirmed in our survey are reproduced in cosmological simulations of star-forming halos at high redshifts; the simulated halos have a 60--70\% probability of collapsing into galaxy clusters, implying that the two structures in G073 are genuinely protoclusters.}

   \keywords{large-scale structure of Universe --
             Submillimetre: galaxies --
             Galaxies: star formation --
             Galaxies: clusters: general
               }
   \titlerunning{PLCK\,G073.4$−$57.5: two protoclusters}
   \authorrunning{Hill et al.}

   \maketitle
%

\section{Introduction}
\label{introduction}

The Universe on the largest scales forms a `cosmic web', with galaxy clusters at the intersections of the filaments \citep{Bond1996}. These are the largest gravitationally bound objects in the Universe, and we believe that they play an important role in the evolution of the galaxies embedded within them because in the local Universe, cluster galaxies are more likely to be massive, red ellipticals with old stellar populations \citep[e.g.,][]{ellis1997,andreon2003,muzzin2012}. These observations effectively point to a process that primarily operates in galaxy clusters, where star formation produces massive elliptical galaxies early on, followed by a quenching phase.

Understanding how the large-scale environment led to a different star formation process that resulted in such a stark differentiation between cluster galaxies and field galaxies is an open question. Many models have been put forward; on the one hand, galaxies could experience accelerated growth through increased merger rates in overdense environments \citep[e.g.,][]{kauffmann1996,gottlober2001,fakhouri2009}, while on the other hand gas could be removed from galaxies as they travel through the intracluster medium due to ram pressure stripping \citep[e.g.,][]{gunn1972,gavazzi2001,boselli2019}, or galaxies might be unable to accrete the new gas required to form more stars due to the hot intracluster medium temperatures present in clusters \citep[e.g.,][]{larson1980,balogh2000,peng2015}. To understand the contributions of these (and other) scenarios to galaxy evolution requires observations of cluster galaxies well before the cluster has fully virialized, which means finding examples of galaxy clusters early in their formation phase (i.e. protoclusters).

Protoclusters are now common in the literature, having been found in large Lyman-$\alpha$ surveys \citep[e.g.,][]{steidel2000,chiang2015,jiang2018}, or around rare and extremely massive galaxies such as quasars \citep[e.g.,][]{dannerbauer2014,noirot2018,hennawi2015} and submillimetre (submm) galaxies (SMGs; e.g., \citealt{chapman2009,casey2015,oteo2018}). However, such a diversity of selection techniques means that it is impossible to compare different protoclusters with one another and draw conclusions about overall populations.

Nonetheless, thanks to cosmic microwave background (CMB) experiments large, uniform samples of protoclusters are being compiled. CMB surveys operate at submm and millimetre wavelengths over very large areas, inevitably detecting the brightest extragalactic foreground objects (relative to the CMB), many of which are now known to be protoclusters \citep[e.g.,][]{flores-cacho2015,miller2018,koyama2021,polletta21}. In particular, {\it Planck} has identified over 2000 extragalactic objects with spectral energy distributions (SEDs) peaking between 350 and 850\,$\mu$m, meaning that the emission is most likely thermal and originating from copious amounts of dust produced by elevated star-formation rates (SFRs) in galaxies shortly before they quench; these objects are listed in the {\it Planck} High-$z$ (PHz) catalogue \citep{planckXXXIX}. 

PLCK\,G073.4$-$57.5 (hereafter G073) is one such object. While G073 is not in the final PHz catalogue due to being located just within the final Milky Way mask, it was part of a preliminary catalogue (with a less conservative mask) and was followed up with the {\it Herschel\/} Spectral and Photometric Imaging REceiver \citep[SPIRE;][]{planckXXVII} and the Submillimetre Common-User Bolometer Array 2 \citep[SCUBA-2;][]{mackenzie2017}. Following these observations, the Atacama Large Millimeter/submillimeter Array (ALMA) was used to observe eight red SPIRE sources in Band~6 (1\,mm), detecting 18 galaxies in the continuum \citep{kneissl19}. While 18 millimetre-bright galaxies is already 8--30 times higher than expected from a random field distribution, the distribution of photometric redshifts suggests that the field contained overdensities around $z\,{\approx}\,1.5$ and $z\,{\approx}\,2$ (although with large uncertainties), while two bright lines were also serendipitously detected at the edge of one of the sidebands and later identified as CO(5--4) at $z\,{=}\,1.5$, coinciding with the first peak in the photometric redshift distribution.

G073 therefore presents an opportunity to study overdense environments at a crucial epoch of cluster and galaxy formation. Around $z\,{=}\,$2, galaxy clusters are beginning to virialize, and cluster galaxies often still have elevated SFRs compared to field galaxies \citep[e.g.,][]{hilton10,hayashi11,brodwin13,tran15,alberts16,nantais17,mei23}. For most of these systems, the star formation activity is significant down to the dense core region \citep[e.g.,][]{hilton10,hayashi10,tran10,fassbender11,tadaki12}. Since star formation will often heavily obscure a galaxy in dust, these key populations can be missed by optical observations, but long-wavelength observations of molecular gas lines (primarily CO) are completely unaffected. Furthermore, since the molecular gas is the main fuel for star formation, such observations can be used to trace gas masses and provide insight into the evolutionary state of cluster and protocluster galaxies. A small number of (proto)cluster galaxies have been observed in CO around $z\,{=}\,$2, with some results suggesting higher gas fractions ($M_{\rm gas}/M_{\rm star}$) and shorter depletion timescales ($M_{\rm gas}/$SFR) compared to field galaxies \citep[e.g.,][]{noble17,hayashi18}, while others do not find such a difference \citep[e.g.,][]{aravena12,rudnick17}.

The main issue to overcome is that our current observations are limited to only a few inhomogeneously selected clusters and protoclusters. The PHz catalogue of over 2000 (uniformly selected) protocluster candidates thus offers the opportunity to find statistically significant results with simple and understandable observational biases. To this end, we present a new ALMA survey of G073, searching for more $z\,{=}\,1.5$ galaxies and spectroscopically confirming the redshift of the $z\,{=}\,2$ structure. Ultimately, a full characterization of this object will help us understand how to carry out future studies of large samples of PHz objects. 

This paper is arranged as follows.
In Sect.~\ref{observations} we describe our new ALMA observations and all ancillary data, in Sect.~\ref{analysis} we outline our data analysis pipeline, and in Sect.~\ref{sec:sed_fitting} we discuss how we derived physical properties. Section~\ref{sec:results} presents our results, which we interpret in Sect.~\ref{discussion}. The paper concludes in Sect.~\ref{sec:conclusion}. Throughout this work we adopt a \citet{chabrier03} initial mass function (IMF) and a flat $\Lambda$ cold dark matter ($\Lambda$CDM) model with cosmological parameters $\Omega_{\Lambda}\,=\,0.685$, $\Omega_{\mathrm{M}}\,=\,0.315$, and $H_{{0}}\,=\,67.4$\,\kms\,Mpc$^{-1}$ \citep{planck2018I}. 


\section{Observations}
\label{observations}

\subsection{Target selection}

\begin{figure*}[htbp!]
\includegraphics[width=\textwidth]{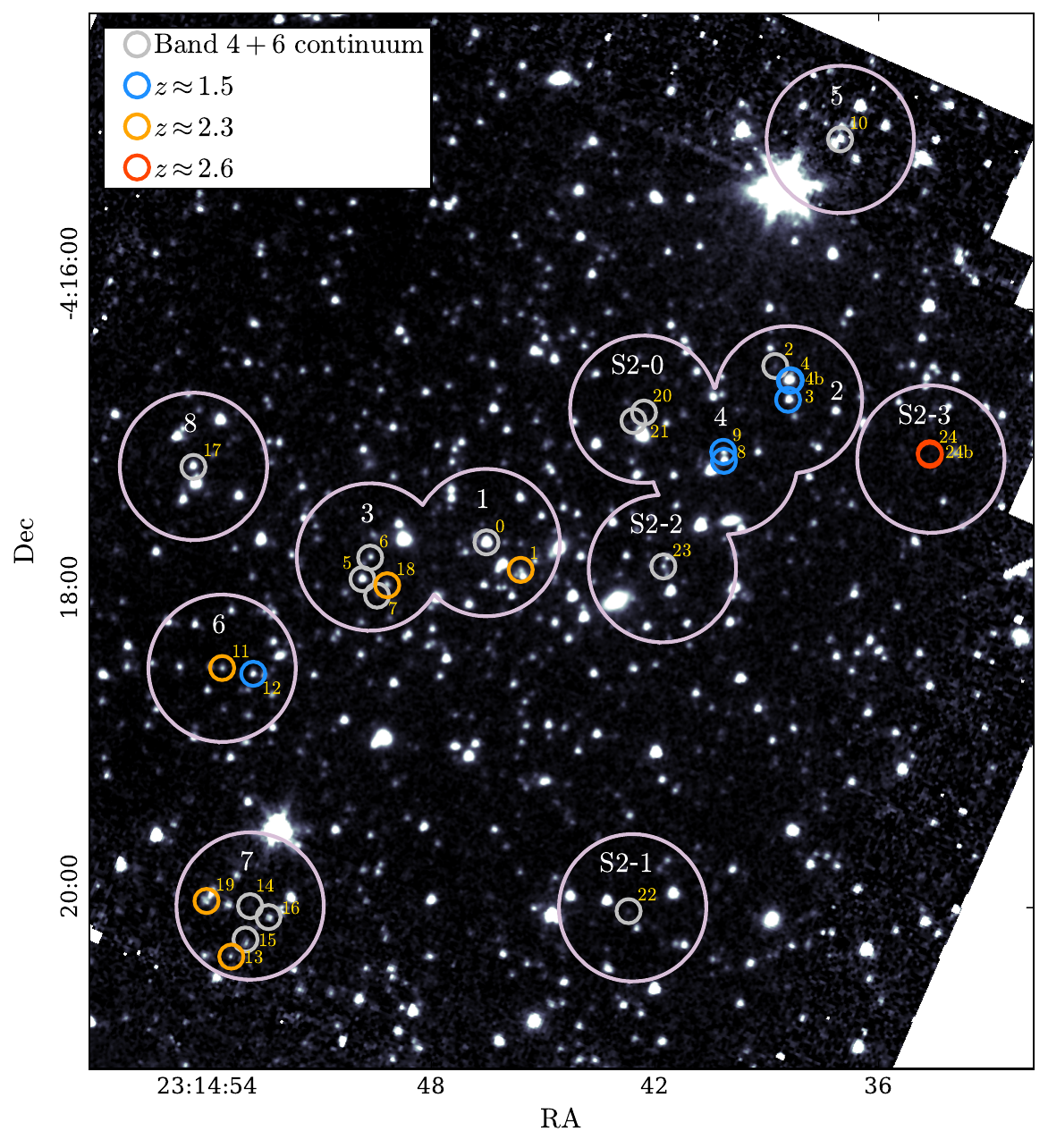}
\caption{Field of G073.4$-$57.5. In the background we show {\it Spitzer} 3.6-$\mu$m imaging, and the purple circles represent ALMA Band~4 follow-up pointings taken in Cycle~7. The ALMA pointings labelled 1 through 8 were also observed by ALMA in Band~6 in Cycle~2; these pointings were selected as bright, unresolved {\it Herschel\/-}SPIRE sources with red colours \citep[see][]{kneissl19}. The ALMA pointings labelled S2-0 through S2-3 were selected as the highest S/N SCUBA-2 850\,$\mu$m sources \citep[see][]{mackenzie2017}. Galaxies with only continuum detections in at least one of the ALMA Band~4 and 6 data are indicated by grey circles, while the galaxies in the $z\,{\approx}\,1.5$ structure are shown as blue circles and galaxies in the $z\,{\approx}\,2.3$ structure as orange circles. The two SCUBA-2-selected galaxies for which we detected line emission are shown as red circles, tentatively assigned to $z\,{=}\,2.6$ (see Sect.~\ref{analysis}).}
\label{all_field}
\end{figure*}

Our ALMA Band~4 survey targeted the eight red SPIRE sources (in eight separate pointings) previously followed up by ALMA in Cycle~2 \citep{kneissl19}. These eight targets form a complete sample of SPIRE sources found within the {\it Planck} beam and with colours satisfying 250, 350 and 500\,$\mu$m flux density ratios $S_{350}\,{/}\,S_{250}\,{>}\,0.7$ and $S_{500}\,{/}\,S_{350}\,{>}\,0.6$, which is expected to filter out low-$z$ interlopers. Since the ALMA primary beam is larger at 2\,mm than at 1\,mm, our observations covered all 18 galaxies detected by \citet{kneissl19}. In addition to these eight SPIRE sources, we targeted the four highest signal-to-noise ratio (S/N) sources detected by SCUBA-2 at 850\,$\mu$m (again within the {\it Planck} beam; see \citealt{mackenzie2017}), bringing the total number of ALMA pointings to 12. The distribution of these pointings is shown in Fig.~\ref{all_field}.

\subsection{Cycle~6 ALMA observations}

We received in total 3.23 hours of on-source observing time on G073 with ALMA in Cycle~6 (Project ID 2018.1.00562.S, PI R. Hill), taken in four executions on 2018 4, 8, 11, and 12 November. \COfive\ and \COtwo\ were previously confirmed in galaxy ID 3 at $z\,{=}\,1.5423$, so we tuned our Band~4 observations to centre the redshifted CO(3--2) line around 136\,GHz in the lower sideband (LSB). The upper sideband (USB) was centred around 149\,GHz in order to search for C{\sc i}(1--0) in the $z\,{\approx}\,2$ structure and to measure dust continuum at 2.1\,mm.

In the LSB data, the redshift range covered by \COthree\ was 1.49--1.56, while in the USB data C{\sc i}(1--0) covered a redshift range of 2.23--2.34. The LSB could also contain \COfour\ between $z\,{=}\,2.32$ and $z\,{=}\,2.41$, making a single line detection degenerate between the two expected structures; these situations are discussed on a case-by-case basis below. There are other potential contaminating lines such as \COtwo\ at $z\,{<}\,0.7$ and higher-$J$ lines at $z\,{>}\,2.8$, which are strongly disfavoured given the photometric redshift distribution from \citet{kneissl19}. There could additionally be C{\sc i}(1--0) in the LSB around $z\,{\approx}\,2.6$, \COthree\ in the USB around $z\,{\approx}\,1.3$, and \COfour\ in the USB around $z\,{\approx}\,2.1$, but since multiple lines are detected in both the $z\,{\approx}\,1.5$ and $z\,{\approx}\,2$ structures (as shown below), we assumed that there are only two redshift possibilities for all of the lines we detected. However, for the SCUBA-2 targets we considered all possible redshifts because the original {\it Planck\/} selection would not have included objects bright at 850\,$\mu$m.

The Band~4 spectral set-up used four 1.78-GHz-wide continuum spectral windows (with channel widths of 15.6\,MHz) around the central frequency of 143.1\,GHz (2.1 mm). The spectral windows were divided into the two receiver sidebands, separated by 12\,GHz (i.e., their central frequencies were 136.1, 138.1, 148.1, and 150.1\,GHz).
45--48 antennas were available in the nominal C-5 array configuration with baseline lengths of 15--1400\,m. This achieved synthesized beam sizes (using natural weighting) of around 0.65\arcsec\,${\times}$\,0.55\arcsec full width at half maximum (FWHM) in the lower side band and 0.58\arcsec\,${\times}$\,0.50\arcsec in the USB.

The observatory standard calibration was used. J0006$-$0623, a grid-monitoring source, was the bandpass and flux calibrator, with a flux density of 1.58\,Jy at the central frequency. All pointings in the datasets shared the same phase calibrator, J2323$-$0317. The central continuum sensitivity in each of the 12 pointings was approximately 15--20\,$\mu$Jy\,beam$^{-1}$. 
The reduction was based on the calibration provided by the ALMA Pipeline using standard \texttt{CASA} tasks \citep{2022PASP..134k4501C}, using natural weighting and pixel sizes of 0.077\arcsec.

\subsection{Cycle~2 ALMA observations}

We made use of existing ALMA observations of G073 taken in Cycle~2. These observations are presented in detail in \citet{kneissl19}; briefly, G073 was observed for 0.4 hours in Band~6 using the standard Band~6 tuning centred at about 1.3\,mm. A total of eight pointings were carried out, targeting the same eight {\it Herschel\/-}SPIRE-detected sources described above (see Fig.~\ref{all_field}). Based on these initial observations, a total of 18 individual galaxies were detected in the continuum, while line emission was detected in two of the 18 galaxies (IDs 3 and 8).

\subsection{Ancillary data}

G073 has been observed in the near-infrared (NIR; 1.25\,$\mu$m [$J$] and 2.15\,$\mu$m [$K_{\rm s}$]) with the Wide-field Infrared Camera (WIRCam) on the Canada France Hawaii Telescope and in the mid-infrared (MIR; 3.6\,$\mu$m and 4.5\,$\mu$m) with the Infrared Array Camera (IRAC) on the {\it Spitzer} Space Telescope. The observations and the photometric data have already been presented and described in~\citet{kneissl19}.  The NIR and MIR counterparts and fluxes for two ALMA sources, IDs 4 and 15, have been revised because they were wrongly assigned to a foreground galaxy located about 1.1\arcsec and 1.6\arcsec from the millimetre-emitting sources, respectively. Since ID 4 is much fainter than the nearby source, it is not possible to obtain reliable flux estimates for it. This is also true at submm/millimetre wavelengths where the SED is consistent with a low redshift (i.e. $z\approx0.2$) source, suggesting that the nearby source dominates also at these wavelengths. The measured NIR and MIR flux densities $S_J$, $S_{K_{\rm s}}$, $S_{3.6}$, and $S_{4.5}$ are listed in Table~\ref{tab:ir_fluxes}.

Since multiple ALMA detections are often found within a single SPIRE source, we assigned de-blended submm flux densities as done in \citet{kneissl19}.  The redshifts used in the de-blending procedure are all consistent with the revised redshifts (see Sect.~\ref{analysis}), with the exception of ID 9 for which the revised redshift is smaller (i.e. $z\,{=}\,1.51$ compared to the previous value, $z\,{=}\,2.21$). Since this source is detected at 850\,$\mu$m and 1.2\,mm, and the de-blended flux density estimates are derived only for 350\,$\mu$m and 500\,$\mu$m (this source was not detected at 250\,$\mu$m), we assumed that its far-infrared (FIR) SED and estimated FIR luminosity are not significantly affected by the SPIRE flux densities derived from the de-blending procedure using a slightly higher redshift.

\section{Data analysis}
\label{analysis}

\subsection{CO and C{\sc i} line search for protocluster galaxies}

\addtolength{\tabcolsep}{-1.5pt}
\begin{table}[htbp!]
\caption{Source properties.}\label{table:cont}
\begin{center}
\begin{tabular}{lcccc}
\hline\hline
\noalign{\vskip 1pt}
Field & ID & RA Dec & $z$ & $S_{2100}$ \\
 &  & [J2000] &  & [mJy] \\
\hline
\noalign{\vskip 1pt}
1 & 0 & 23:14:46.52 $-$4:17:33.4 & \dots & $<$0.097 \\
  & 1 & 23:14:45.60 $-$4:17:44.5 & 2.3060 & 0.325$\pm$0.072 \\
\hline
\noalign{\vskip 1pt}
2 & 2 & 23:14:38.78 $-$4:16:22.6 & \dots & 0.225$\pm$0.050 \\
  & 3 & 23:14:38.43 $-$4:16:36.2 & 1.5424 & 0.122$\pm$0.059 \\
  & 4 & 23:14:38.36 $-$4:16:28.3 & 1.5390 & $<$0.042 \\
  & 4b & 23:14:38.38 $-$4:16:28.6 & 1.5494 & $<$0.034 \\
\hline
\noalign{\vskip 1pt}
3 & 18 & 23:14:49.18 $-$4:17:50.6 & 2.3370 & $<$0.098 \\
  & 5 & 23:14:49.85 $-$4:17:48.2 & \dots & 0.117$\pm$0.042 \\
  & 6 & 23:14:49.64 $-$4:17:39.5 & \dots & $<$0.041 \\
  & 7 & 23:14:49.45 $-$4:17:54.8 & \dots & $<$0.129 \\
\hline
\noalign{\vskip 1pt}
4 & 8 & 23:14:40.15 $-$4:17:00.7 & 1.5125 & 0.203$\pm$0.064 \\
  & 9 & 23:14:40.17 $-$4:16:57.1 & 1.5055 & $<$0.079 \\
\hline
\noalign{\vskip 1pt}
5 & 10 & 23:14:37.04 $-$4:14:51.7 & \dots & $<$0.100 \\
\hline
\noalign{\vskip 1pt}
6 & 11 & 23:14:53.61 $-$4:18:23.9 & 2.3058 & 0.166$\pm$0.042 \\
  & 12 & 23:14:52.78 $-$4:18:26.1 & 1.5065 & $<$0.073 \\
\hline
\noalign{\vskip 1pt}
7 & 13 & 23:14:53.36 $-$4:20:19.6 & 2.3395 & 0.129$\pm$0.064 \\
  & 14 & 23:14:52.86 $-$4:19:59.4 & \dots & 0.108$\pm$0.038 \\
  & 15 & 23:14:52.98 $-$4:20:12.8 & \dots & $<$0.049 \\
  & 16 & 23:14:52.37 $-$4:20:03.8 & \dots & $<$0.044 \\
  & 19 & 23:14:54.04 $-$4:19:57.3 & 2.2762 & $<$0.055 \\
\hline
\noalign{\vskip 1pt}
8 & 17 & 23:14:54.38 $-$4:17:03.1 & \dots & 0.091$\pm$0.035 \\
\hline
\noalign{\vskip 1pt}
S2-0 & 20 & 23:14:42.29 $-$4:16:41.3 & \dots & 0.502$\pm$0.062 \\
  & 21 & 23:14:42.58 $-$4:16:44.7 & \dots & 0.421$\pm$0.067 \\
\hline
\noalign{\vskip 1pt}
S2-1 & 22 & 23:14:42.71 $-$4:20:01.4 & \dots & 1.326$\pm$0.098 \\
\hline
\noalign{\vskip 1pt}
S2-2 & 23 & 23:14:41.77 $-$4:17:43.1 & \dots & 0.174$\pm$0.053 \\
\hline
\noalign{\vskip 1pt}
S2-3 & 24 & 23:14:34.63 $-$4:16:57.6 & 2.6127 & 0.377$\pm$0.050 \\
  & 24b & 23:14:34.63 $-$4:16:58.3 & 2.6337 & $<$0.061 \\
\hline
\end{tabular}\\
\end{center}
\end{table}
\addtolength{\tabcolsep}{1.5pt}

\begin{table*}[htbp!]
\caption{Properties of line detections.}\label{table:line}
\begin{center}
\begin{tabular}{lcccccc}
\hline\hline
\noalign{\vskip 1pt}
Field & ID & Line & $F$ & $L$ & $L^{\prime}$ & FWHM \\
 &  &  & [Jy\,km\,s$^{-1}$] & [10$^{6}$\,L$_{\odot}$] & [10$^{9}$\,K\,km\,s$^{-1}$\,pc$^2$] & [km\,s$^{-1}$] \\
\hline
\noalign{\vskip 1pt}
1 & 1 & C{\sc i}(1--0) & 1.22$\pm$0.26 & \pz68$\pm$15 & 17.8$\pm$3.9 & 580$\pm$100 \\
\hline
\noalign{\vskip 1pt}
2 & 3 & $\mathrm{CO(3}$-$\mathrm{2)}$ & 2.38$\pm$0.10 & 45.6$\pm$1.9 & 34.4$\pm$1.4 & 420$\pm$17\pz \\
  & 4 & $\mathrm{CO(3}$-$\mathrm{2)}$ & 0.57$\pm$0.07 & 10.8$\pm$1.4 & \pz8.2$\pm$1.0 & 480$\pm$60\pz \\
  & 4b & $\mathrm{CO(3}$-$\mathrm{2)}$ & 0.34$\pm$0.08 & \pz6.5$\pm$1.6 & \pz4.9$\pm$1.2 & 720$\pm$180 \\
\hline
\noalign{\vskip 1pt}
3 & 18 & $\mathrm{CO(4}$-$\mathrm{3)}$ & 0.33$\pm$0.04 & 17.9$\pm$2.3 & \pz5.7$\pm$0.7 & 146$\pm$18\pz \\
 &  & C{\sc i}(1--0) & 0.20$\pm$0.06 & 11.3$\pm$3.7 & \pz3.0$\pm$1.0 & 230$\pm$70\pz \\
\hline
\noalign{\vskip 1pt}
4 & 8 & $\mathrm{CO(3}$-$\mathrm{2)}$ & 1.94$\pm$0.11 & 35.9$\pm$2.1 & 27.1$\pm$1.6 & 720$\pm$50\pz \\
  & 9 & $\mathrm{CO(3}$-$\mathrm{2)}$ & 0.35$\pm$0.06 & \pz6.4$\pm$1.1 & \pz4.8$\pm$0.8 & 320$\pm$50\pz \\
\hline
\noalign{\vskip 1pt}
6 & 11 & C{\sc i}(1--0) & 0.51$\pm$0.15 & 28.6$\pm$8.6 & \pz7.5$\pm$2.3 & 670$\pm$160 \\
  & 12 & $\mathrm{CO(3}$-$\mathrm{2)}$ & 1.28$\pm$0.12 & 23.4$\pm$2.1 & 17.7$\pm$1.6 & 570$\pm$50\pz \\
\hline
\noalign{\vskip 1pt}
7 & 13 & $\mathrm{CO(4}$-$\mathrm{3)}$ & 0.52$\pm$0.08 & 27.9$\pm$4.1 & \pz8.9$\pm$1.3 & 218$\pm$32\pz \\
 &  & C{\sc i}(1--0) & 0.15$\pm$0.08 & \pz8.7$\pm$4.5 & \pz2.3$\pm$1.2 & 140$\pm$40\pz \\
  & 19 & C{\sc i}(1--0) & 0.99$\pm$0.12 & 54.2$\pm$6.3 & 14.2$\pm$1.6 & 340$\pm$50\pz \\
\hline
\noalign{\vskip 1pt}
S2-3 & 24 & C{\sc i}(1--0) & 0.49$\pm$0.11 & 33.6$\pm$7.6 & \pz8.8$\pm$2.0 & 470$\pm$100 \\
  & 24b & C{\sc i}(1--0) & 0.32$\pm$0.06 & 22.1$\pm$3.9 & \pz5.8$\pm$1.0 & 330$\pm$60\pz \\
\hline
\end{tabular}\\
\end{center}
\end{table*}

To conduct our line search in both the LSB and USB of our ALMA observations, we first used the publicly available tool {\tt LineSeeker} \citep{gonzalez-lopez2017,gonzalez-lopez2019}. This takes as input a primary beam-uncorrected data cube, and convolves the cube along the spectral axis with Gaussians of varying width, searching for significant peaks. The noise per channel is estimated iteratively by computing the standard deviation of all the pixels in a given channel, then recomputing the standard deviation of all the pixels whose absolute values are lower than 5 times the initial standard deviation. After convolution with a Gaussian of a given size, pixels whose spectra show peak S/N values above a chosen threshold are then returned to the user, and the procedure is repeated to search for pixels containing negative S/N peaks.

We ran {\tt LineSeeker} on all primary beam-uncorrected LSB and USB data cubes; for reference, the solid angle of each cube is about 0.76\arcmin, while the bandwidth is about 3.5\,GHz, corresponding to a velocity range of about 8000\,km\,s$^{-1}$. We set the maximum width of the spectral convolution kernel to be 1000\,km\,s$^{-1}$, which sets an upper limit to the width of line emission features in our search.

From the catalogues of positive and negative peaks returned by {\tt LineSeeker}, we found that the most significant negative peak across all of the data cubes had an S/N of 6.1; thus, we used this as our cutoff for identifying real positive peaks. There were a total of eight unique positive peaks in the LSB data with S/N values greater than 6.1; six of these peaks are spatially coincident with Band~6 continuum-detected galaxies from \citet{kneissl19}, one peak is located in a field previously observed by ALMA but at a position with no continuum counterpart (ID 18), and one peak is in the S2-3 pointing (ID 24; see Fig.~\ref{all_field}). Similarly, we found three unique positive peaks in the USB data above the S/N threshold, with two peaks coincident with a Band~6 continuum-detected galaxy, and one peak with no previously known counterpart (ID 19).

For completeness, we also placed 1.5$^{\prime\prime}$ apertures around the previously detected galaxies in G073 from \citet{kneissl19} with no line detections from {\tt LineSeeker} -- this aperture size was chosen to be slightly smaller than the average aperture size constructed for the detected sources, as described in Sect.~\ref{sec:measurements}. Since for these galaxies we have prior knowledge of their positions, we systematically searched their spectra for fainter peaks down to 5$\sigma$ (without any Gaussian convolution), but this did not return any additional lines.

\subsection{ALMA Band~4 continuum search}

We also searched our primary beam-uncorrected data cubes for previously undiscovered (i.e., not reported in \citealt{kneissl19}) submm continuum sources. To do this, we calculated the noise-weighted mean of each pointing (LSB+USB), where the noise per channel was estimated as the standard deviation of all the pixels after masking out the 18 known galaxies from \citet{kneissl19}, and obviously bright galaxies in our SCUBA-2 pointings. We searched each continuum map for both positive and negative S/N peaks, finding that the most significant negative peak across all 12 maps was 5.1. Following our line search, we set this as our continuum search threshold. This search did not yield any new detections in the pointings centred on the positions of the eight previous ALMA observations (fields 1 through 8), but in the four new pointings centred on SCUBA-2 sources, we detected five sources (one in fields S2-1, S2-2, and S2-3, and two in S2-0).

\subsection{Redshifts, line measurements and continuum measurements}\label{sec:measurements}

Our photometric catalogue of submm-detected galaxies in G073 consists of 11 galaxies with line detections, plus an additional 14 continuum-detected galaxies (detected in Band~4 or in Band~6) that may still be at a protocluster redshift, but whose line emission was too faint to be detected in our exposure. Here we outline the various Band~4 measurements made on this catalogue.

First, for each position in our catalogue, we created 3\arcsec$\,{\times}\,3$\arcsec continuum-subtracted cutouts in our Band~4 primary beam-corrected data cubes using the {\tt CASA} task {\tt imcontsub}. For the continuum subtraction, at each pixel we fit a zeroth-order polynomial (i.e. a constant) across frequency space to the channels containing no line emission, and subtracted the fit.

Next, we generated LSB and USB spectra for each galaxy using the continuum-subtracted data cubes by manually placing elliptical apertures at the positions returned by either our line search or our continuum search, and summing the pixels within the aperture. The radius and ellipticity of the apertures are set by the size of the 2$\sigma$ contours in each cutout after calculating the average over all available channels. For reference, the average aperture semi-major axis size was found to be 1.6$^{\prime\prime}$. For sources 4 (located in field 2) and 24 (located in field S2-3), two significant emission lines were detected in the spectrum. For both cases, averaging over the channels corresponding to each line, showed that the two moment-0 maps were spatially offset by about 0.5\arcsec, with one peak spatially coincident with the bright continuum source and the other having no continuum counterpart. We checked that there were no pairs of millimetre lines able to explain the two peaks assuming they came from the same galaxy (e.g. CO(7--6) and [C{\sc ii}](2--1) around $z\,{\approx}\,5$). We therefore classify the two secondary sources as companion galaxies (designated 4b and 24b), bringing the total number of line detections to 13. The resulting spectra are provided in Appendix~\ref{appendix0}. 

As discussed above, we assumed that there are two possible CO transitions for the lines detected in the LSB, while the lines in the USB are C{\sc i}(1--0) at $z\,{\approx}\,2.3$. To establish which CO transition we observed, we note that for lines detected between about 137.8 and 139.1\,GHz we would expect \COfour\ in the LSB and C{\sc i}(1--0) in the USB; this is the case for five sources (IDs 8, 9, 12, 13, and 18). For these sources, while C{\sc i}(1--0) was not detected in the USB with S/N$\,{>}\,6.1$, lines are still possible at a lower significance. We note that \COfour\ typically has a peak flux density brighter by a factor of 2 compared to C{\sc i}(1--0) in SMGs \citep[e.g.,][]{birkin2021,hagimoto2023}, while for each of these sources the line detected in the LSB has a peak flux density about 3--5 times the spectrum rms.

We therefore performed a likelihood ratio test. We first fit a single Gaussian with three free parameters (amplitude, mean, and standard deviation) to the LSB line and calculate the likelihood function (here given by $\mathcal{L}\,{=}\,\exp(-\chi^2/2)$). We then performed a second fit for two Gaussians, where only the amplitude of the second Gaussian is a free parameter as the mean is fixed to the expected frequency of C{\sc i}(1--0) and the standard deviation is fixed to be equal to the first Gaussian. We recalculated the likelihood function and then calculated $\Delta \mathcal{L}\,{=}\,2(\log \mathcal{L}_2 - \log \mathcal{L}_1)$. The difference in degrees of freedom between the models is 1, so $\Delta \mathcal{L}$ is expected to follow a $\chi^2$ distribution with 1 degree of freedom and the relevant statistic ($t$) is the $\chi^2$ survival function (or 1 minus the cumulative distribution function) evaluated at $\Delta \mathcal{L}$. This statistic can be interpreted as the probability that the null hypothesis (in this case the model with one Gaussian) describes the data better than the alternative, and so we applied a threshold of 0.05 below which we rejected the null hypothesis.

We find that we cannot reject the null hypothesis for IDs 8, 9, and 12 ($t\,{=}\,0.77$, $t\,{=}\,1.00$, and $t\,{=}\,0.99$, respectively); thus, we interpret the line emission as \COthree. For IDs 13 and 18 we reject the null hypothesis with ${>}\,$95\% confidence ($t\,{=}\,0.038$ and $t\,{=}\,2.3\,{\times}\,10^{-5}$, respectively), so we assumed that we have detected both \COthree\ and C{\sc i}(1--0). \citet{kneissl19} discussed a tentative (S/N$\,{\approx}\,3$) CO(5--4) line detection in ID 8 at $z\,{=}\,1.545$, but the CO(3--2) line we have identified with an S/N of about 17 places this galaxy at $z\,{=}\,1.513$ -- we conclude that the possible \COfive\ line was in fact a noise excursion. Comparing our other results to \citet{kneissl19}: the spectroscopic redshift of ID 1 is 2.306, compared to the photometric redshift of 2.42$\pm$0.15; 1.506 compared to 2.21$\pm$0.22 for ID 9; 2.306 compared to 2.43$\pm$0.29 for ID 11; 1.507 compared to 1.40$\pm$0.10 for ID 12; and 2.340 compared to 2.63$\pm$0.25 for ID 13. Finally, due to the revised NIR and MIR counterparts for ID 4, the photometric redshift was not reliable and cannot be compared to our spectroscopic redshift.

For IDs 4 and 24 (plus their companion galaxies) the line emission is observed at a frequency where we cannot perform this test. ID 4 (and its companion 4b) happens to lie behind a large $z\,{<}\,0.5$ foreground galaxy and therefore we do not have good NIR continuum measurements; thus, we just assumed that it is at the same redshift as ID 3. For ID 24 we fit the available NIR and mm continuum photometry to a range of possible redshifts assuming the detected line is \COthree, \COfour, and C{\sc i}(1--0), then calculate $\chi^2$ for each fit. We find that a redshift of 2.6, corresponding to the C{\sc i}(1--0) line, minimizes $\chi^2$, so we adopted this as the redshift of this galaxy and its companion. Lastly, the line detected at the location of ID 3 is unambiguously \COthree, since two other CO lines have previously been detected. 

For the 13 galaxies with line emission, we fit single Gaussian functions to the observed lines in order to estimate various line properties. From the fit, line strengths are estimated by summing all channels within ${\pm}3\sigma$ of the mean of the fit, while the line FWHM are taken directly from the standard deviation of the fit, and redshifts are calculated using the mean of each fit. The line luminosities are calculated in $L^{\prime}$ units following~\citet{solomon97} as

\begin{equation}
L'_{\rm line} = \frac{c^2}{2 k} S \left( \Delta V\right) \nu^{-2}_{\rm obs} D^2_{\rm L} (1+z)^{-3},
\end{equation}
\noindent

where $S \left( \Delta V\right)$ is the line intensity derived from the Gaussian fit, $\nu_{\rm obs}$ the observed frequency of the line (see Table~\ref{table:cont}) and $D_{\rm L}$ is the usual cosmological luminosity distance. We also calculated the line luminosities in solar luminosity units following the same integration range.

Following this, continuum flux densities are calculated using the same apertures by averaging over all of the channels outside the ${\pm}3\sigma$ range, or for galaxies with no line emission detected, by averaging over all of the available Band~4 channels. Since many of the galaxies in our catalogue come from detections in Band~6 data, we are effectively performing forced aperture photometry at these positions of interest; therefore, we set a low continuum flux density S/N threshold of 2 for our continuum detections. We find that we are able to measure Band~4 continuum flux densities above this level in eight out of the 18 galaxies reported \citet{kneissl19}, with an additional five continuum flux density detections coming from new galaxies.

The results are summarized in Tables~\ref{table:cont} and \ref{table:line}. In Appendix~\ref{appendix0} we provide spectra and moment-0 maps (where lines are detected) for all of the galaxies in the sample.
                                       
\section{Physical properties of protocluster galaxies}\label{sec:sed_fitting}

\subsection{Stellar masses and star-formation rates}\label{sec:main_sequence}

We estimated the stellar mass of the line-detected sources by modelling their SEDs with the Code Investigating GAlaxy
Emission ~\citep[\texttt{CIGALE};][]{boquien19}.  This code models galaxy SEDs using stellar and dust components, and offers the advantage of modelling simultaneously the NIR part of the SED and the submm/millimetre emission in a self-consistent way by preserving the energy balance between the stellar radiation that is absorbed by the dust and its re-emission at longer wavelengths (see \citealt{buat19} for a discussion on this assumption).
 
\begin{figure*}[htbp!]
\centering
\includegraphics[width=\linewidth]{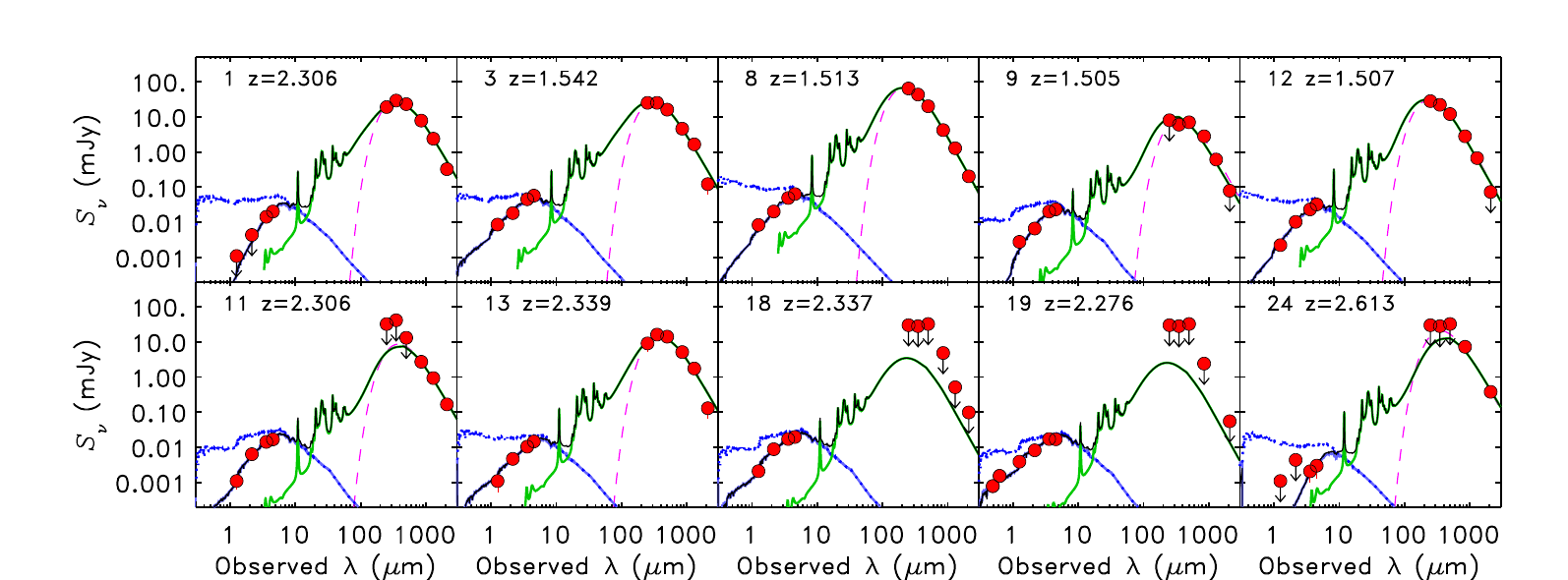}
\caption{{NIR-through-millimetre SEDs (CFHT/WIRCam, \spitzer/IRAC, \herschel/SPIRE, SCUBA-2, and ALMA at 1.3 and 2.2\,mm) of the sources with line detections (full red circles). Downward arrows are 5$\sigma$ upper limits. The \texttt{CIGALE} best-fit model is shown with a solid black line. The dotted blue line shows the stellar light before dust attenuation, and the solid blue line shows the attenuated stellar light. The green line shows the dust component. The best-fit modified blackbody model to the submm/millimetre data is shown with a dashed magenta line. The source identifier is annotated and listed in Table~\ref{tab:properties}. }}
\label{fig:cigale_sed}
\end{figure*}

For the reference model, we assumed a delayed star-formation history with an optional burst, a \citet{chabrier03} IMF, the stellar population models of~\citet{bruzual03}, the dust models from \citet{draine14}, the \citet{calzetti00} attenuation law, and solar metallicity ($Z$\,=\,0.02).  The best-fit is assessed through $\chi^2$, and the best-fit parameters and associated uncertainties are estimated as the likelihood-weighted means and standard deviations.  The \texttt{CIGALE} best-fit models, available for seven out of nine CO-emitters with available NIR photometry, are shown in Fig.~\ref{fig:cigale_sed}.  The stellar masses and SFRs (instantaneous, SFR$_{\rm inst}$, and averaged over the past 100\,Myrs, SFR$_{100}$) derived from the \texttt{CIGALE} best-fits are reported in Table~\ref{tab:properties}. 

We also derived SFR$_{\rm IR}$, the SFR from the total infrared luminosity ($L_{\rm IR}$) assuming the relation in \citet{kennicutt12} corrected for a Chabrier IMF (i.e. ${\rm SFR_{\rm IR}}\,=\,1.40\times 10^{-10}\times L_{\rm IR}$; see Table~\ref{tab:properties}). The infrared luminosity was obtained from fitting the submm/millimetre data with a modified blackbody using the \texttt{cmcirsed} package \citep{casey12c}, and assuming the CO-derived redshift and a dust emissivity-index $\beta$ equal to 1.8 \citep{cortese14,pokhrel16}, leaving the dust temperatures as free parameters. The resulting average dust temperature of the sample was found to be (27$\pm$3)\,K. Since SFR$_{\rm inst}$ and SFR$_{100}$ can differ significantly, by up to a factor of 3, we  use SFR$_{\rm IR}$ in the following because it is usually intermediate between the two values provided by the \texttt{CIGALE} best fit. In addition, it is directly comparable with the values reported in the literature where it is commonly used, in particular for millimetre-selected galaxies. Compared to SFR$_{\rm IR}$, SFR$_{\rm inst}$ can be twice as high, and SFR$_{100}$ four times smaller. We note that \citet{kneissl19} estimated the SFRs for the same galaxy sample using the same \texttt{cmcirsed} package; however, they lacked spectroscopic redshifts and the additional 2-mm flux densities ($S_{2000}$). Nonetheless, we find comparable results. We also tested running \texttt{CIGALE} with the \texttt{cmcirsed} dust model, which includes shorter wavelength data. This resulted in similar or slightly higher SFRs, owing to the fact that the model includes a power-law component in the MIR that is not used when fitting the FIR/millimetre data alone; however, the differences were within the uncertainties.

\begin{figure}[htbp!]
\centering
\includegraphics[width=\linewidth]{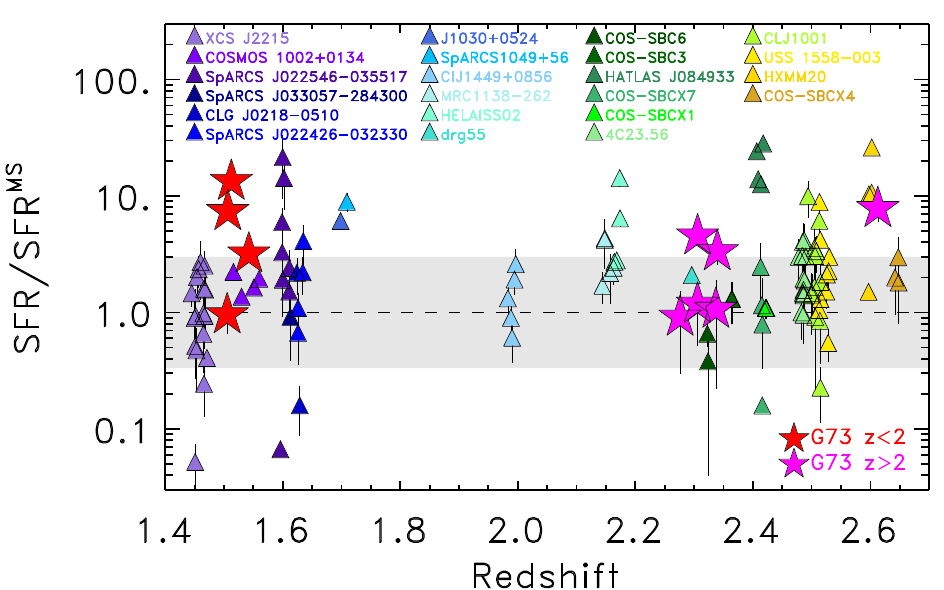}
\caption{Offset from the MS of star formation \citep{popesso23} as a function of redshift. The grey region represents 0.5\,dex scatter around the MS. Filled stars show the CO-emitters in G073 with a stellar mass estimate (red are those with $z\,{<}\,2$ and magenta those with $z\,{>}\,2$). Coloured triangles represent CO-detected cluster and protocluster members from the literature at $1.4\,{<}\,z\,{<}\,2.65$: XCS\,J2215 at $z=1.46$ from \citet{hayashi18}; COSMOS\,1002$+$0134 at $z=1.55$ from \citet{aravena12}; SpARCS\,J022546$-$035517 at $z\,{=}\,1.59$ from \citet{noble19}; SpARCS\,J033057$-$284300 at $z\,{=}\,1.613$ from \citet{noble17}; CLG\,J0218$-$0510 at $z\,{=}\,1.613$ from \citet{rudnick17}; SpARCS\,J022426$-$032330 at $z\,{=}\,1.613$ from \citet{noble17}; J1030$+$0524 at $z\,{=}\,1.694$ from \citet{damato20,damato21};
 SpARCS1049$+$56 at $z\,{=}\,1.710$ from \citet{webb15b};
 ClJ1449$+$0856 at $z\,{=}\,1.990$ from \citet{coogan18};
 MRC1138$-$262 at $z\,{=}\,2.160$ from \citet{dannerbauer17,tadaki19};
 HELAISS02 at $z\,{=}\,2.171$ from \citet{gomez19};
 DRG55 at $z\,{=}\,2.290$ from \citet{chapman15};
 COS$-$SBC6 at $z\,{=}\,2.323$;
 COS$-$SBC3 at $z\,{=}\,2.365$ from \citet{sillassen24};
 HATLAS\,J084933 at $z\,{=}\,2.410$ from \citet{ivison13};
 COS$-$SBCX7 at $z\,{=}\,2.415$;
 COS$-$SBCX1 at $z\,{=}\,2.422$ from \citet{sillassen24};
 4C23.56 at $z\,{=}\,2.490$ from \citet{tadaki19,lee17};
 CLJ1001 at $z\,{=}\,2.510$ from \citet{wang18};
 USS\,1558$-$003 at $z\,{=}\,2.530$ from \citet{tadaki14,tadaki19};
 HXMM20 at $z\,{=}\,2.602$ from \citet{gomez19};
 COS$-$SBCX4  at $z\,{=}\,2.646$ from \citet{sillassen24}.}
\label{fig:main_sequence}
\end{figure}

In Fig.~\ref{fig:main_sequence} we show the offset from the average star-forming main sequence (MS) as parameterized by \citet{popesso23} using each source's redshift and stellar mass. The grey shaded region corresponds to the scatter of 0.3\,dex around the MS. Following \citet{rodighiero11}, we classified a source as a starburst if the SFR is four times higher than expected according to the MS for a galaxy with the same stellar mass and at the same redshift, and as a normal star-forming galaxy (SFG) if the SFR is lower than this threshold. Four G073 sources (IDs 1, 8, 12, and 24) are considered starbursts based on the MS offset and the remaining six are consistent with the MS. In the figure we also show the position of CO-detected cluster and protocluster members at $1.40\,{<}\,z\,{<}\,2.65$ from the literature (see the caption for detailed references). Compared to other clusters at similar redshifts, the G073 sources are on average more star-forming, but sources with similarly high SFRs are observed in other structures.

\setlength\tabcolsep{2pt}
\begin{table*}[htbp!]
\caption{Physical properties derived from line detections and \texttt{CIGALE}.\label{tab:properties}}
\begin{center}
\begin{tabular}{cc cccc cc cc ccc}
\hline\hline
\noalign{\vskip 2pt}
ID  &    $z$     & $M_{\rm star}$ & SFR$_{\rm inst}$ & SFR$_{100}$ & SFR$_{\rm IR}$    & $M^{\rm CO}_{\rm gas}$ & $M^{\rm CI}_{\rm gas}$ & $\mu_{\rm gas}$ & $\tau_{\rm depl}$ & SFR/SFR$^{\rm MS}$ & $\tau_{\rm depl}/\tau^{\rm MS}_{\rm depl}$ & $\mu_{\rm gas}/\mu^{\rm MS}_{\rm gas}$ \\
   &            & (10$^{10}$\,\msun)&  \multicolumn{3}{c}{(\msun\,yr$^{-1}$)} & (10$^9$\,\msun) & (10$^9$\,\msun) &    & (Gyr) & & & \\
\noalign{\vskip 1pt}
\hline 
\noalign{\vskip 1pt}
 {\bf 1}  &   2.3060   &     8.9$\pm$3.6  &  609$\pm$ 84  &    280$\pm$48   &   442$^{+48}_{-42}$   &  \nodata   &  404$\pm$86  &    4.52$\pm$2.07   &   0.91$\pm$0.22  &   4.6$\pm$0.5   &    1.9$\pm$0.4  &  5.4$\pm$2.5  \\
    3  &   1.5424   &     7.8$\pm$0.5  &  301$\pm$ 21  &    172$\pm$ 9   &   174$^{+14}_{-14}$   & 237$\pm$10 &  \nodata     &    3.06$\pm$0.26   &   1.37$\pm$0.12  &   3.2$\pm$0.3   &    2.3$\pm$0.2  &  4.9$\pm$0.4  \\
    4  &   1.5390   &    \nodata  &  \nodata  &    \nodata   &   331$^{+25}_{-23}$   &  57$\pm$ 7 &  \nodata     &  \nodata           &  \nodata         &  \nodata        &  \nodata        & \nodata       \\
   4b  &   1.5494   &    \nodata  &  \nodata  &    \nodata  &   \nodata   &  34$\pm$ 8 &  \nodata     &  \nodata           &  \nodata         &  \nodata        &  \nodata        & \nodata       \\
   18  &   2.3370   &    11.6$\pm$1.7  &  116$\pm$ 92  &    102$\pm$32   &    \nodata   &  72$\pm$ 9 &   68$\pm$20  &    0.62$\pm$0.12   &   0.62$\pm$0.50  &   1.1$\pm$0.8   &    1.4$\pm$1.1  &  0.9$\pm$0.2  \\
{\bf 8}  &   1.5125   &     5.4$\pm$1.1  &  873$\pm$102  &    264$\pm$33   &   609$^{+63}_{-57}$   & 186$\pm$11 &  \nodata     &    3.48$\pm$0.75   &   0.31$\pm$0.03  &  13.4$\pm$1.3   &    0.5$\pm$0.1  &  4.6$\pm$1.0  \\
    9  &   1.5055   &     6.4$\pm$1.2  &   54$\pm$ 19  &     52$\pm$ 7   &    47$^{+16}_{-13}$   &  33$\pm$ 6 &  \nodata     &    0.52$\pm$0.13   &   0.72$\pm$0.26  &   1.0$\pm$0.3   &    1.1$\pm$0.4  &  0.8$\pm$0.2  \\
   11  &   2.3058   &    10.0$\pm$3.3  &  245$\pm$233  &    136$\pm$66   &   120$^{+84}_{-50}$   &  \nodata   &  169$\pm$50  &    1.69$\pm$0.76   &   1.41$\pm$0.90  &   1.2$\pm$0.7   &    3.0$\pm$1.9  &  2.1$\pm$1.0  \\
{\bf 12}   &   1.5065   &     2.7$\pm$0.7  &  400$\pm$ 51  &    131$\pm$22   &   240$^{+32}_{-29}$   & 122$\pm$11 &   51$\pm$27  &    4.44$\pm$1.19   &   0.51$\pm$0.08  &   7.4$\pm$0.9   &    0.6$\pm$0.1  &  4.1$\pm$1.1  \\
   13  &   2.3395   &     4.4$\pm$1.9  &  360$\pm$ 91  &    154$\pm$35   &   237$^{+25}_{-49}$   & 114$\pm$17 &  \nodata     &    2.59$\pm$1.18   &   0.48$\pm$0.13  &   3.3$\pm$0.8   &    0.8$\pm$0.2  &  2.0$\pm$0.9  \\
   19  &   2.2762   &     8.6$\pm$1.1  &   86$\pm$ 58  &     76$\pm$23   &     \nodata   &  \nodata   &  320$\pm$39  &    3.73$\pm$0.66   &   3.73$\pm$2.56  &   0.9$\pm$0.6   &    7.6$\pm$5.2  &  4.4$\pm$0.8  \\
{\bf 24}  &   2.6127   &     2.2$\pm$1.6  &  194$\pm$ 95  &     77$\pm$35   &   
  413$^{+108}_{-86}$  &  \nodata   &  201$\pm$45  &    9.23$\pm$7.10   &   0.49$\pm$0.16  &   7.8$\pm$1.8   &    0.6$\pm$0.2  &  4.1$\pm$3.2  \\
  24b  &   2.6337   &   \nodata  &  \nodata  &    \nodata   &    \nodata   &  \nodata   &  133$\pm$25  &  \nodata           &  \nodata         &  \nodata        &  \nodata        & \nodata       \\
\hline
\end{tabular}\\
\end{center}
\tablefoot{Source IDs in boldface are classified as starbursts based on the main-sequence offset. The stellar mass ($M_{\rm star}$), the instantaneous star-formation rate (SFR$_{\rm inst}$) and the SFR averaged over the past 100\,Myr (SFR$_{100}$) were obtained from the \texttt{CIGALE} SED best-fit model (see Sect.~\ref{sec:sed_fitting}). The \texttt{CIGALE} parameters are not available for IDs 4, 4b, and 24b because they are not detected in the NIR and MIR images. SFR$_{\rm IR}$ was derived from the total IR luminosity ($L_{\rm IR}$) assuming the relation in \citet{kennicutt12} corrected for a Chabrier IMF (i.e., ${\rm SFR}\,=\,1.40\times 10^{-10}\times L_{\rm IR}$) and $L_{\rm IR}$ was obtained from fitting the sub-mm/mm data with a modified blackbody using the \texttt{cmcirsed} package \citep{casey12c}. SFR$_{\rm IR}$ was not derived for IDs 4b, 18, 19, and 24b because of lack of submm/mm detections. The molecular gas mass, $M^{\rm CO}_{\rm gas}$, was derived from \LpCOone\ assuming $\alpha_{\rm CO}=4.36$ \msun\,pc$^{-2}$ (K\,\kms)$^{-1}$ \citep{bolatto13,genzel15}, and $M^{\rm CI}_{\rm gas}$ as 5.2$\alpha_{\rm CO}$\LpCI\ \citep{birkin2021}. The gas-to-stellar mass ratio is given by $\mu_{\rm gas}=M_{\rm gas}/M_{\rm star}$, and the gas depletion time by $\tau_{\rm depl}=M_{\rm gas}$/SFR$_{\rm IR}$ (here we use CO-derived gas masses when both CO and C{\sc i} are available). The main sequence offset (SFR/SFR$^{\rm MS}$) is derived using SFR$_{\rm IR}$, or SFR$_{\rm inst}$ when the former is not available (i.e., for IDs 18 and 19), and the SFR predicted by the relation in \citet{popesso23} assuming the redshift and the stellar mass of each source. The relative quantities $\tau_{\rm depl}/\tau^{\rm MS}_{\rm depl}$ and $\mu_{\rm gas}/\mu^{\rm MS}_{\rm gas}$ are derived by assuming the scaling relations for $\mu^{\rm MS}_{\rm gas}$ and $\tau^{\rm MS}_{\rm depl}$ as a function of stellar mass and redshift from \citet{liu19}.}.
\end{table*}                                                                                                

\subsection{Molecular gas masses}\label{sec:mol_masses}

Molecular gas masses are derived from CO luminosities using a single scaling factor as
\begin{equation}
    M_{\rm gas}^{\rm CO}=\alpha_\mathrm{CO}\LpCOone,
\end{equation}
where $\alpha_\mathrm{CO}$ is the scaling parameter. To derive gas masses from C{\sc i} we used a similar equation, namely
\begin{equation}
    M_{\rm gas}^{\rm CI}=\alpha_\mathrm{CI}\LpCI.
\end{equation}
We note that since we have not observed the C{\sc i}(2--1) transition we cannot take into account the gas excitation temperature, which is commonly used to calculate the partition function and obtain gas mass estimates with less uncertainty \citep[e.g.,][]{bothwell2017,dunne2021,gururajan2023}. This simple parameterization was calibrated using a large sample of SMGs with both C{\sc i}(1--0) and \COfour\ detections by \citet{birkin2021}, finding an average $\alpha_\mathrm{CI}/\alpha_\mathrm{CO}$ of 5.2$\,{\pm}\,$1.3, which we used here.

With this calibration, the only conversion factors we need to assume are $\alpha_\mathrm{CO}$ and the scaling of the various CO luminosities to \LpCOone. $\alpha_\mathrm{CO}$ can range from around 0.2 to 10\,\msun\,pc$^{-2}$\,(K\,\kms)$^{-1}$~\citep{tacconi08,casey14}, depending on the physical properties of the galaxy such as   metallicity~\citep{genzel12,genzel15,amorin16,tacconi18,inoue21} or SFR, with respect to that expected for a galaxy with the same stellar mass and redshift on the MS relation \citep[see equation~2 in][]{castignani20}. Galaxies above the MS, like starburst galaxies, can have smaller $\alpha_{\rm CO}$ values than those on the MS. Similarly, galaxies with lower metallicities can have smaller $\alpha_{\rm CO}$ values. The $\alpha_{\rm CO}$--metallicity relation, however, is valid only for massive ($M_{\rm star}{>}$10$^{10}$\,\msun) SFGs~\citep{genzel15}. For the galaxies in G073 with $M_{\rm star}{>}$10$^{10}$\,\msun\ the derived $\alpha_\mathrm{CO}$ values range from 4.0 and 6.7~\citep{genzel15,tacconi18}. To be consistent with previous works, we adopted a conservative $\alpha_\mathrm{CO}$\,=\,4.36\,\msun\,pc$^{-2}$\,(K\,\kms)$^{-1}$~\citep{bolatto13,genzel15} for all sources, independent of their MS-based classification. This $\alpha_\mathrm{CO}$ conversion factor is commonly used for the Milky Way and for normal SFGs with solar metallicities \citep{bolatto13,genzel15}, and it includes a correction factor for helium. Regardless of the choice of $\alpha_\mathrm{CO}$, there are large systematic uncertainties involved in estimating the gas mass, which we ignore here for simplicity.

As a last step, the measured CO luminosities were converted to \COone\ luminosities, \LpCOone, using the \LpCOthree\ and \LpCOfour\ to \LpCOone\ ratios $r_{3,1}\,{=}\,0.60\,{\pm}\,0.11$ and $r_{4,1}\,{=}\,0.32\,{\pm}\,0.05$ \citep{birkin2021}. The derived gas masses are listed in Table~\ref{tab:properties}.

For comparison, we also tested standard scaling relations between the Rayleigh-Jeans dust continuum luminosity and gas mass. Since we only have dust continuum estimates for our entire sample from the ALMA Band 6 observations as 1.3\,mm (rest frame 390--520\,$\mu$m; see \citealt{kneissl19}), we scaled these to rest-frame 850\,$\mu$m following Equations 9 and 10 from \citet{liu19} assuming a modified blackbody SED with $\beta\,{=}\,2$ and a dust temperature of 35\,K, then apply the dust continuum-to-gas mass scaling factor of $\alpha_{\rm RJ,mol}\,{=}\,(6.7\pm1.7)\,{\times}\,10^{19}\,$[erg\,s$^{-1}$\,Hz$^{-1}$\,M$_{\odot}^{-1}$] from \citet{scoville2016}. We find similar results, with a mean ratio of the  dust continuum gas mass over the CO/C{\sc i} gas mass of 1.04. For the following comparisons with literature values we used our CO/C{\sc i}-derived gas masses since they are not correlated with our SFR estimates, which rely significantly on the same dust continuum flux densities; however our results are not strongly influenced by our choice of gas  mass estimator.

\section{Results}
\label{sec:results}

\subsection{Redshift distribution}

\begin{figure}
\includegraphics[width=\linewidth]{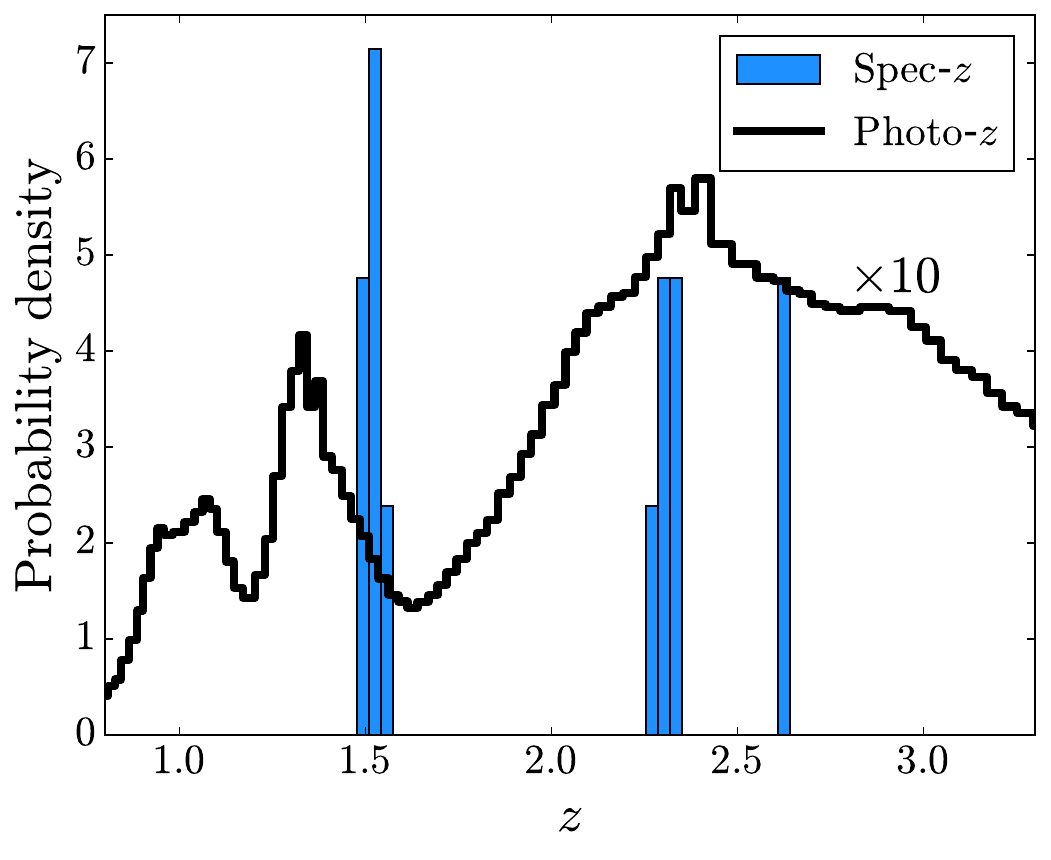}
\caption{Distribution of the 13 spectroscopic redshifts found in our ALMA survey compared to the total photometric redshift distribution of all 18 continuum-selected galaxies in \citet{kneissl19}. The photometric redshift distribution has been scaled by a factor of 10 to be more easily comparable with the spectroscopic redshift distribution.}
\label{fig:redshift_histo}
\end{figure}

In Fig.~\ref{fig:redshift_histo} we show the distribution of the 13 galaxy redshifts we detected in our ALMA survey, alongside the combined photometric redshift probability distributions originally determined for all of the 18 galaxies from \citet{kneissl19}. Here the combined probability density was calculated by summing the probability densities estimated by {\tt EAZY} over the whole sample. The two structures initially predicted by \citet{kneissl19} at $z\,{\approx}\,1.5$ and $z\,{\approx}\,2$ are seen as the two broad peaks in the probability distribution. Since this distribution includes the photometric redshift uncertainties, the widths of the peaks correspond to the uncertainties in the redshifts of the protocluster structures -- roughly $\pm$0.2 for the $z\,{\approx}\,1.5$ structure, and $\pm$0.5 for the $z\,{\approx}\,1.5$ structure. The redshifts we have spectroscopically confirmed agree with these two peaks. For the $z\,{\approx}\,1.5$ structure, we find six galaxies with an average redshift of $z\,{=}\,1.526$, while for the $z\,{\approx}\,2$ structure we find five galaxies (ignoring the SCUBA-2 galaxies) with an average redshift of $z\,{=}\,2.313$. Evidently none of the SCUBA-2 sources we targeted are within the structures identified by {\it Planck\/} and {\it Herschel}, which we discuss further below.

\begin{figure} 
\centering
\includegraphics[width=\linewidth]{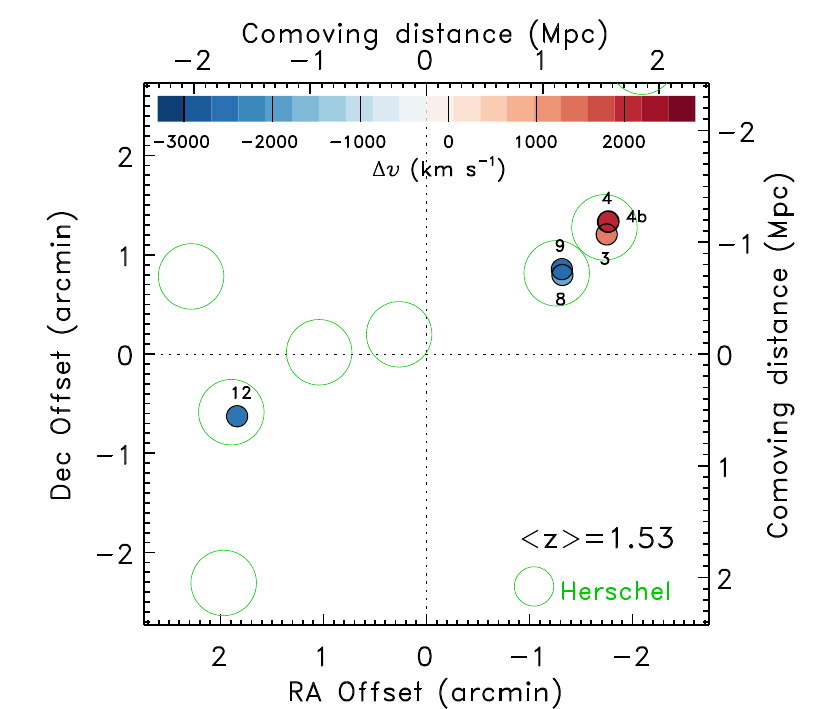}
\includegraphics[width=\linewidth]{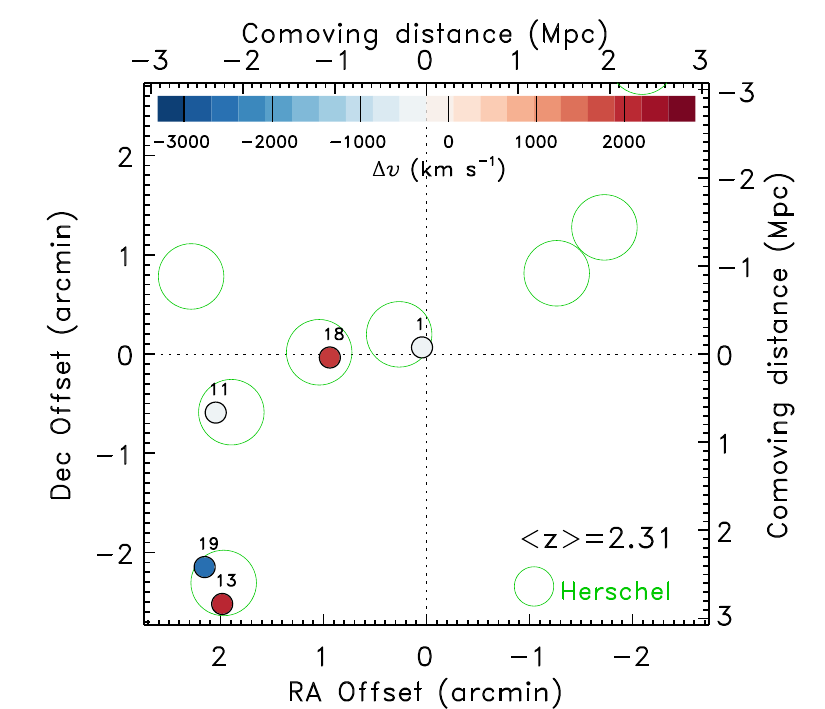}
\caption{{Spatial distribution of the CO-detected sources at $1.50\,{<}\,z\,{<}\,1.55$ (top panel) and at $2.27\,{<}\,z\,{<}\,2.34$ (bottom panel) in the G073.4$-$57.5 field (full circles) with colour corresponding to the redshift offset from the average values of $\left\langle z\right\rangle$\,=\,1.53 and 2.31, respectively, expressed in terms of velocity offset as indicated by the horizontal bar on the top. Large green circles represent the positions of the \textit{Herschel} sources in the field.}}
\label{fig:deltav}
\end{figure}

Figure~\ref{fig:deltav} shows the spatial distribution of the two structures, colour-coded according to their line-of-sight velocity relative to the mean redshift of each group. We find that most of the $z\,{=}\,1.5$ galaxies are spatially concentrated in the north-west corner of the field of view initially defined by the {\it Planck\/} beam, while the $z\,{=}\,2.3$ galaxies are concentrated in the south-east corner, about 1\arcmin from the lower-redshift structure. This distribution points to the conclusion that the original {\it Planck\/} data has selected regions on the sky where two protoclusters are aligned along the line of sight to within about 3\arcmin of one another, relative to the 5\arcmin {\it Planck\/} beam.

\subsection{Depletion timescales}

\begin{figure}[htbp!]
\centering
\includegraphics[width=\linewidth]{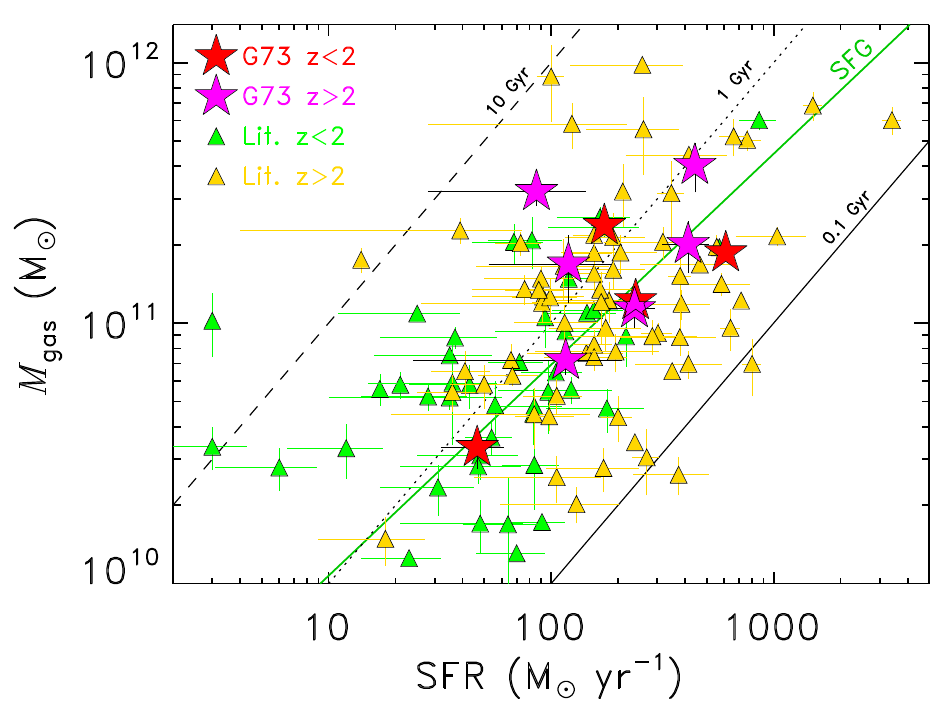}
\caption{Molecular gas mass as a function of SFR for the G073 sources (stars; red for $z\,{<}\,2$ and magenta for $z\,{>}\,2$). Small triangles represent the cluster and protocluster members from the literature shown in Fig.~\ref{fig:main_sequence} (green for 1.4$\,{<}\,z\,{<}\,$2; yellow for $2.0\,{<}\,z\,{<}\,2.65$). The solid green lines represent the average relation for normal SFGs \citep{sargent14}. The black lines represent constant gas depletion times (solid, 0.1\,Gyr; dotted, 1\,Gyr; and dashed, 10\,Gyr).
}
\label{fig:sfe}
\end{figure}

The depletion timescale is given by $\tau_{\rm depl}\,{=}\,M_{\rm gas}/$SFR (thus it has units of time), which provides a crude estimate for how long it would take a galaxy to convert all of its gas into stars, assuming the SFR remains constant. We therefore plot the molecular gas mass versus the SFR in Fig.~\ref{fig:sfe}, along with lines of constant depletion time and the relations that describe normal SFGs, as parameterized by~\citet{sargent14}. 

We also show for comparison the values of the CO-detected cluster and protocluster galaxies described in Sect.~\ref{sec:main_sequence}. The gas masses from the literature have been derived from various CO transitions and $\alpha_{\rm CO}$ values -- for example $\alpha_{\rm CO}=4.36$ in \citet{noble17}, \citet{noble19}, and \citet{rudnick17} and $\alpha_{\rm CO}=0.8$ in \citet{ivison13} -- and in some cases a stellar mass-dependent correction was applied resulting in final $\alpha_{\rm CO}$ values ranging from 4.16 to 6.09 \citep{hayashi18}.

The conversion factor adopted by \citet{sargent14} varies with the metallicity following \citet{wolfire10} and is 4.4 at solar metallicity and always above 3. The gas masses in \citet{liu19} are derived from the continuum. We corrected all the gas mass estimates from this literature cluster sample to have the same $\alpha_{\rm CO}$ conversion factor of 4.36 (when possible), which is the value we used for G073. The scaling relations used to represent normal SFGs in the field are consistent with the adopted conversion.

We find that the depletion timescales of the G073 CO emitters are scattered around the values observed in normal SFGs ($0.3\,{<}\,\tau_{\rm depl}/{\rm Gyr}\,{<}\,3.71$). The median depletion times for the G073 sources and for the cluster members are ($0.7\,{\pm}\,0.6$)\,Gyr and ($0.8\,{\pm}\,1.2$)\,Gyr, respectively; thus, we find no significant difference.

\subsection{Gas-to-stellar mass ratio}

\begin{figure}[htbp!]
\centering
\includegraphics[width=\linewidth]{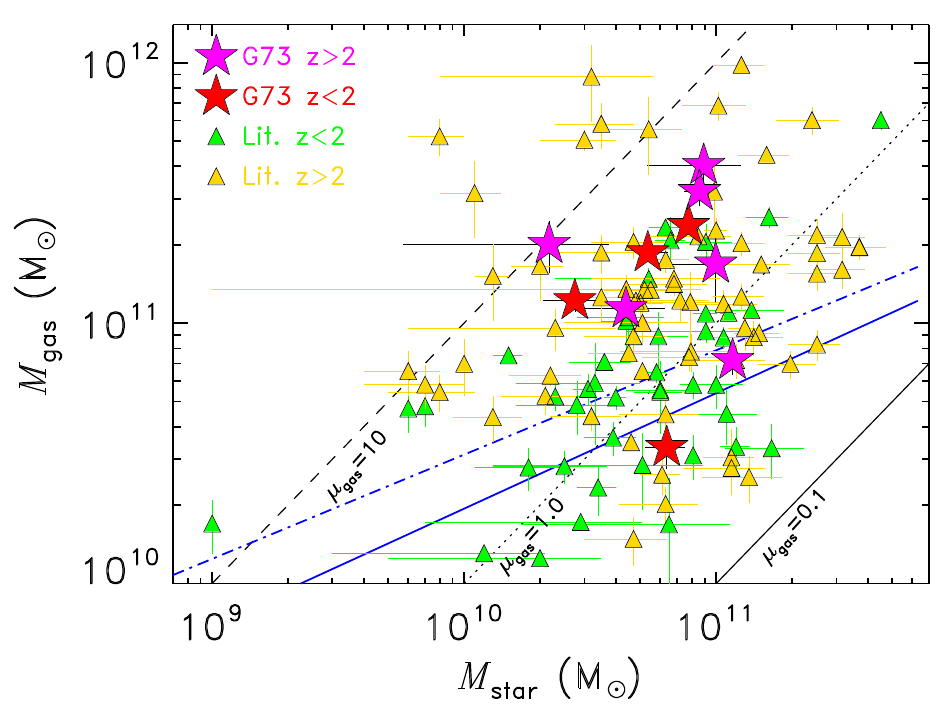}
\caption{Gas mass of the galaxies in G073 as a function of stellar mass. Symbols are as in Fig.~\ref{fig:sfe}. The solid, dotted and dashed black lines represents $\mu_{\rm gas}\,{=}\,M_{\rm gas}/M_{\rm star}\,{=}\,0.1$, 1.0, and 10, respectively. 
The solid and dot-dashed blue lines represent the scaling relations from \citet{liu19} at $z\,{=}\,$1.53, and 2.3, respectively.  
}
\label{fig:mgas_mstar}
\end{figure}

The gas-to-stellar mass ratio, defined as $\mu_{\rm gas}=M_{\rm gas}/M_{\rm star}$, is of interest because it provides insight into the fraction of baryonic matter available to be converted to stars. In the limit where a galaxy does not accrete additional gas throughout its lifetime, the gas-to-stellar mass ratio is a crude estimate of a galaxy's maturity. Figure~\ref{fig:mgas_mstar} shows the gas masses as a function of stellar masses for the galaxies in G073, along with the CO-detected cluster and protocluster galaxies from the literature and the scaling relations expected for field galaxies from \citet{liu19}. 
We also include several curves of constant gas-to-stellar mass ratio for reference.

Our sample of galaxies in G073, along with a collection from the literature, mostly lies above the field galaxy curve. Quantitatively, the G073 sample has a median gas-to-stellar mass ratio of $\mu_{\rm gas}\,{=}\,3.5^{+3.1}_{-1.6}$, while our comparison sample has a median gas-to-stellar mass ratio of $\mu_{\rm gas}\,{=}\,1.6^{+2.3}_{-0.9}$. This is contrary to what we might have expected given the higher SFRs of our sample; according to the MS, higher SFRs should lead to higher stellar masses, and therefore lower gas-to-stellar mass ratios. Since the opposite is observed in our case, the relatively high SFRs must be compensated by relatively high gas masses. Again, this might be a selection effect due to the fact that our CO survey is sensitive to galaxies with the highest gas masses (we discuss this possibility more in Sect.~\ref{discussion}).

\section{Discussion}\label{discussion}

\subsection{The large-scale structure probed by G073}

The primary result of our ALMA survey of G073 is the spectroscopic confirmation of two protoclusters within a single PHz object, one at $z\,{=}\,1.53$, and the other at $z\,{=}\,2.31$. These two overlapping structures were initially suggested by \citet{kneissl19} based on photometric redshifts with large uncertainties, but now with improved knowledge of the systems we can further understand the nature of this PHz object (and hence others), including using simulations.

\citet{gouin2022} compared the $z\,{\approx}\,1.5$ and $z\,{\approx}\,2$ structures in G073 to the (300\,Mpc)$^3$ TNG300 simulation using primarily photometric redshift data from \citet{kneissl19}; at the time, only two galaxies had spectroscopic redshifts around $z\,{\approx}\,1.5$ (IDs 3 and 8, although the redshift for ID 8 has been updated by our new observations). \citet{gouin2022} identified three additional galaxies as being part of the $z\,{\approx}\,1.5$ based on their photometric redshifts (IDs 5, 6, and 12), but we have instead found that IDs 4, 9, and 12 are in this structure. At $z\,{\approx}\,2$ they identified IDs 1, 11, 13, 14, and 15, whereas we found IDs 1, 11, 13, 18, and 19 are in the structure. The handful of misidentified galaxies reflects the uncertain nature of photometric redshifts, although the discrepancy is not particularly large.

For the simulation comparison, \citet{gouin2022} chose the redshift snapshot closest to the actual redshifts of the G073 structures, selected 30 of the most star-forming halos (defined to include all gas cells in a friends-of-friends group), then accounted for observational biases by including galaxies within a cylinder (centred on a given star-forming halo) with a diameter of 2.4\arcmin and a length encompassing ${\pm}0.17$ in redshift for the $z\,{\approx}\,1.5$ structure, and ${\pm}0.26$ in redshift for the $z\,{\approx}\,2$ structure. Lastly, an SFR cut of 30\,M$_{\odot}$\,yr$^{-1}$ was applied to the galaxies within the simulated cylinders. This is comparable to our observational sensitivity -- our 5$\sigma$ line luminosity threshold is about 0.3\,Jy\,km\,s$^{-1}$, corresponding to a gas mass of $3\,{\times}\,10^{10}\,$M$_{\odot}$ (assuming CO(3--2) at $z\,{=}\,1.5$) or a SFR of 30\,M$_{\odot}$\,yr$^{-1}$ using the scaling relation from \citet{sargent14}. For C{\sc i}(1--0) at $z\,{=}\,2.3$, this limit corresponds to a gas mass of $10\,{\times}\,10^{10}\,$M$_{\odot}$, or a SFR of 100\,M$_{\odot}$\,yr$^{-1}$.

\begin{figure*}[htbp!]
\centering
\fbox{
\parbox{0.48\textwidth}{
\includegraphics[width=\linewidth]{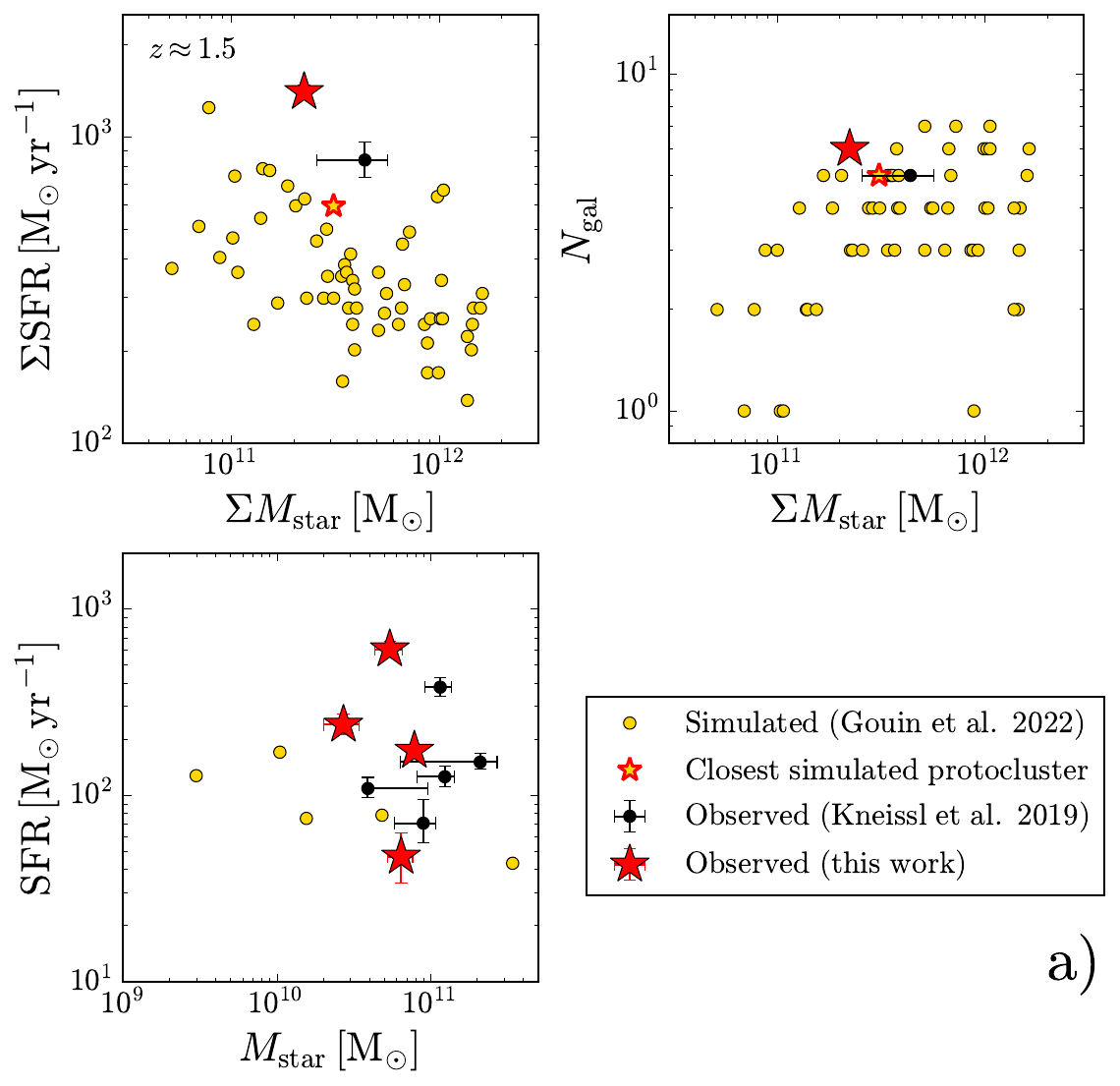}}}
\centering
\fbox{
\parbox{0.48\textwidth}{
\includegraphics[width=\linewidth]
{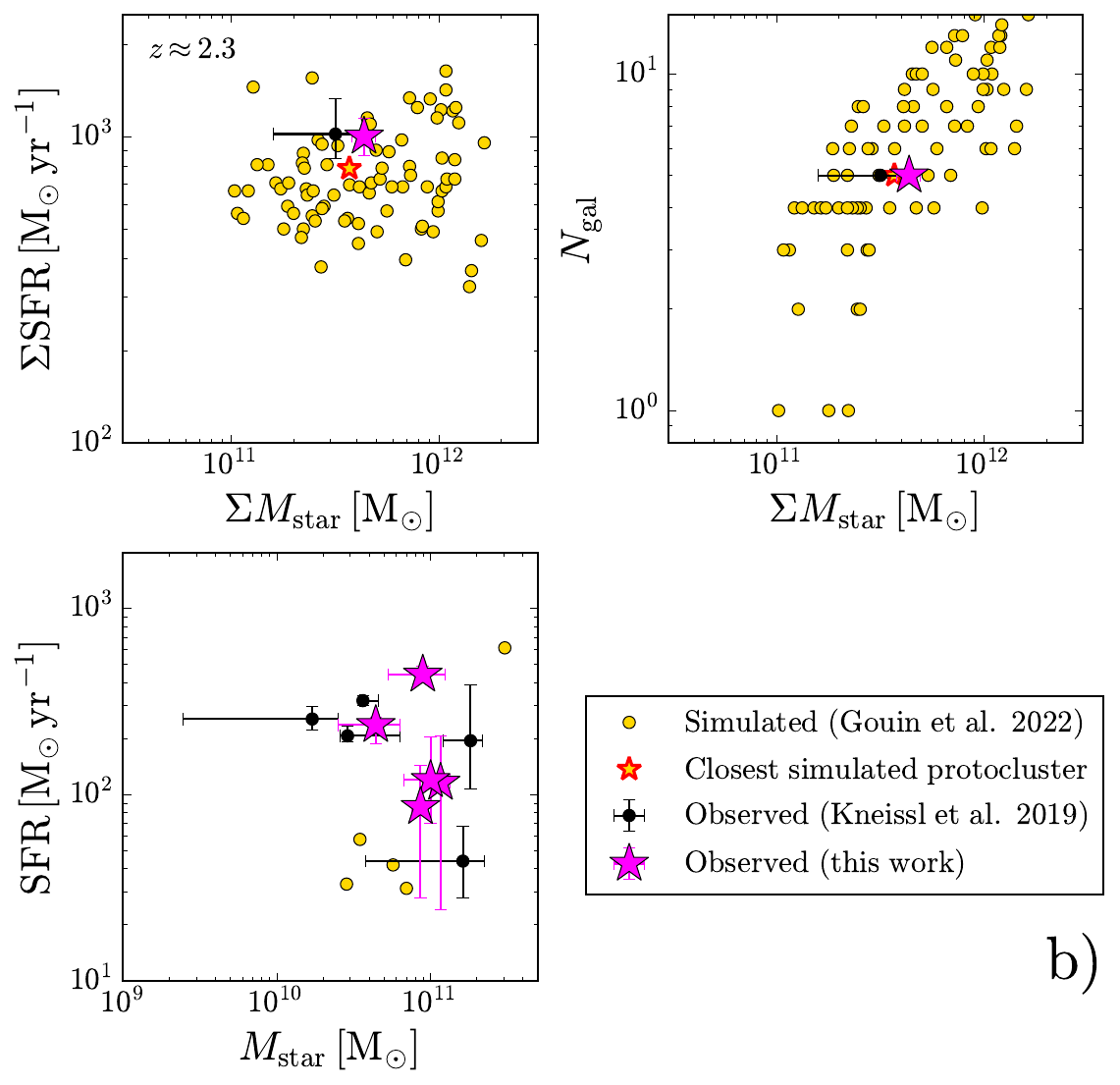}}}
\caption{Comparison of our results to the TNG300 simulation \citep{gouin2022} at (a) $z\,{\approx}\,1.5$ and (b) $z\,{\approx}\,2.3$. Top left in both panels: Total protocluster SFR versus stellar mass for 30 simulated protoclusters (yellow points), compared to previous observations based primarily on photometric redshifts (black point; \citealt{kneissl19}) and our new observations based only on spectroscopic redshifts (red stars). The closest matching simulated halo selected by \citet{gouin2022} is highlighted in red. Top right in both panels: Same as previous panel but showing the total number of protocluster galaxies with SFR$\,{>}\,$10\,M$_{\odot}$\,yr$^{-1}$ versus stellar mass. Bottom in both panels: SFR versus stellar mass for the individual galaxies in the closest-matching simulated halo (yellow points), compared to previous observations \citep{kneissl19} and this work.}
\label{fig:simulation_gouin}
\end{figure*}

In Fig.~\ref{fig:simulation_gouin} we reproduce the panels in figures~6 and 7 from \citet{gouin2022} relating to the $z\,{\approx}\,1.5$ and $z\,{\approx}\,2$ structures in G073; we show the total SFR versus total stellar mass (i.e., the sum of all the galaxies within the cylinder) of all 30 simulated structures, and the total number of galaxies versus total stellar mass for the same simulated sample. We also show the same quantities for the observed galaxies as reported in \citet{kneissl19}. \citet{gouin2022} selected one of the 30 simulated structures that most closely matched the observed total stellar mass, SFR, and number of SFGs (highlighted in red), and we show individual SFRs and stellar masses of these specific simulated structures compared to the galaxies from \citet{kneissl19}. \citet{gouin2022} conclude that the TNG300 simulation is able to reproduce the PHz selection, and that the simulated structures are comparable to the observations.

With our new ALMA observations spectroscopically confirming six galaxies around $z\,{\approx}\,1.5$ and five galaxies around $z\,{\approx}\,2$, we would like to ask if the conclusions from \citet{gouin2022} concerning G073 have changed. In Fig.~\ref{fig:simulation_gouin} we therefore update the observed data points with values derived in this work. We find that the total observed SFR within the $z\,{\approx}\,1.5$ structure increases while the total stellar mass decreases, although not substantially, while the $z\,{\approx}\,2$ structure is in good agreement. The properties of the individual galaxies within G073 also remain consistent with the simulated structures selected by \citet{gouin2022} to be the most similar to the observations from \citet{kneissl19}. It is worth noting that while there still appear to be some discrepancies between the simulated protocluster galaxies and the observations (e.g. higher observed stellar masses and SFRs compared to the selected simulation at $z\,{\approx}\,1.5$), deciding which simulation to focus on for the comparison is subjective; a more statistically robust comparison would marginalize over all of the simulated protoclusters, and so we emphasize here that this comparison only qualitatively illustrates broad agreement.

Regarding the fates of the most star-forming halos found in the TNG300 simulation, \citet{gouin2022} found that 60--70\% of the simulated structures (which spanned a redshift range of 1.3 to 3.0) evolved into ${>}\,10^{15}\,$M$_{\odot}$ galaxy clusters by redshift 0. Moreover, they found that protoclusters with more SFGs and with that star formation evenly spread across the galaxies (i.e., systems without a single extremely luminous galaxy alongside a number of faint galaxies) are more likely to become clusters by $z\,{=}\,0$. This is roughly in line with our observations, so we conclude that the results from \citet{gouin2022} suggest a high probability that both structures in G073 will ultimately evolve into a galaxy clusters.

Lastly, we can speculate on the current masses of the G073 systems by scaling our estimated stellar masses to dark matter halo masses and adding them up. \citet{behroozi2013} provide a fitting function scaling halo mass to stellar mass as a function of redshift, calibrated to a large body of observational constraints. Using their best-fit function, for the $z\,{=}1.5$ structure we find $M_{\rm star}/M_{\rm halo}\,{=}\,0.020$--0.025 and a total dark matter halo mass of ${\approx}\,1\,{\times}\,10^{13}\,$M$_{\odot}$. For the $z\,{=}2$ structure we find $M_{\rm star}/M_{\rm halo}\,{=}\,0.007$--0.026 and a total dark matter halo mass of ${\approx}\,4\,{\times}\,10^{13}\,$M$_{\odot}$.

While these values are nowhere near the mass of a redshift 0 galaxy cluster, it is worth emphasizing that our line luminosity limits are modest. For example, \citet{vanderburg2013} fit Schechter functions to a sample of 10 spectroscopically confirmed galaxy clusters around $z\,{\approx}\,1$, splitting the sources into a sample of SFGs and a sample of quiescent galaxies. Using their fit to the SFGs, and assuming we are only sensitive to galaxies with stellar masses ${\gtrsim}\,1\,{\times}\,10^{10}$\,M$_{\odot}$, we find that we would expect to find about 4 times more galaxies with stellar masses ${>}\,10^{9}$\,M$_{\odot}$ than with stellar masses ${>}\,10^{10}$\,M$_{\odot}$ (corresponding to a total of about 25 galaxies in each system). This is again not enough mass to account for 10$^{15}\,$M$_{\odot}$, but growth through accretion outside of the field-of-view is expected.

\subsection{Environmental effects on protocluster galaxies around redshift 1.5}

To investigate possible environmental effects on the gas properties of $z\,{\approx}\,2$ cluster and protocluster galaxies, we derived the field-relative gas depletion timescales and gas-to-stellar mass ratios by dividing the measured values from our sample and the collection from the literature by those of coeval field galaxies obtained through the \citet{liu19} scaling relations. The predicted field values are calculated using the redshift and the stellar mass of each source and ignoring corrections related to offsets to the MS. The broad range of stellar masses (about (0.5--17)$\,{\times}\,10^{10}$\,\msun\, or 1.5\,dex span) and of SFRs (about 3--600\,\msun\,yr$^{-1}$, or 2.3\,dex span) of the G073 sources, combined with the cluster members, allows us to investigate possible dependences with these parameters. The properties relative to field galaxies are shown as a function of SFR and stellar mass in Fig.~\ref{fig:scaling_relations}.

\begin{figure*}[htbp!]
\centering
\includegraphics[width=0.48\linewidth]{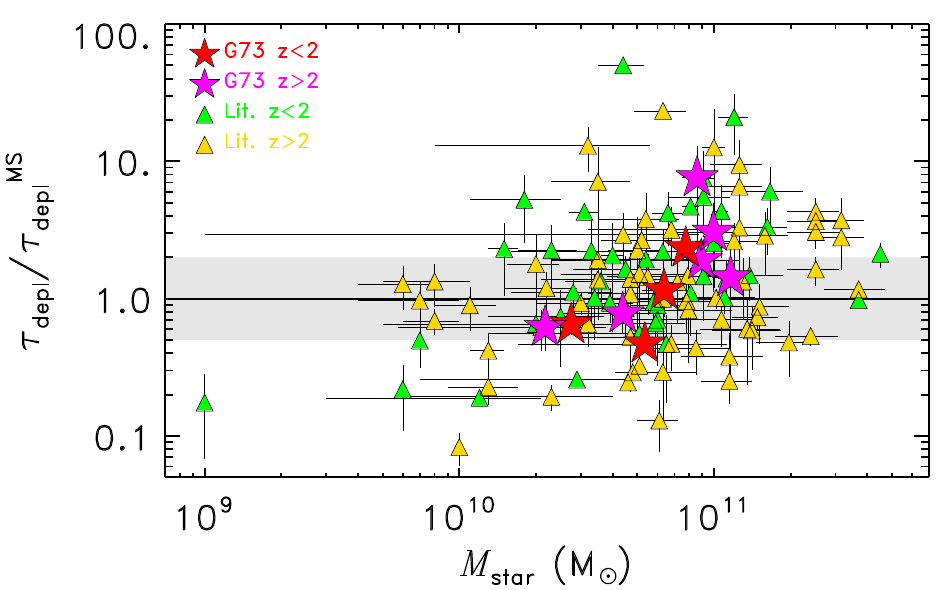}
\includegraphics[width=0.46\linewidth]{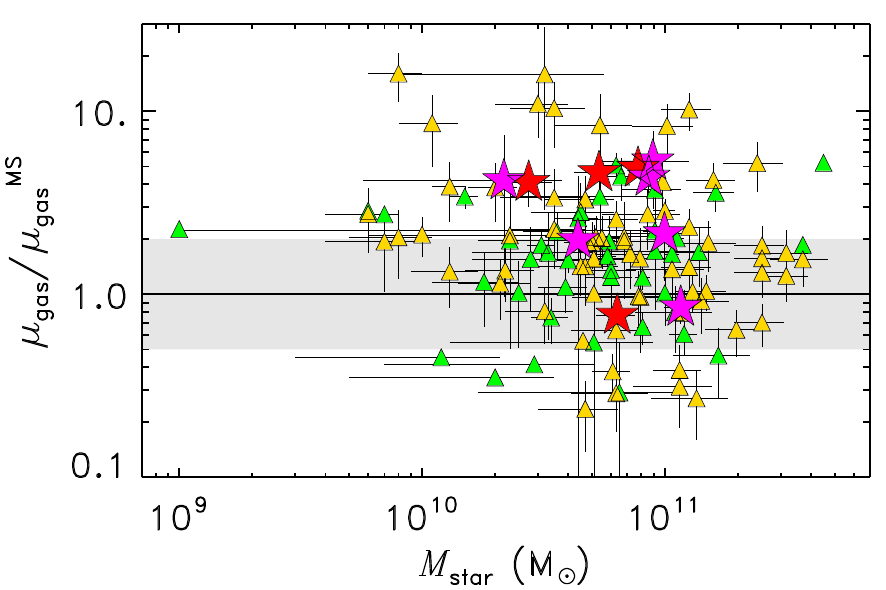}
\includegraphics[width=0.48\linewidth]{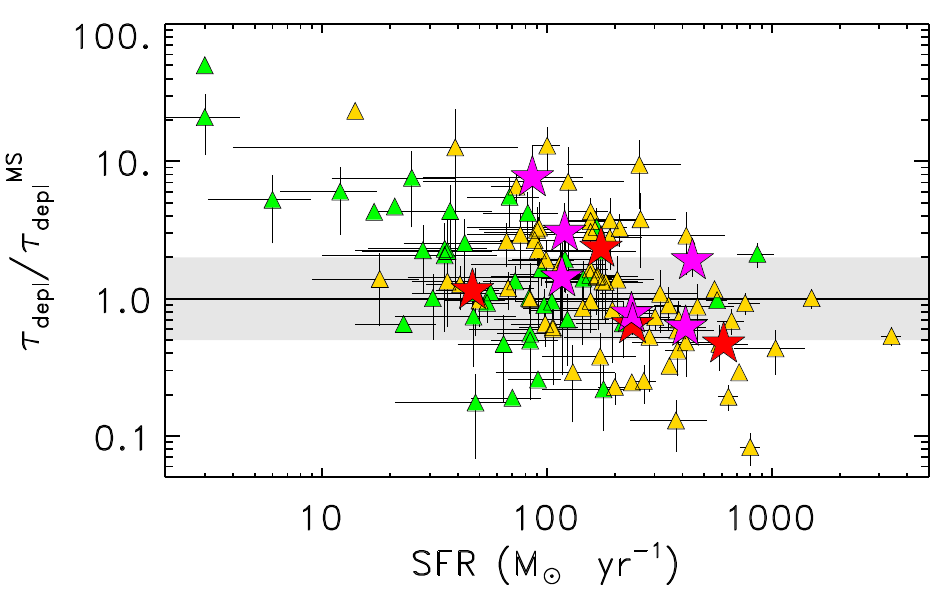}
\includegraphics[width=0.46\linewidth]{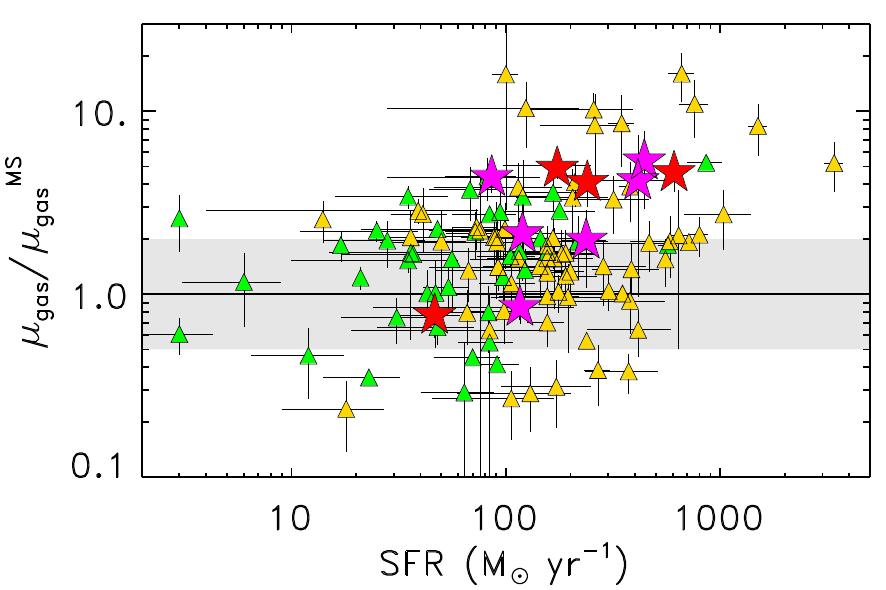}
\caption{Gas depletion timescales (left) and gas-to-stellar mass ratios (right) of cluster
galaxies relative to the predicted gas properties as a function of
stellar mass (top) and of SFR (bottom). The predicted properties are estimated for each individual source given its redshift and stellar mass assuming the 
scaling relations derived for field galaxies by \citet{liu19}.  Symbols are as in Fig.~\ref{fig:sfe}.  The solid black line represents the \citet{liu19}  scaling relation for $\tau_{\rm depl}$ and $\mu_{\rm gas}=M_{\rm gas}/M_{\rm star}$
with a scatter of 0.3 dex (grey shaded regions). About half of the CO-emitters in this work is characterized by higher gas-to-stellar mass ratios, which is also seen in the other cluster samples. On the other hand, most of the galaxies in G073 have depletion times consistent with the field.}
\label{fig:scaling_relations}
\end{figure*}

We find that 51\% of the cluster and protocluster members and 30\% of the G073 sources have gas-to-stellar mass ratios that are consistent with coeval field galaxies, while 39\% and 70\% have higher ratios, respectively.  This is consistent with previous results that find that CO emitters in high-$z$ clusters have field-like or enhanced molecular gas fractions \citep[i.e.,][]{noble17,rudnick17,hayashi18}. Interestingly, the excess in gas fraction is observed at all masses, but there seems to be a mild trend with SFR. Although these results might suffer from a bias in favour of gas rich galaxies, it is expected that high-density environments at high redshift favour cold gas buildup, and the systems with the highest gas fractions are also the most star-forming ones.  

Regarding the relative gas depletion times,  60\% of the G073 sources have shorter depletion times than field galaxies, compared to 47\% for the sample from the literature. There is a trend in relative depletion timescale with SFR, although this is to be expected as the depletion timescale is proportional to the inverse of the SFR. There is also a potential trend in relative depletion timescale with $M_{\rm star}$, where higher stellar mass corresponds to longer depletion timescale. This trend might be explained by the galaxy maturity level, where galaxies in an earlier evolutionary phase have smaller stellar masses, larger star-forming efficiencies and higher gas-to-stellar mass ratios. 

\subsection{The nature of \texorpdfstring{850\,$\mu$m-bright}{850-micron-bright} sources in PHz fields}

It is also worth discussing the fact that no SCUBA-2-selected sources appear to be in the redshift 1.5 or 2 structures. These sources all have $S_{850}\,{>}\,7\,$mJy (and three have $S_{850}\,{>}\,8\,$mJy), and were part of a larger study of 61 PHz fields followed up by SCUBA-2 \citep{mackenzie2017}. Of these 61 fields, 51 were protocluster candidates (so not strong gravitational lenses), and ultimately contained a factor of 6 enhancement in number density for sources above 8\,mJy compared to blank-field surveys. This implied that the PHz objects at least contain projected overdensities of sources at 850\,$\mu$m on the sky. \citet{mackenzie2017} also fitted for photometric redshifts using the three {\it Herschel\/}-SPIRE bands and photometry at 850\,$\mu$m from SCUBA-2, specifically finding photometric redshifts ${\gtrsim}\,2.5$ for the four sources followed up in our programme (although with large uncertainties) -- effectively consistent with one of these sources being at $z\,{=}\,2.6$.

Despite the fact that G073 contains statistically more 850\,$\mu$m-bright sources than the field, they appear to be at higher redshifts than the two structures making up this field that we have studied in this work. This suggests a possibility that there are in fact three (or more) protoclusters along the line of sight, although this remains to be confirmed.

\subsection{The nature of the \planck\ high-$z$ sources}

The PHz sources are the brightest 5\arcmin-scale submm sources on the sky. By selection, their thermal submm SEDs peak between 350 and 850\,$\mu$m, corresponding to redshifts of around 1.5--4, and their brightnesses (if unlensed) imply SFRs ${>}\,$25,000\,\msun\,yr$^{-1}$ \citep{planckXXXIX}. We now know that only a few percent are strong lenses \citep[e.g.][]{canameras2015}, with the remaining objects consisting of groups of submm sources that exhibit extended emission that is absent in randomly selected \herschel\ samples \citep[e.g.][]{lammers2022}.

The PHz selection clearly favours fields where submm sources are aligned along the line of sight, but the physical associations (i.e. the redshift distributions) remain debated. Nonetheless, observations are routinely identifying two or more galaxy groups along the line of sight (PLCK\,G95.5$-$61.6 at $z=1.7$ and $z=2.0$ \citep{flores-cacho2015}; PLCK\,G237.01$+$42.50 at $z=2.16$ and $z=2.195$ \citep{polletta21}; and PLCK\,G073.4$−$57.5 at $z=1.50$ and $z=2.3$, as confirmed here and initially found by \citealt{kneissl19}), or a single galaxy group with numerous galaxies in the foreground and background (PLCK\,G176.60+59.01 at $z=2.75$; PLCK\,DU\,G059.1$+$37.4 at $z=2.36$;  PLCK\,DU\,G124.1$+$68.8 at $z=2.15$; and PHz\,G191.24$+$62.04 at $z=2.42$ -- see \citealt{polletta22,polletta24}).

Ultimately, based on the limited and biased spectroscopic follow-up programmes that have been carried out so far, there seems to be evidence that the \planck\ objects consist of sometimes one, but often more, structures with redshifts between about 2 and 3, each traced by several \herschel\ sources. There could of course be an observational bias against fields containing no protoclusters (meaning that all of the line of sight galaxies are at different redshifts) as securing the redshifts of numerous unassociated galaxies is challenging, but simulations seem to suggest otherwise \citep[e.g.,][]{negrello17,gouin2022}. However the question still remains as to how massive these structures are, requiring deep spectroscopic observations sensitive to both dusty obscured sources and optically luminous sources over wide areas.

,
\section{Summary and conclusions}
\label{sec:conclusion}

We used ALMA to search for $z\,{\approx}\,1.5$ and $z\,{\approx}\,2$ protocluster members in the {\it Planck\/}-selected field G073. In eight pointings targeting bright and red {\it Herschel\/}-SPIRE sources and four pointings targeting high S/N SCUBA-2 sources across the field, we have identified 27 individual galaxies through a combination of continuum and line emission, confirming six spectroscopic redshifts around $z\,{\approx}\,1.5$ and five spectroscopic redshifts around $z\,{\approx}\,2$. We have also found that the SCUBA-2 sources are not part of either of these protoclusters; they likely lie in a more distant structure (or structures) at  $z\,{>}\,2.5$.

By fitting NIR (${\sim}\,1\,\mu$m) and FIR (up to 2\,mm) photometry to our sources, we derived SFRs and stellar masses, and we estimated gas masses using the CO and C{\sc i} line detections from our ALMA programme. Four galaxies are considered starbursts (having SFRs a factor of 4 higher than the MS), and the average gas-to-stellar mass ratio ($\mu_{\rm gas}=M_{\rm gas}/M_{\rm star}$) of our protocluster galaxies is higher than typical scaling relations for field galaxies.

We compared our more detailed view of G073 to the TNG300 simulation by updating the results from \citet{gouin2022} with spectroscopic redshift information. Our observations remain consistent with the simulated star-forming halos at high redshifts selected from the simulation. Since about 60--70\% of the simulated halos collapse into galaxy clusters by redshift zero, we conclude that the two structures uncovered in G073 are comparably likely to become clusters as well. We estimate the total dark matter halo mass in observed galaxies to be around $10^{13}\,\textrm{M}_{\odot}$ in each structure, although we speculate that there are still a factor of 4 more galaxies with stellar masses an order of magnitude below our detection threshold of ${\sim}\,10^10\,$M$_{\odot}$.

We show that the two structures spectroscopically detected by our observations are separated by about 3\arcmin in projection on the sky. Relative to {\it Planck\/}'s 5\arcmin beam originally used to select this field, this is in line with previous claims that at least some of the PHz objects are line-of-sight alignments \citep[e.g.][]{flores-cacho2015}, and indeed the presence of an overdensity of SCUBA-2 sources at higher redshifts could indicate that this field contains three or more protoclusters. While this poses an observational challenge for future follow-up studies of a statistically significant sample of PHz objects, it also provides the potential for extracting more information than achievable from typical single-system fields. For example, the upcoming ALMA receiver upgrade \citep{carpenter2023} will boost the facility's spectral coverage by a factor of 4, meaning that with short observations of single PHz fields (such as G073) we will be able to spectroscopically identify several protoclusters. A modest observing programme targeting of order ten PHz objects therefore has the potential to find dozens of protoclusters, easily providing powerful population statistics of a homogeneously selected sample of star-forming protoclusters (and their embedded galaxies) around the peak of the star formation activity of the Universe.

\begin{acknowledgements}
This work was supported by the Natural Sciences and Engineering Research Council of Canada. M.P. thanks the UCSD Astronomy \& Astrophysics department for their warm hospitality. M.P. acknowledges financial support from INAF mini-grant 2023 `Galaxy growth and fuelling in high-z structures'. 
This paper makes use of the following ALMA data: ADS/JAO.ALMA\#2018.1.00562.S and 2013.1.01173.S. ALMA is a partnership of ESO (representing its member states), NSF (USA) and NINS (Japan), together with NRC (Canada), MOST and ASIAA (Taiwan), and KASI (Republic of Korea), in cooperation with the Republic of Chile.  The Joint ALMA Observatory is operated by ESO, AUI/NRAO and NAOJ. The National Radio Astronomy Observatory is a facility of the National Science Foundation operated under cooperative agreement by Associated Universities, Inc.
The \herschel\ spacecraft was designed, built, tested, and launched under a contract to ESA managed by the \herschel/\planck\ Project team by an industrial consortium under the overall responsibility of the prime contractor Thales Alenia Space (Cannes), and including Astrium (Friedrichshafen) responsible for the payload module and for system testing at spacecraft level, Thales Alenia Space (Turin) responsible for the service module, and Astrium (Toulouse) responsible for the telescope, with in excess of a hundred subcontractors.  
The development of Planck has been supported by: ESA; CNES and CNRS/INSU-IN2P3-INP (France); ASI, CNR, and INAF (Italy); NASA and DoE (USA); STFC and UKSA (UK); CSIC, MICINN, JA, and RES (Spain); Tekes, AoF, and CSC (Finland); DLR and MPG (Germany); CSA (Canada); DTU Space (Denmark); SER/SSO (Switzerland); RCN (Norway); SFI (Ireland); FCT/MCTES (Portugal); and PRACE (EU).
This work is based in part on observations made with the {\it Spitzer Space Telescope}, which is operated by the Jet Propulsion Laboratory, California Institute of Technology under a contract with NASA. 
Based in part on observations obtained with WIRCam, a joint project of CFHT, Taiwan, Korea, Canada, France, and the Canada-France-Hawaii Telescope (CFHT) which is operated by the National Research Council (NRC) of Canada, the Institute National des Sciences de l’Univers of the Centre National de la Recherche Scientifique of France, and the University of Hawaii.
Based in part on observations made at James Clerk Maxwell Telescope (JCMT) with SCUBA-2. The JCMT is operated by the Joint Astronomy Centre on behalf of the Science and Technology Facilities Council of the United Kingdom, the Netherlands Organisation for Scientific Research, and the National Research Council of Canada.

\end{acknowledgements}

\section*{Data availability}

All of the data presented in this paper are publicly available.  The ALMA
data can be found under the two programmes 2013.1.01173.S (PI: R. Kneissl)
and 2018.1.00562.S (PI: R.  Hill). The {\it Spitzer} data are from programme ID 11004 (PI: H. Dole) and the CFHT data are from Proposal IDs 13BF12 and 14BF08 (PI: H. Dole).

\bibliography{biblio_G073}

\begin{appendix}
\onecolumn
\section{ALMA spectra}
\label{appendix0}

Here we show continuum maps, moment-0 maps, and spectra of the galaxies observed in G073 (Fig.~\ref{fig:spectra}). For each source, the upper panels show 3\arcsec$\,{\times}\,$3\arcsec cutouts. Continuum images obtained by stacking all channels containing no line emission are shown as the background, and overlaid are corresponding continuum contours starting at 2$\sigma$ and increasing in steps of 3$\sigma$. We also show line emission moment-0 contours from stacking all channels between $-3\sigma$ and $3\sigma$ (where $\sigma$ is the standard deviation of the best-fitting linewidth). These contours also start at 2$\sigma$ and increase in steps of 3$\sigma$.

In the lower panels we show our continuum-subtracted LSB and USB spectra. The best-fitting Gaussian functions are plotted on top of the spectra. The shaded region ranges from $-3\sigma$ to $3\sigma$, corresponding to the range used to calculate line strengths.

\begin{figure*}[h!]
\setcounter{figure}{0}
\fbox{
\parbox{0.31\textwidth}{
\centering
\includegraphics[trim=0 0 16.5cm 0,clip,width=0.15\textwidth]{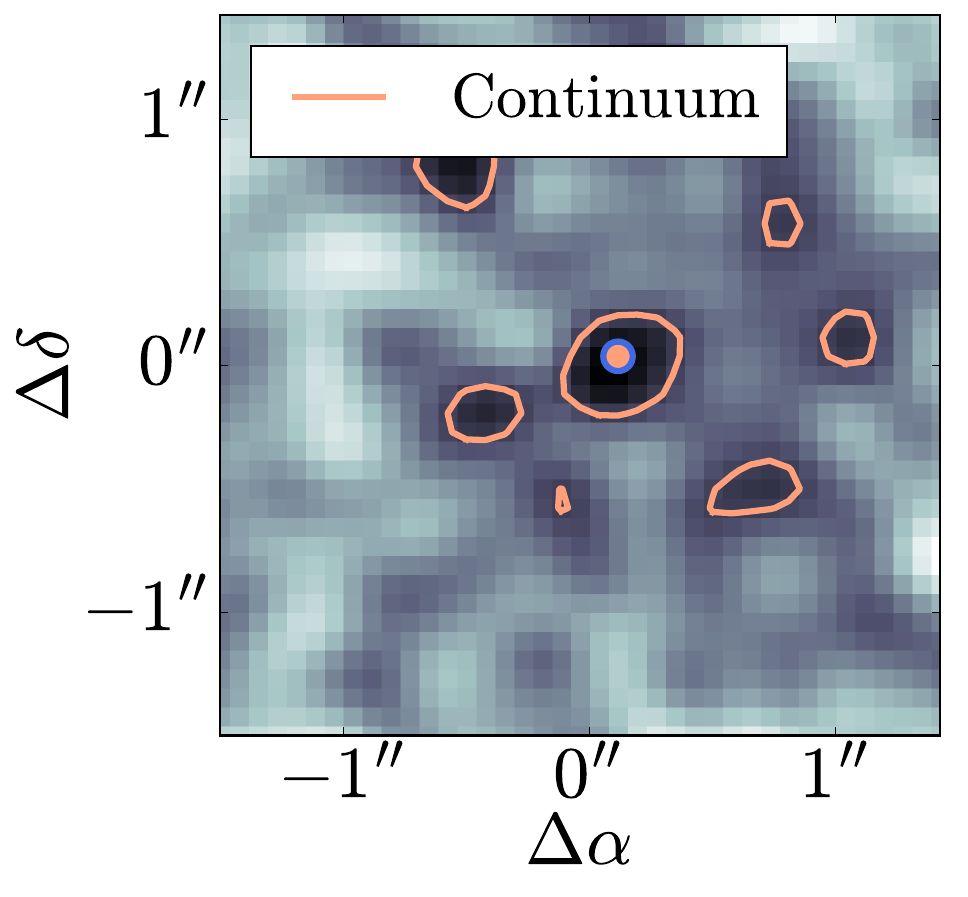} \\
\includegraphics[trim=0 11.2cm 0 0,clip,width=0.30\textwidth]{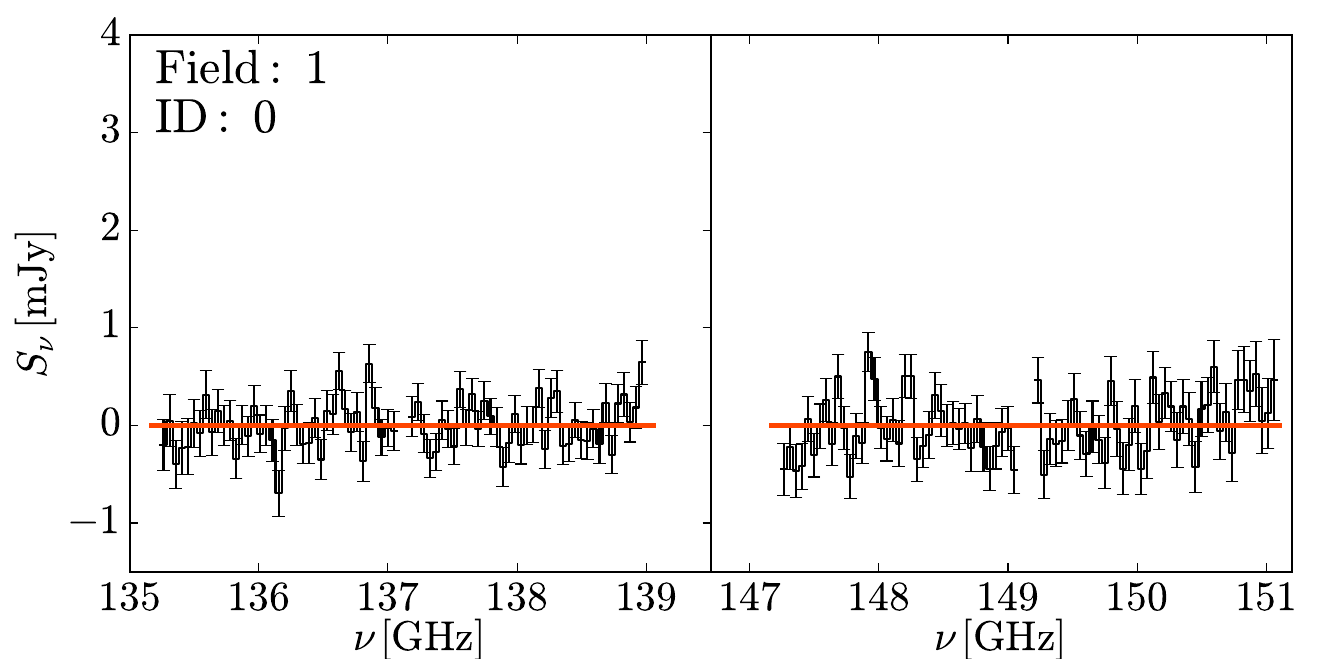}}}
\fbox{
\parbox{0.31\textwidth}{
\centering
\includegraphics[trim=0 0 16.5cm 0,clip,width=0.15\textwidth]{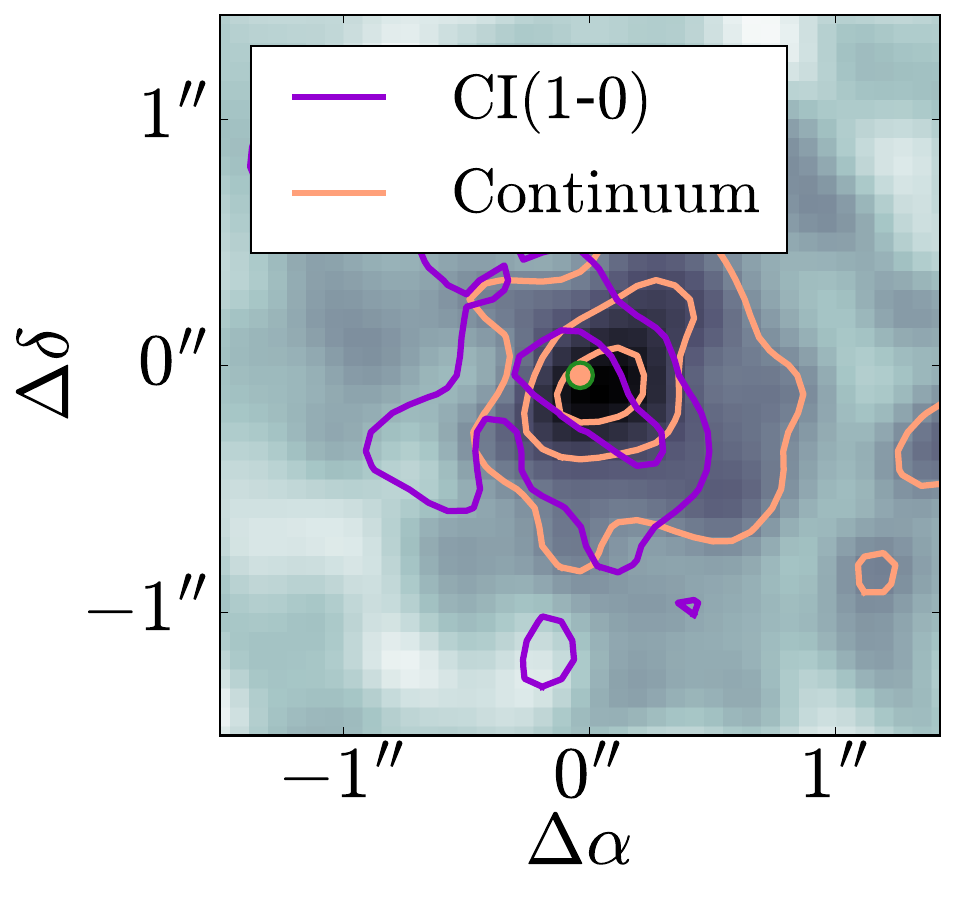} \\
\includegraphics[trim=0 11.2cm 0 0,clip,width=0.30\textwidth]{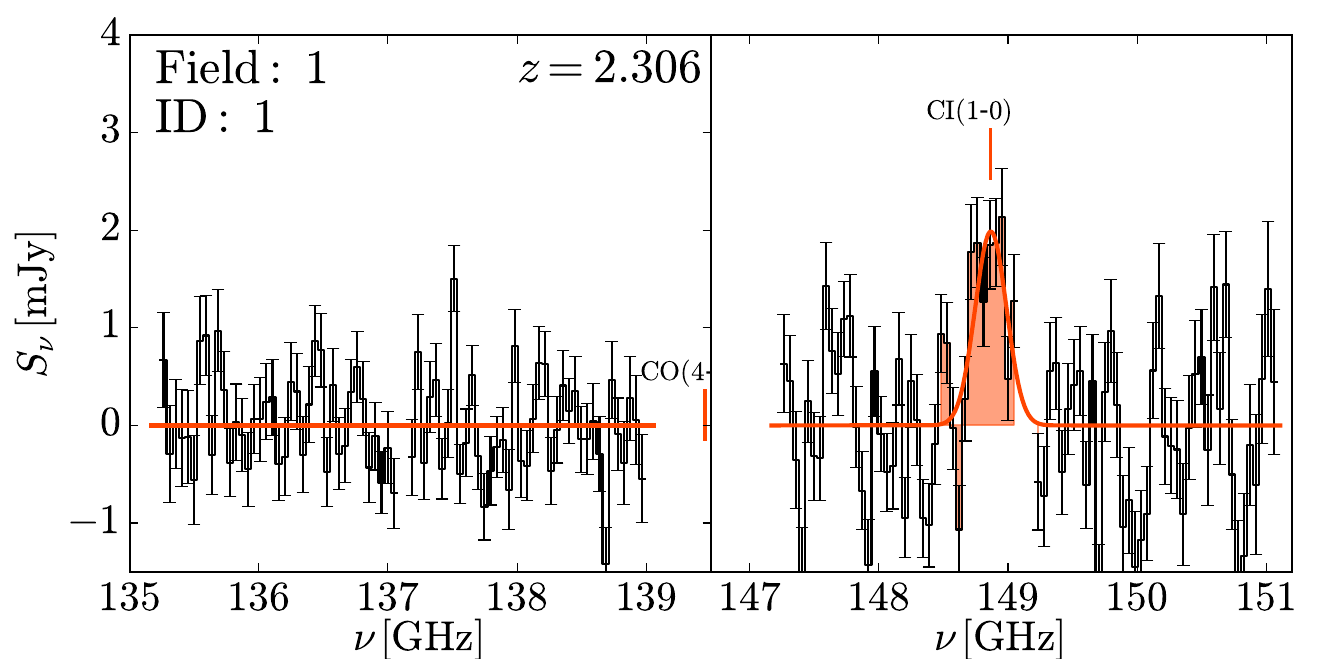}}}
\fbox{
\parbox{0.31\textwidth}{
\centering
\includegraphics[trim=0 0 16.5cm 0,clip,width=0.15\textwidth]{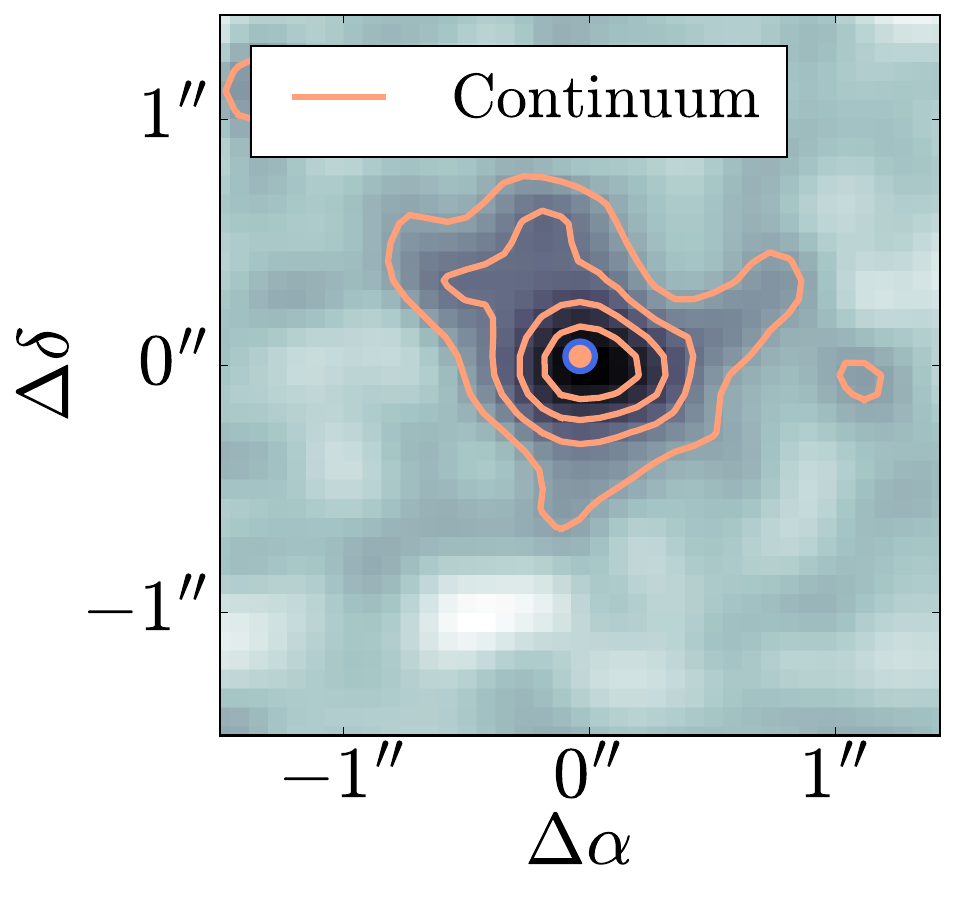} \\
\includegraphics[trim=0 11.2cm 0 0,clip,width=0.30\textwidth]{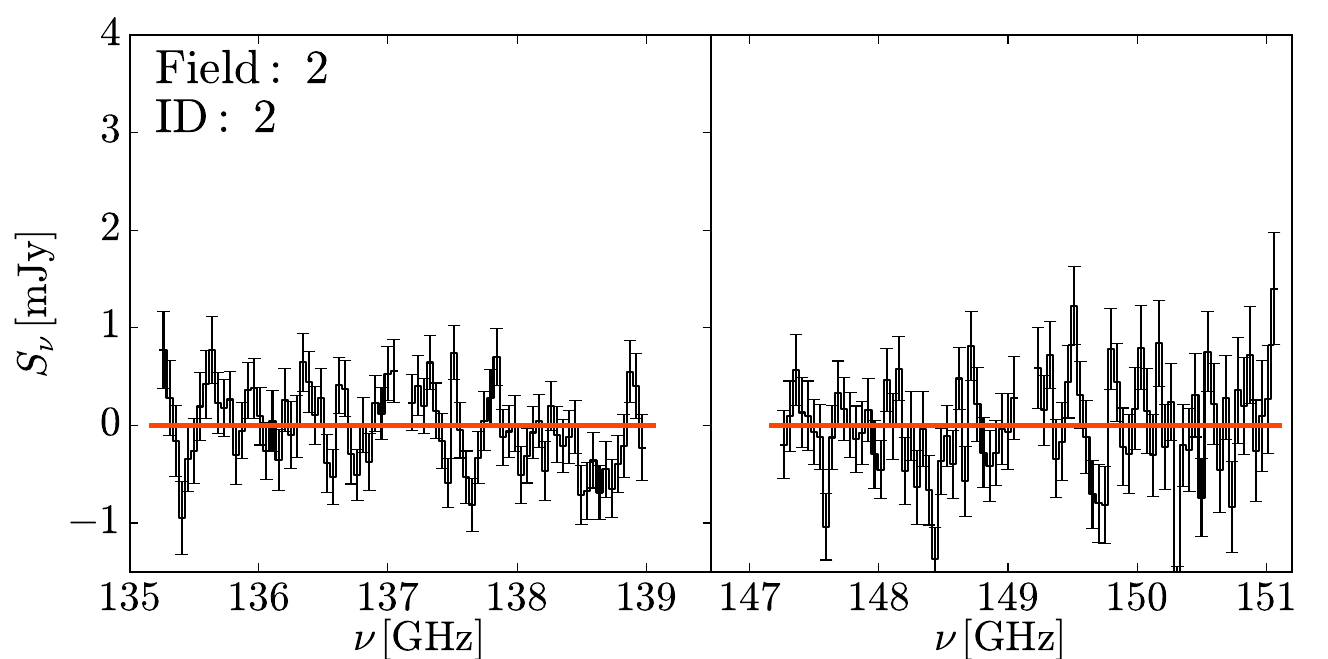}}}
\fbox{
\parbox{0.31\textwidth}{
\centering
\includegraphics[trim=0 0 16.5cm 0,clip,width=0.15\textwidth]{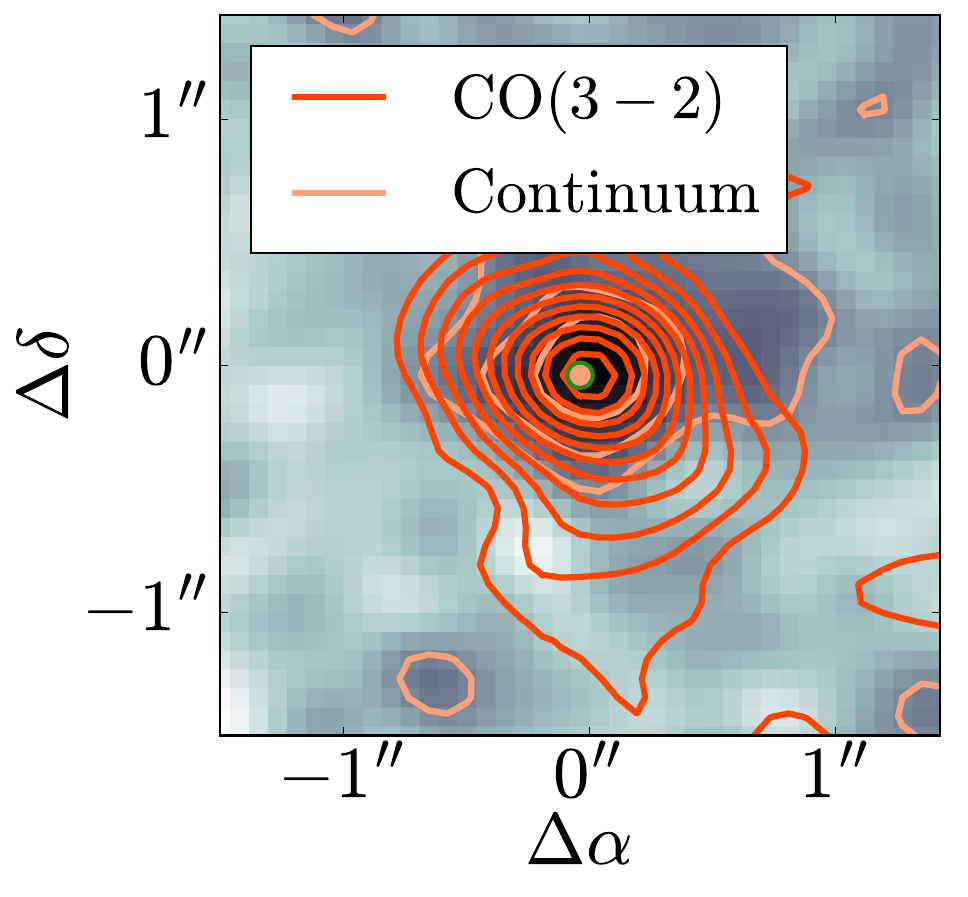} \\
\includegraphics[trim=0 11.2cm 0 0,clip,width=0.30\textwidth]{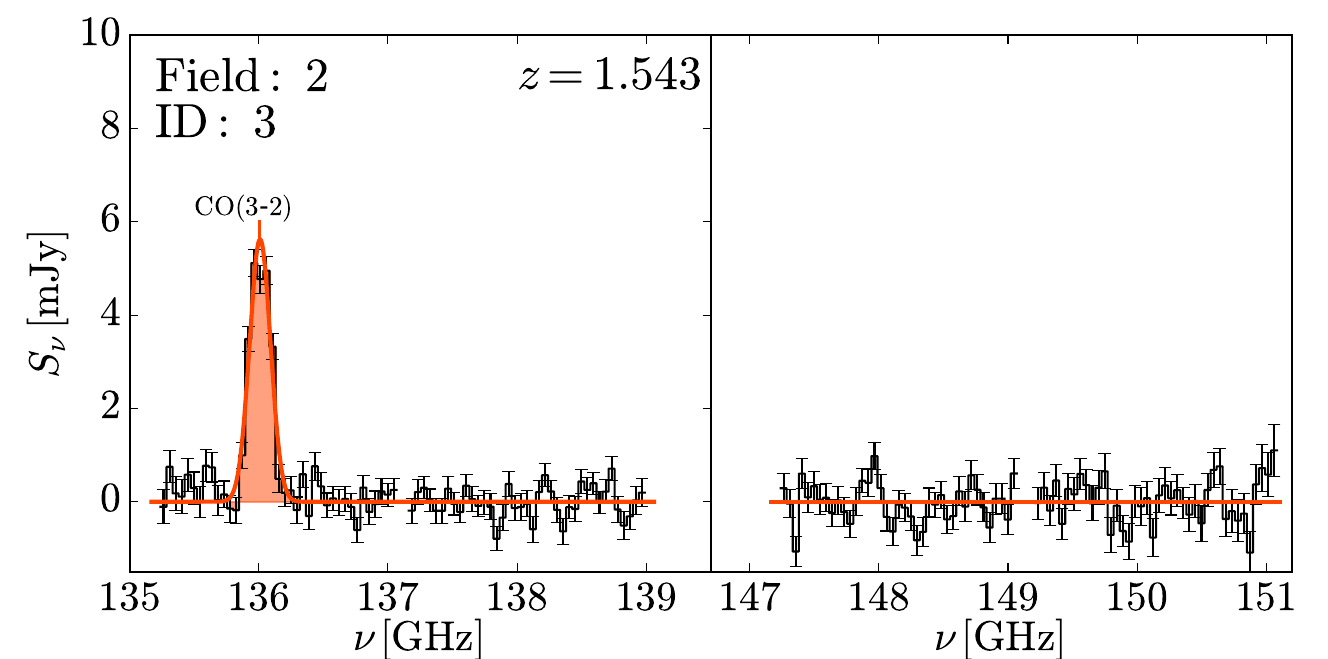}}}
\fbox{
\parbox{0.31\textwidth}{
\centering
\includegraphics[trim=0 0 16.5cm 0,clip,width=0.15\textwidth]{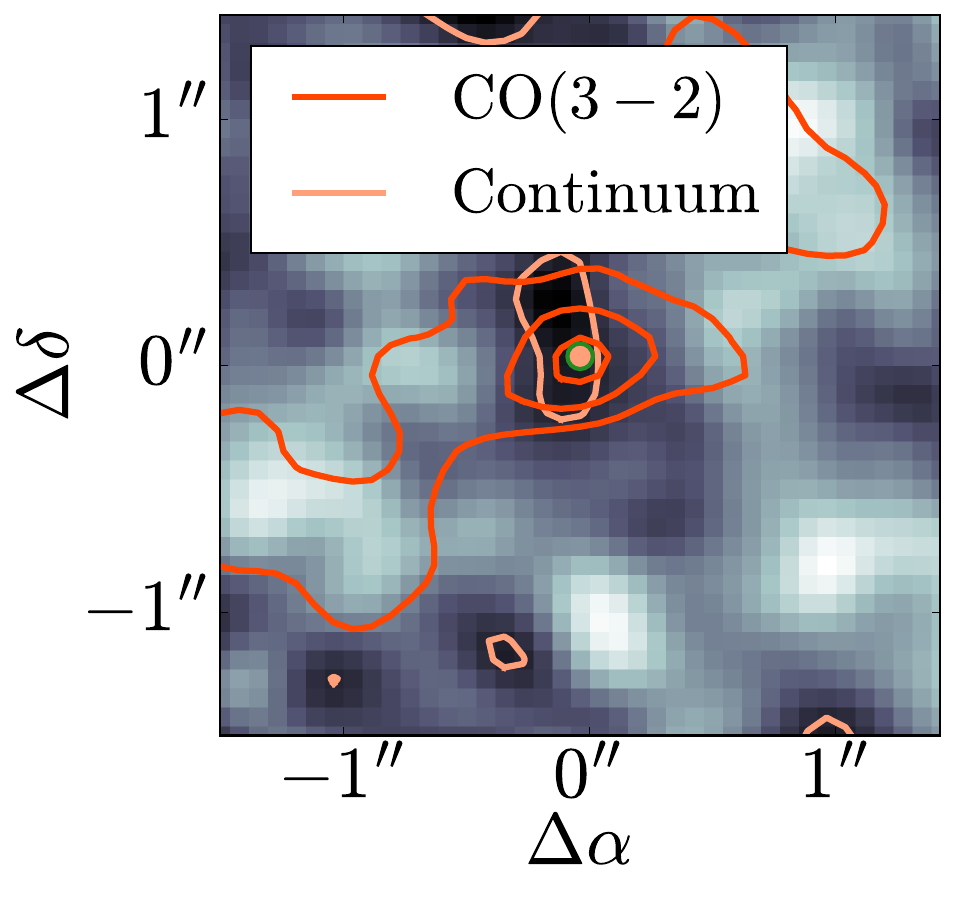} \\
\includegraphics[trim=0 11.2cm 0 0,clip,width=0.30\textwidth]{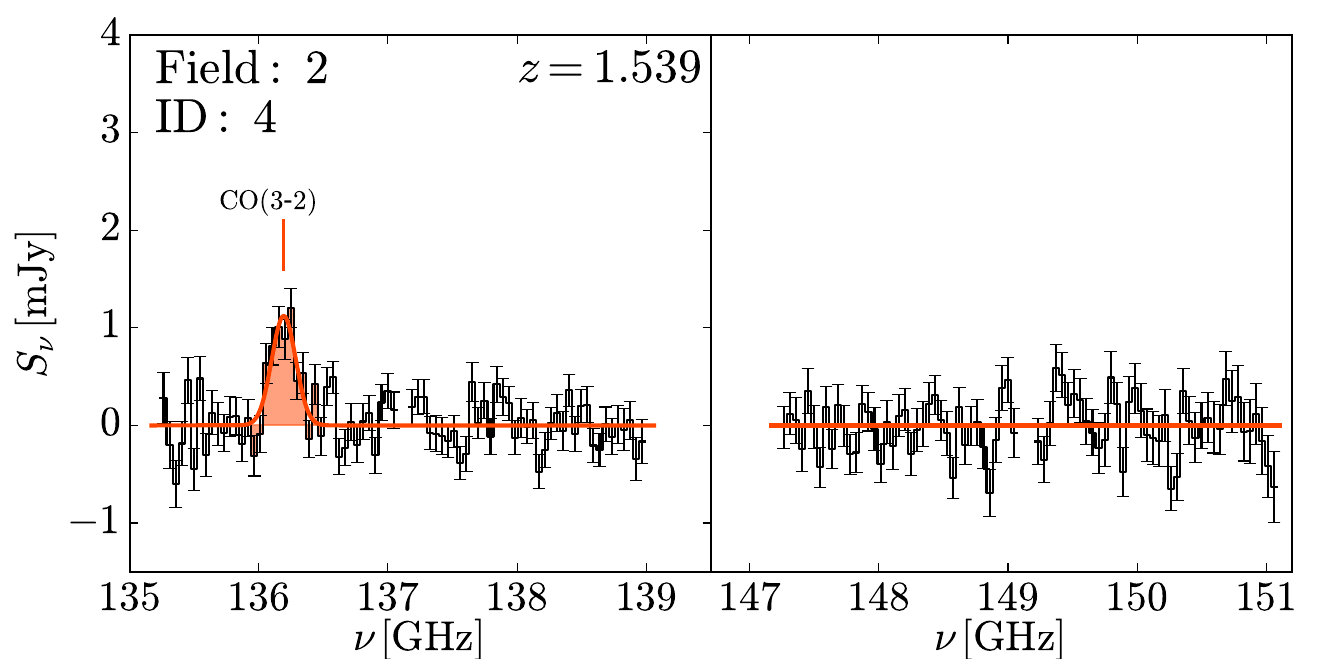}}}
\fbox{
\parbox{0.31\textwidth}{
\centering
\includegraphics[trim=0 0 16.5cm 0,clip,width=0.15\textwidth]{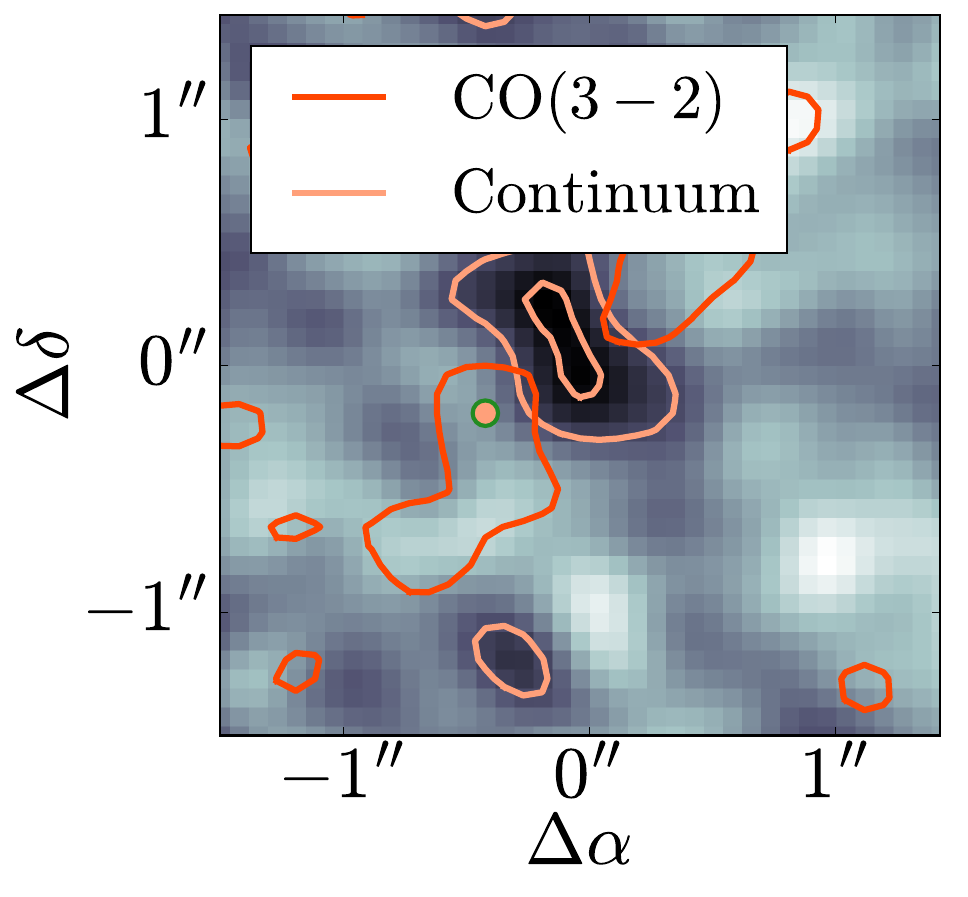} \\
\includegraphics[trim=0 11.2cm 0 0,clip,width=0.30\textwidth]{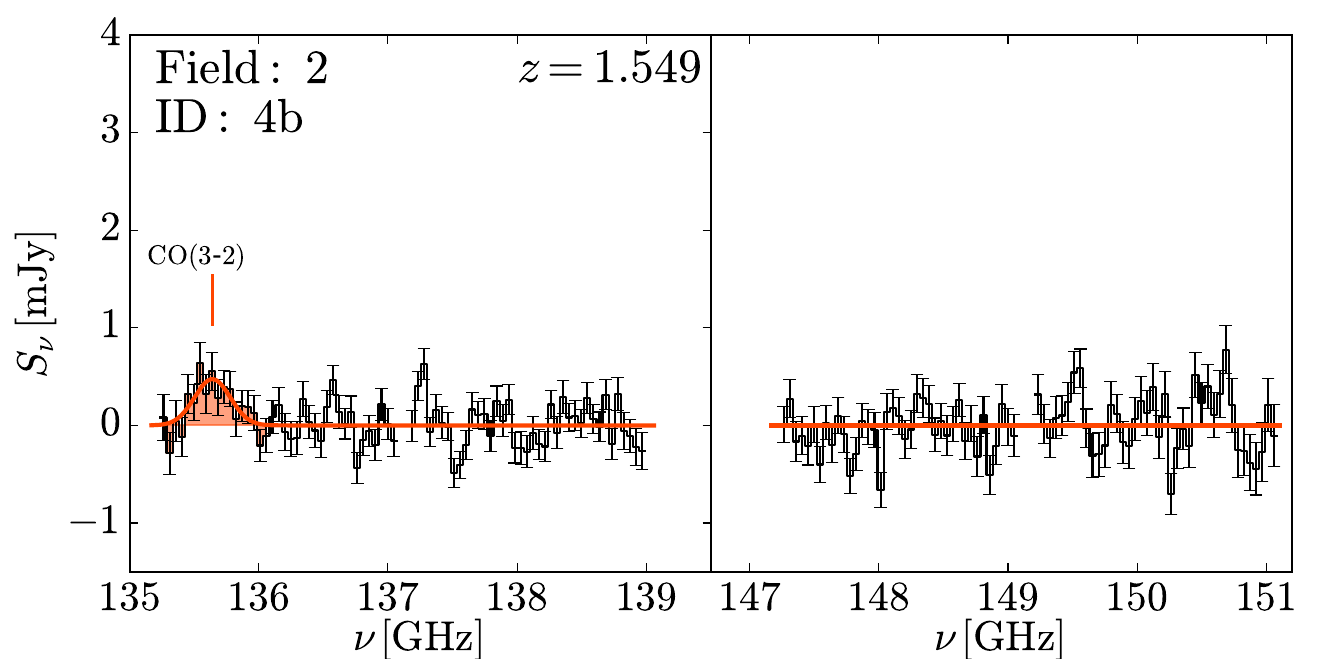}}}
\fbox{
\parbox{0.31\textwidth}{
\centering
\includegraphics[trim=0 0 16.5cm 0,clip,width=0.15\textwidth]{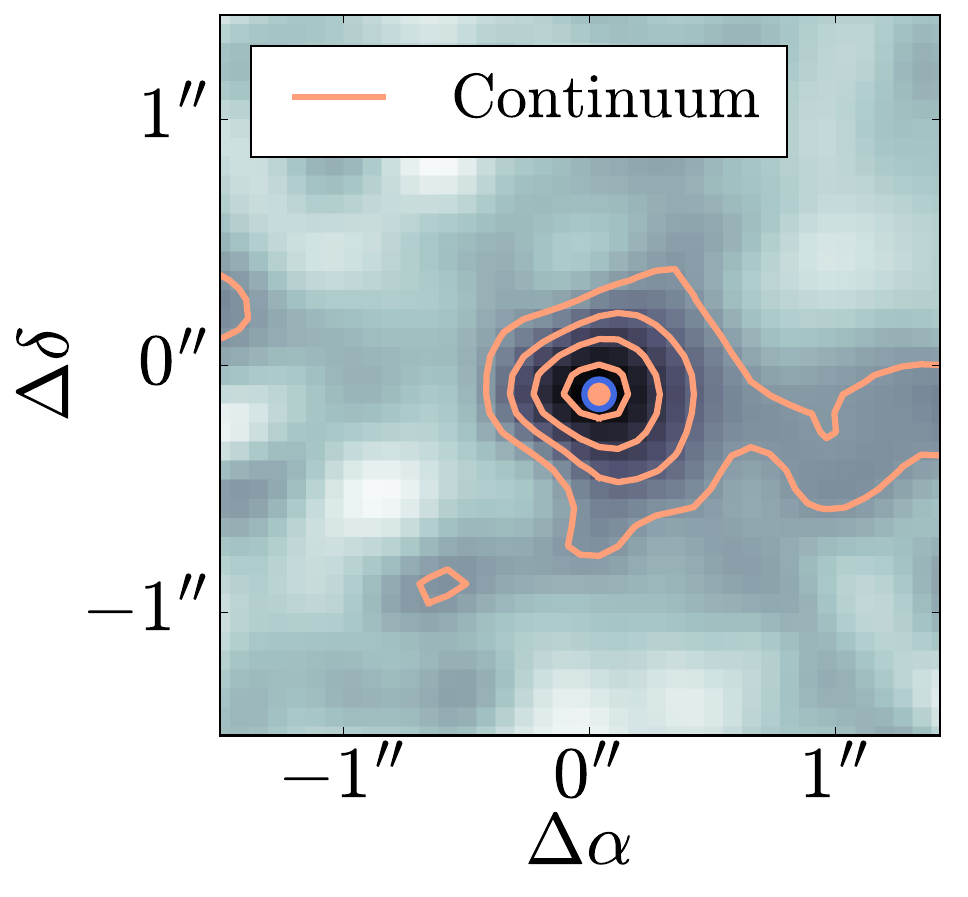} \\
\includegraphics[trim=0 11.2cm 0 0,clip,width=0.30\textwidth]{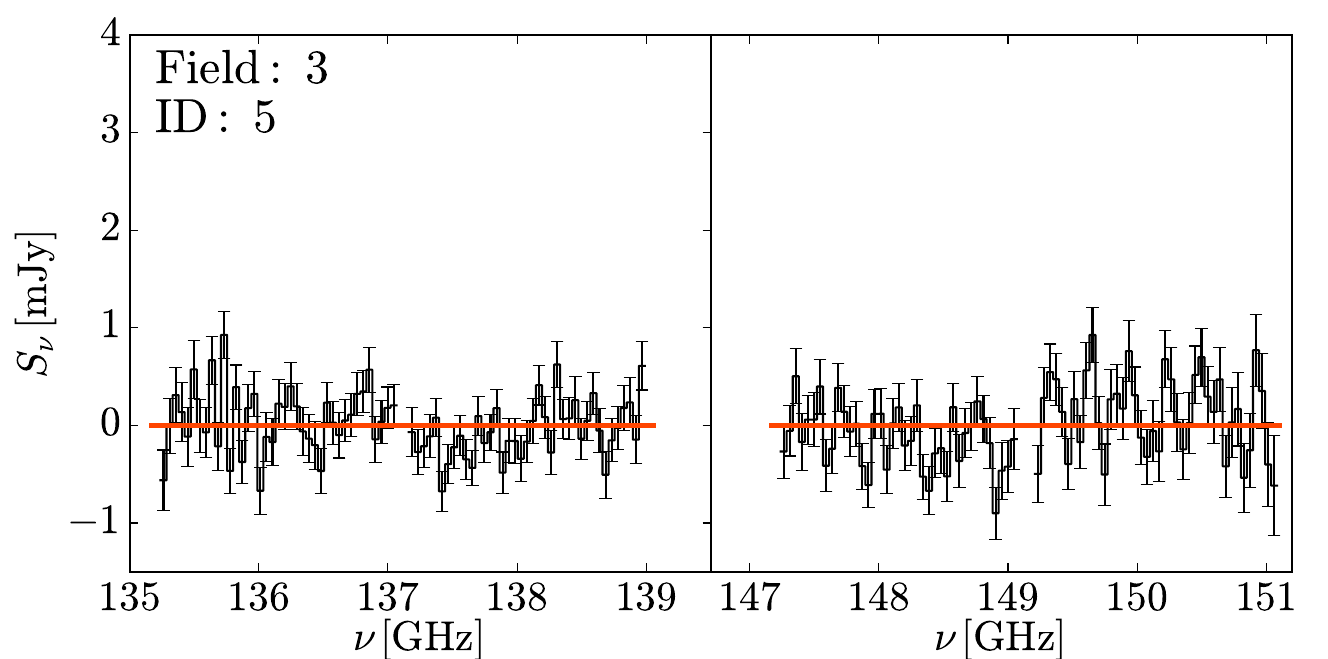}}}
\fbox{
\parbox{0.31\textwidth}{
\centering
\includegraphics[trim=0 0 16.5cm 0,clip,width=0.15\textwidth]{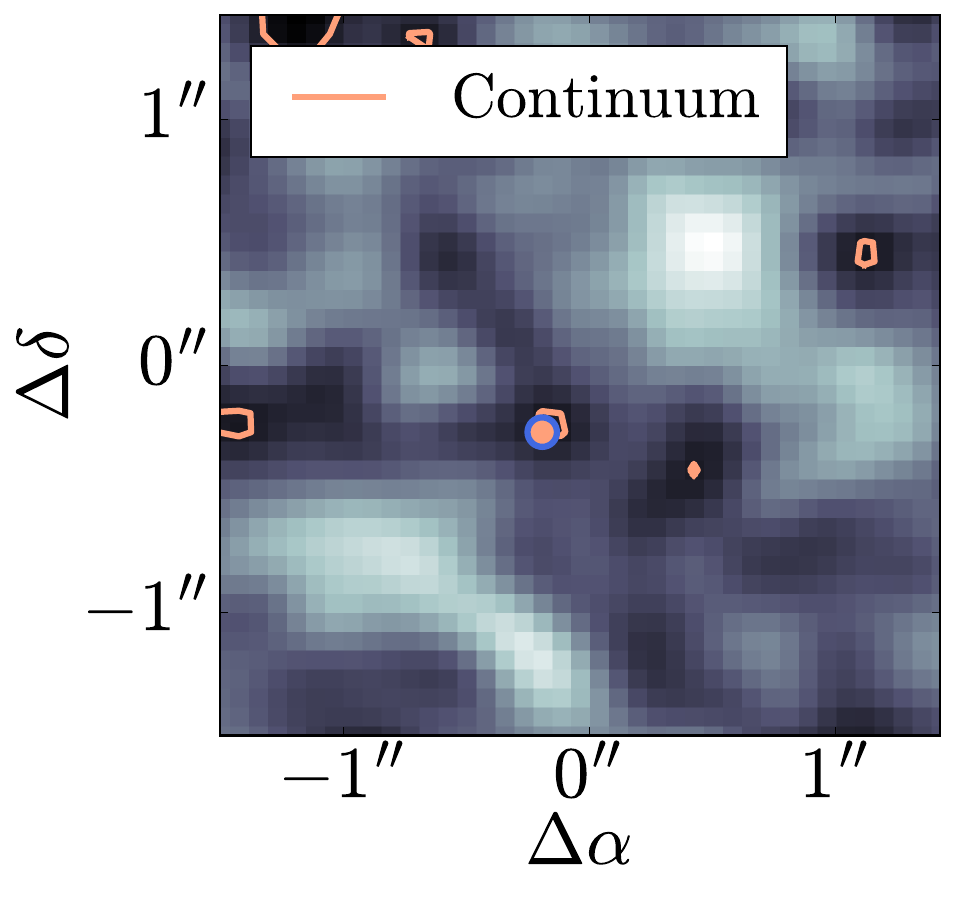} \\
\includegraphics[trim=0 11.2cm 0 0,clip,width=0.30\textwidth]{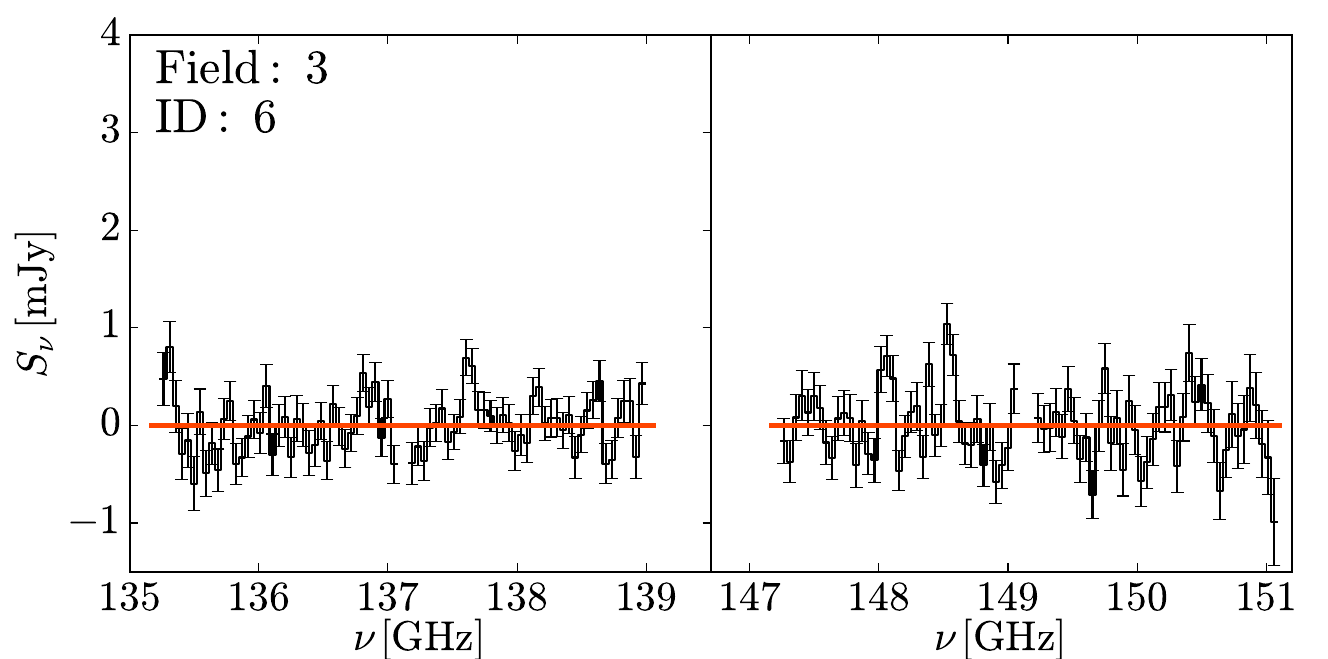}}}
\fbox{
\parbox{0.31\textwidth}{
\centering
\includegraphics[trim=0 0 16.5cm 0,clip,width=0.15\textwidth]{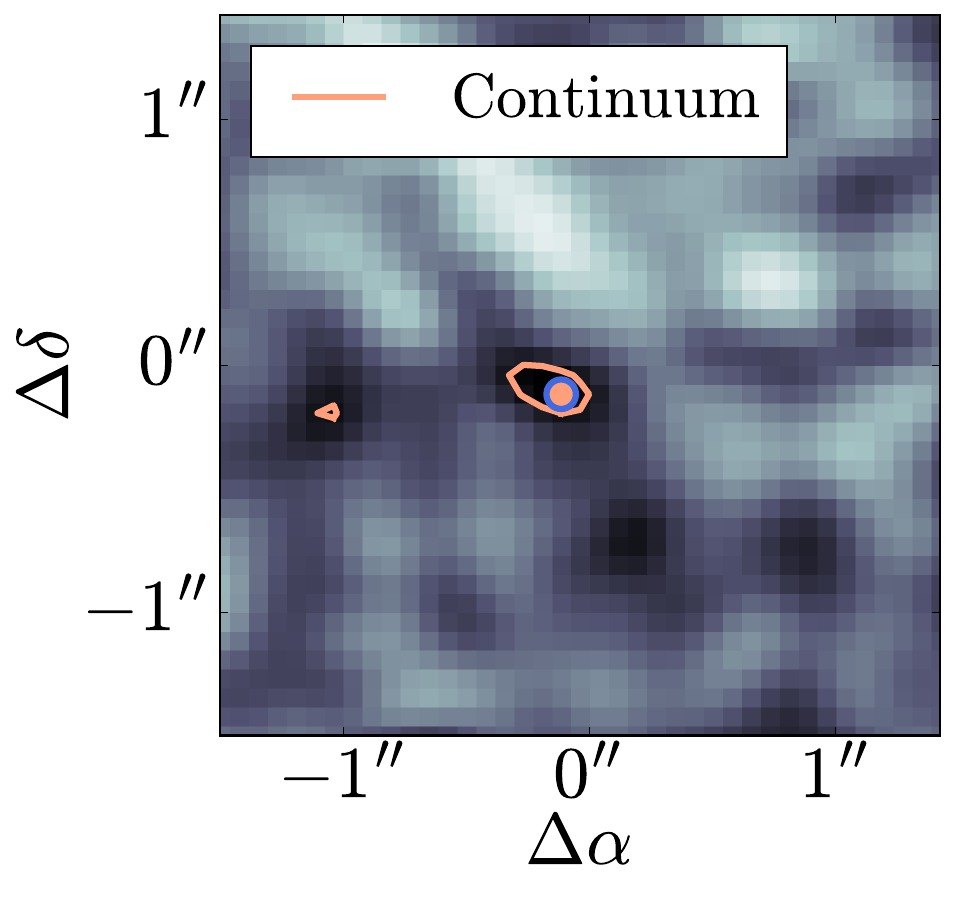} \\
\includegraphics[trim=0 11.2cm 0 0,clip,width=0.30\textwidth]{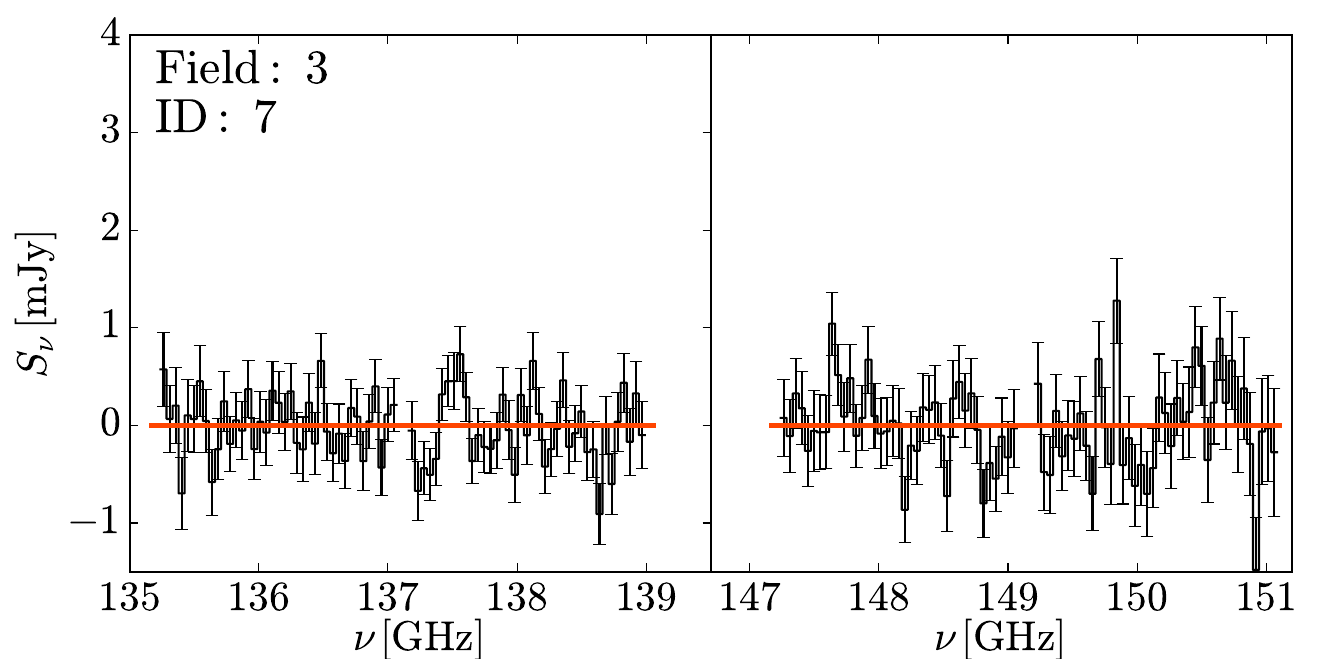}}}
\caption{{ALMA Band~4 cutouts and spectra for all sources in G073. The upper panels show 3\arcsec$\,{\times}\,$3\arcsec cutouts, with continuum images in the background highlighted by orange contours starting at 2$\sigma$ and increasing in steps of 3$\sigma$. For sources with line detections, moment-0 contours (red) are overlaid following the same steps. The lower panels show the continuum-subtracted LSB and USB spectra, with the best-fitting Gaussian functions plotted over the raw data and the shaded regions indicating the integration range used to measure the line strengths.}}
\label{fig:spectra}
\end{figure*}

\begin{figure*}[h!]
\setcounter{figure}{0}
\fbox{
\parbox{0.31\textwidth}{
\centering
\includegraphics[trim=0 0 16.5cm 0,clip,width=0.15\textwidth]{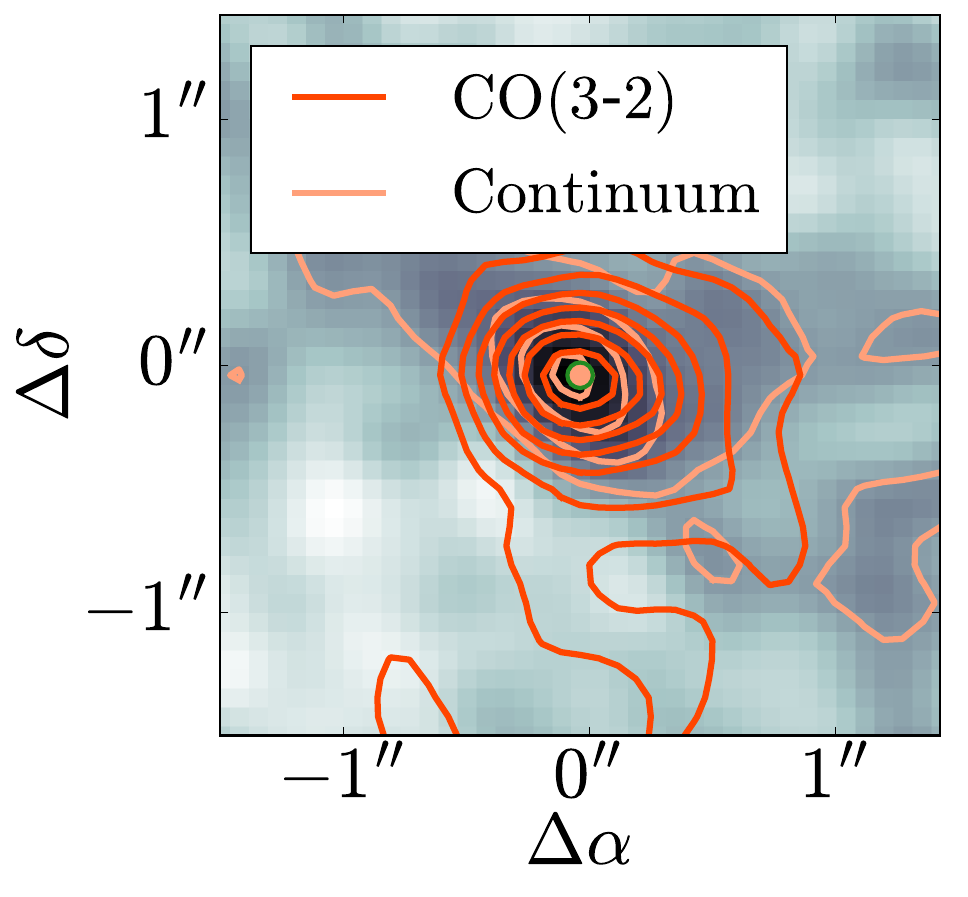} \\
\includegraphics[trim=0 11.2cm 0 0,clip,width=0.30\textwidth]{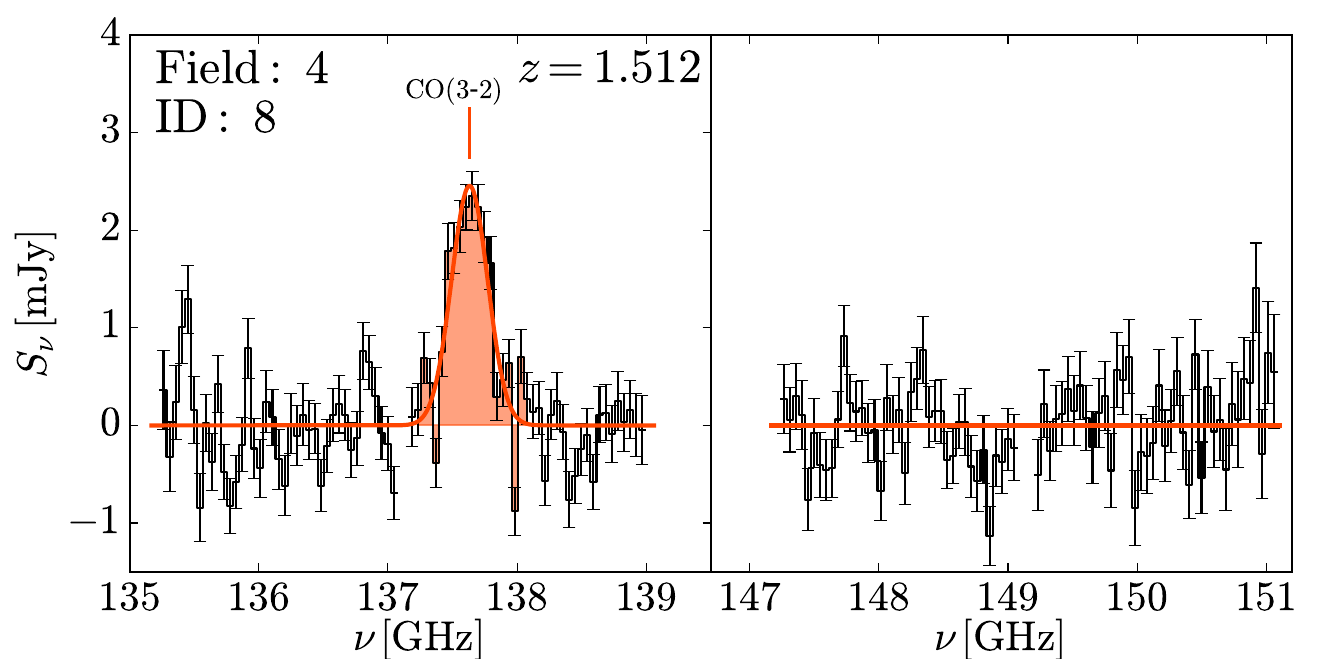}}}
\fbox{
\parbox{0.31\textwidth}{
\centering
\includegraphics[trim=0 0 16.5cm 0,clip,width=0.15\textwidth]{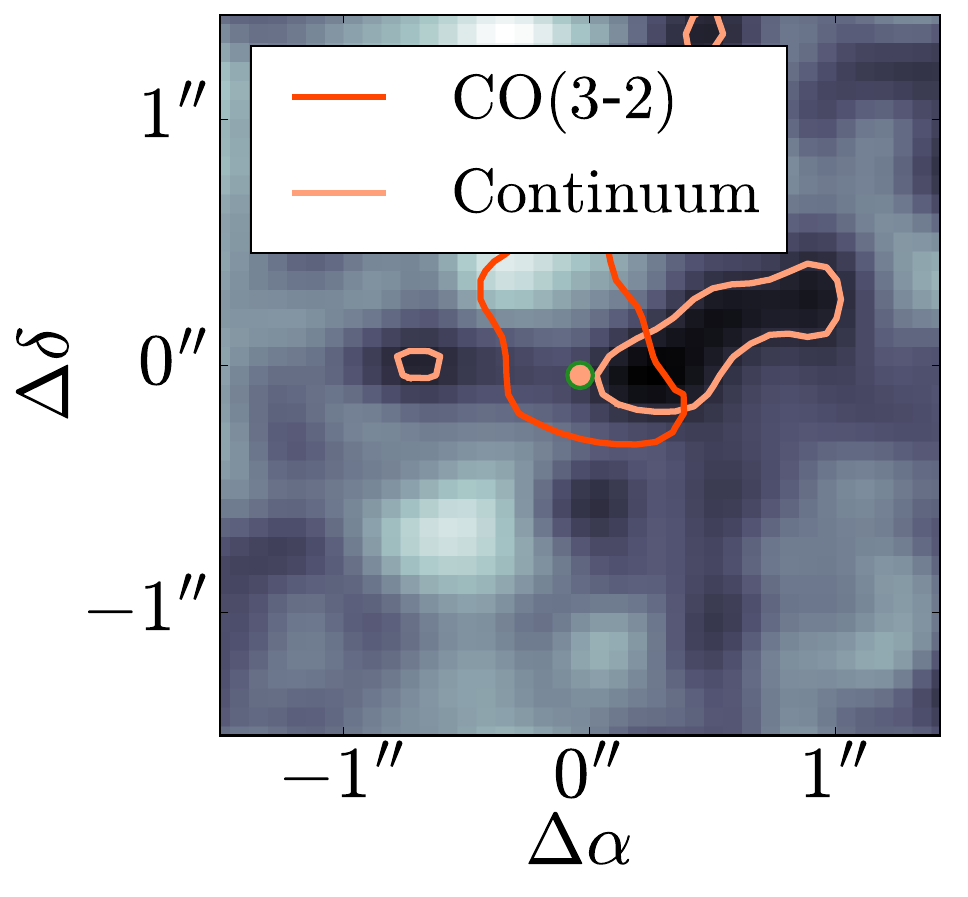} \\
\includegraphics[trim=0 11.2cm 0 0,clip,width=0.30\textwidth]{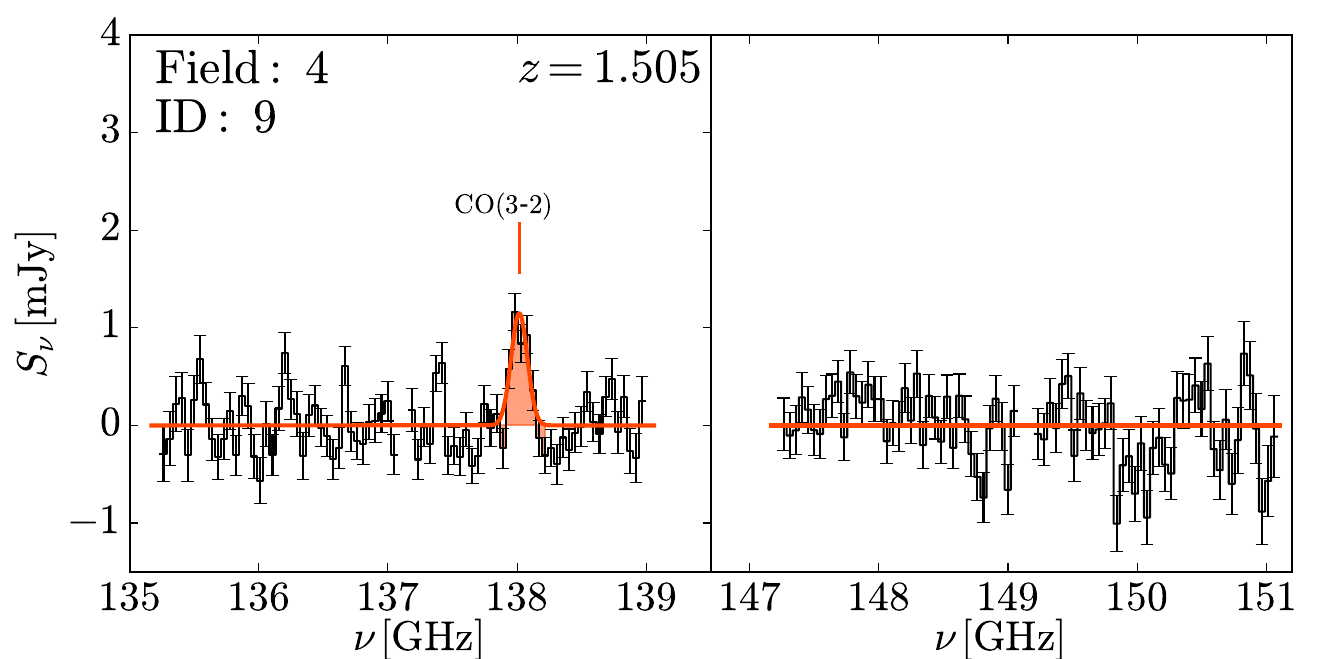}}}
\fbox{
\parbox{0.31\textwidth}{
\centering
\includegraphics[trim=0 0 16.5cm 0,clip,width=0.15\textwidth]{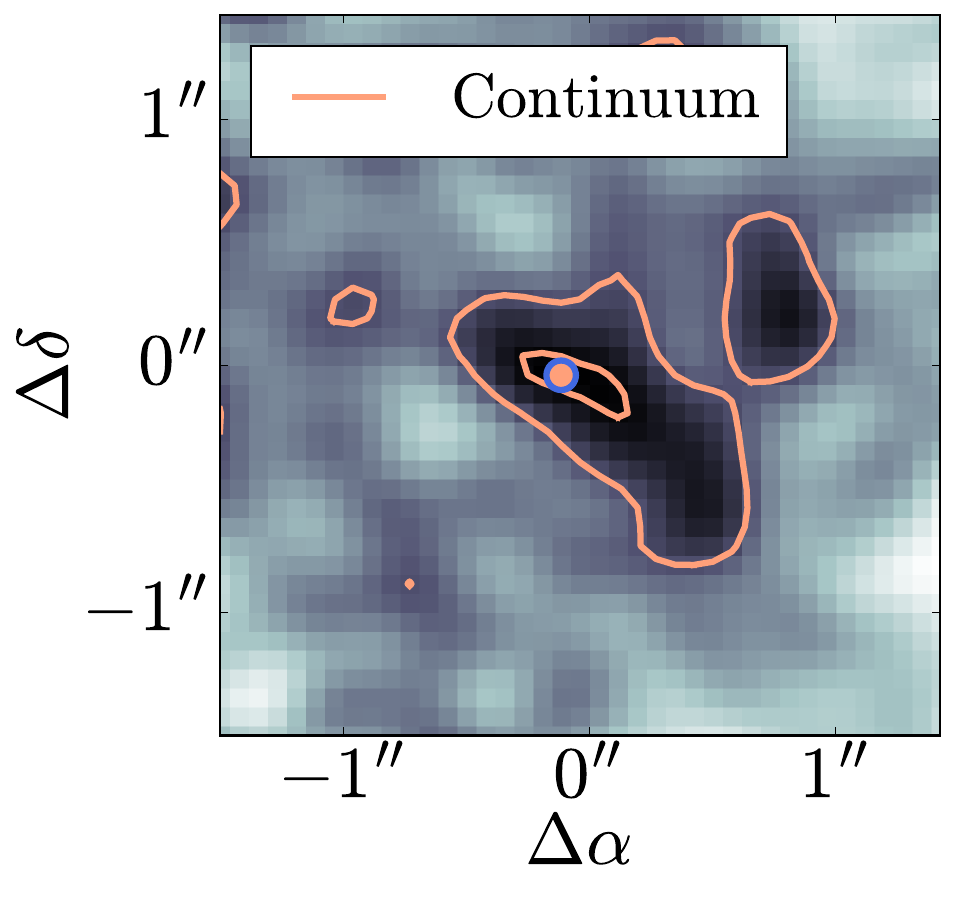} \\
\includegraphics[trim=0 11.2cm 0 0,clip,width=0.30\textwidth]{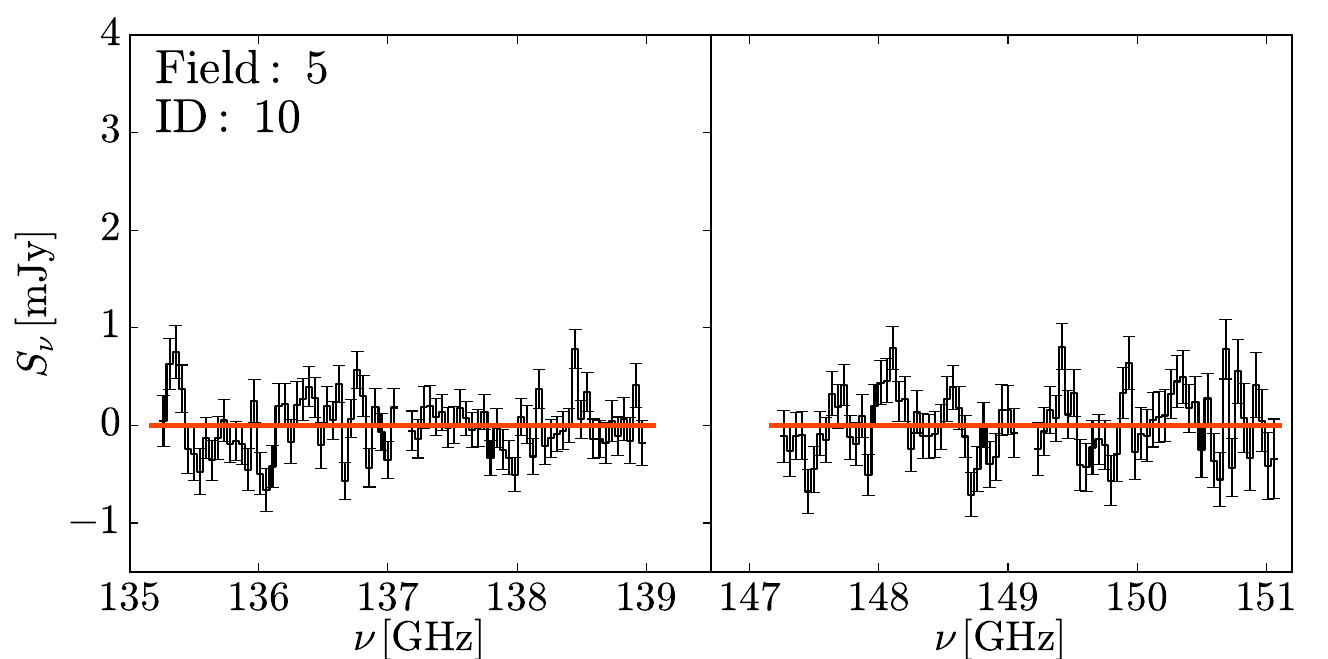}}}
\fbox{
\parbox{0.31\textwidth}{
\centering
\includegraphics[trim=0 0 16.5cm 0,clip,width=0.15\textwidth]{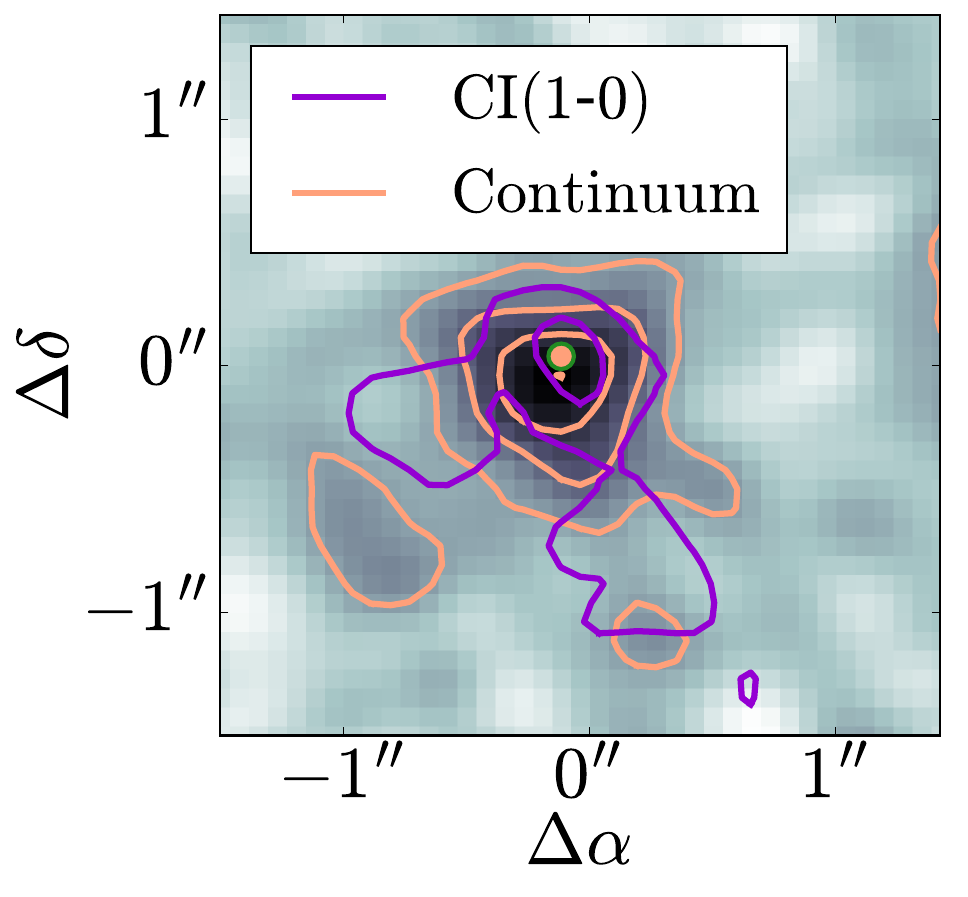} \\
\includegraphics[trim=0 11.2cm 0 0,clip,width=0.30\textwidth]{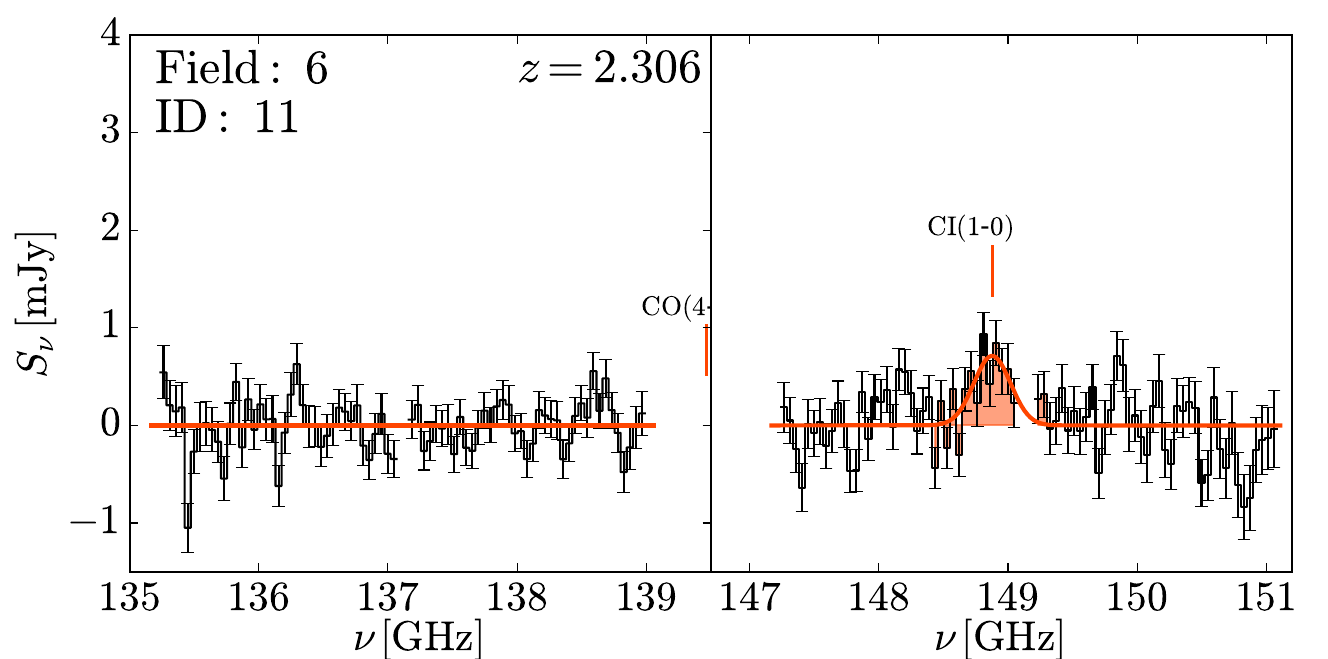}}}
\fbox{
\parbox{0.31\textwidth}{
\centering
\includegraphics[trim=0 0 16.5cm 0,clip,width=0.15\textwidth]{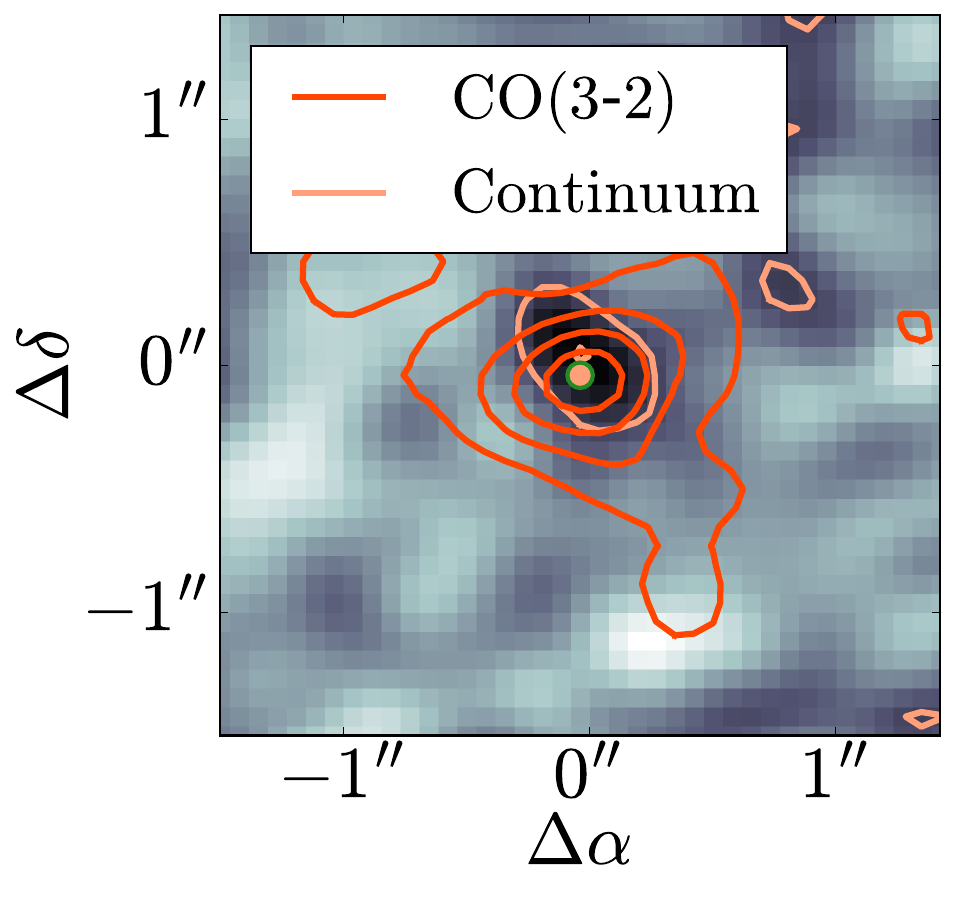} \\
\includegraphics[trim=0 11.2cm 0 0,clip,width=0.30\textwidth]{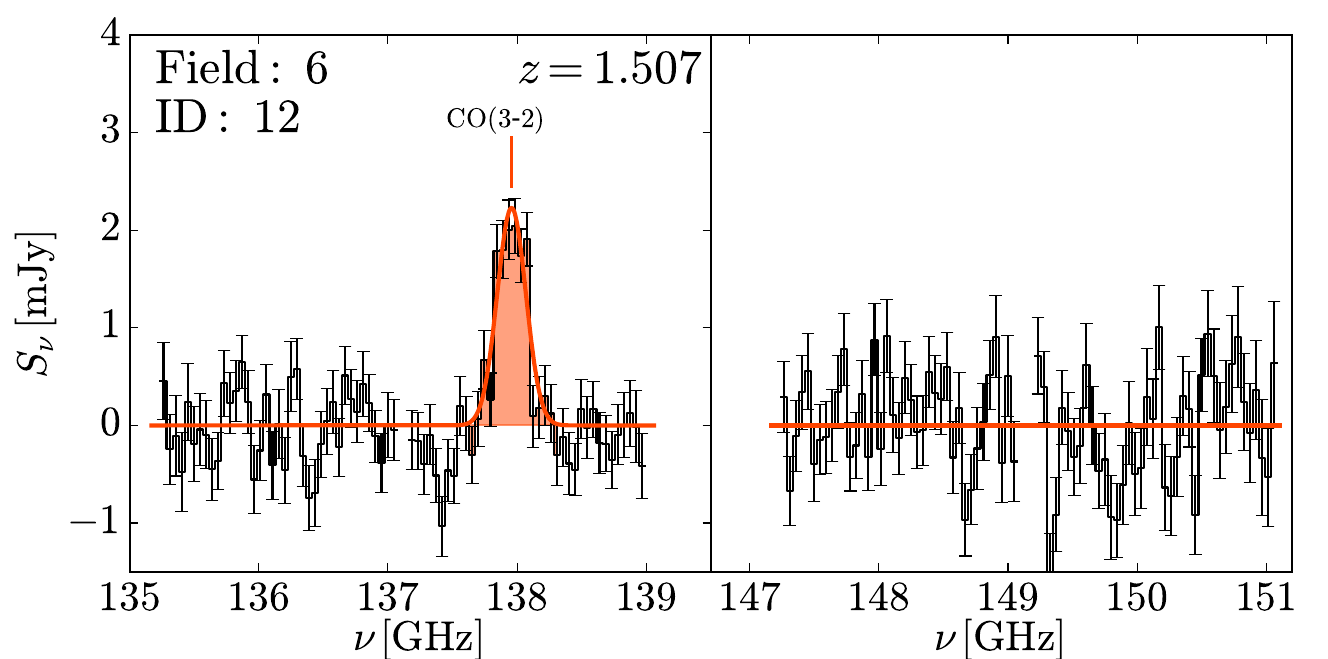}}}
\fbox{
\parbox{0.31\textwidth}{
\centering
\includegraphics[trim=0 0 16.5cm 0,clip,width=0.15\textwidth]{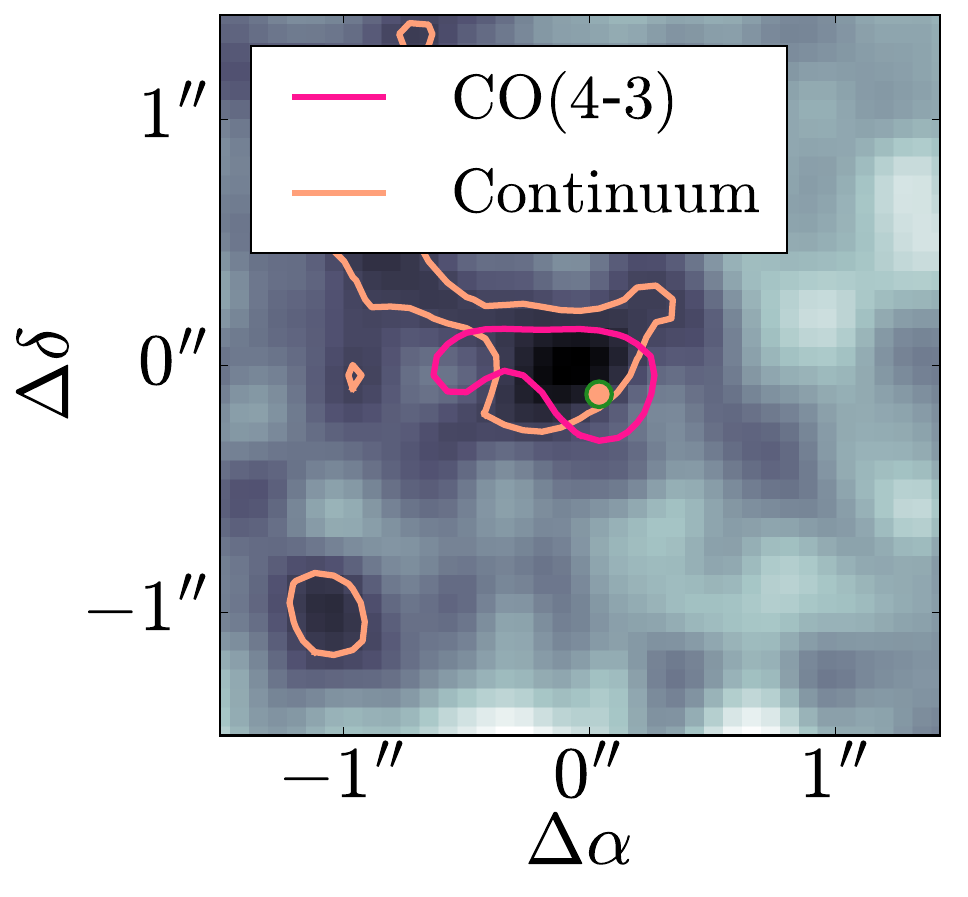}
\includegraphics[trim=0 0 16.5cm 0,clip,width=0.15\textwidth]{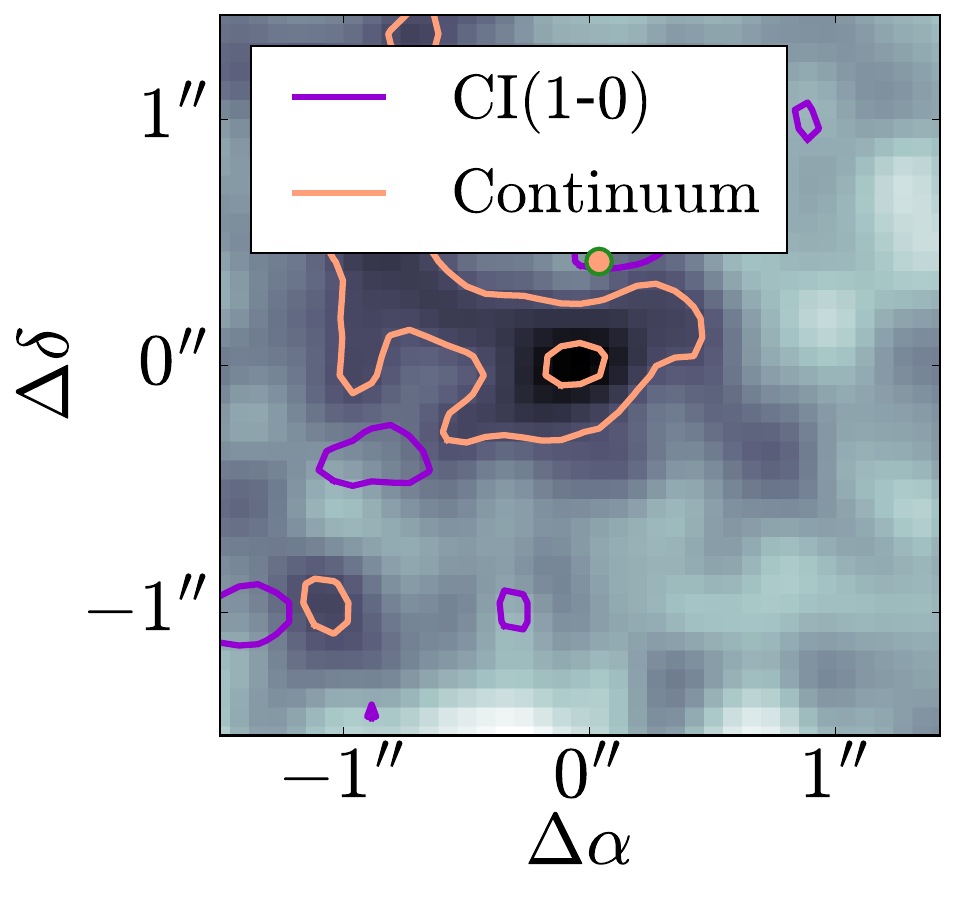} \\
\includegraphics[trim=0 11.2cm 0 0,clip,width=0.30\textwidth]{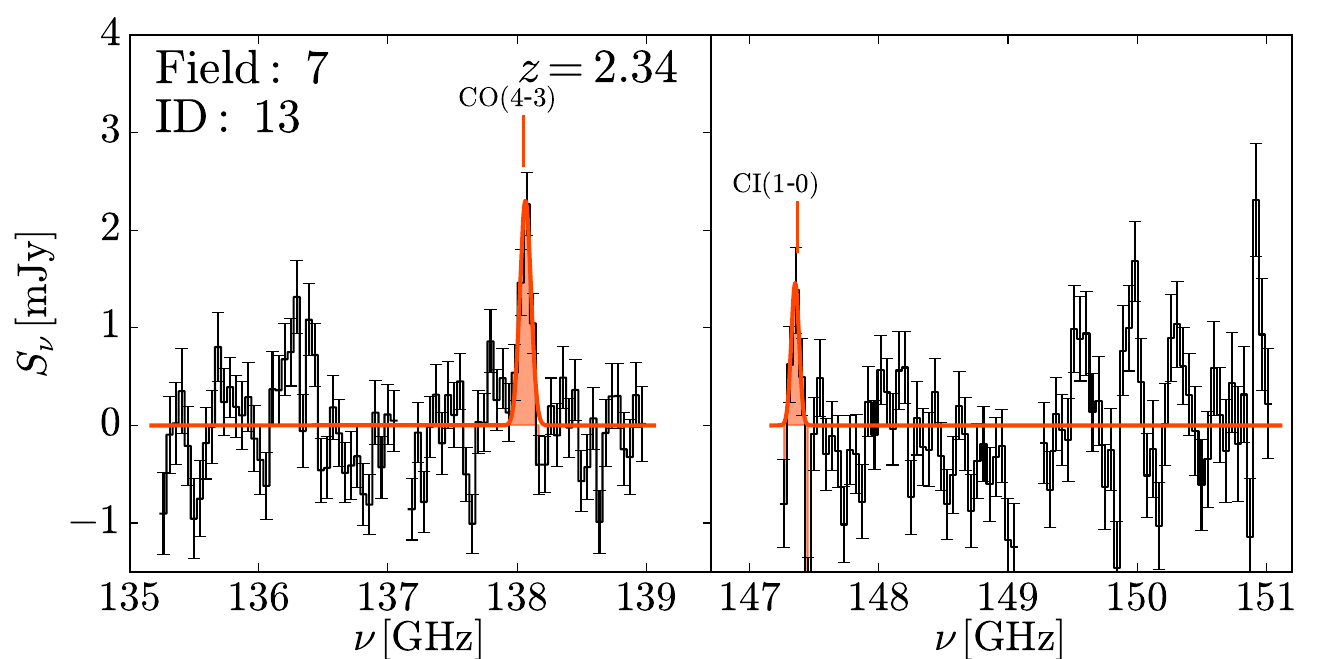}}}
\fbox{
\parbox{0.31\textwidth}{
\centering
\includegraphics[trim=0 0 16.5cm 0,clip,width=0.15\textwidth]{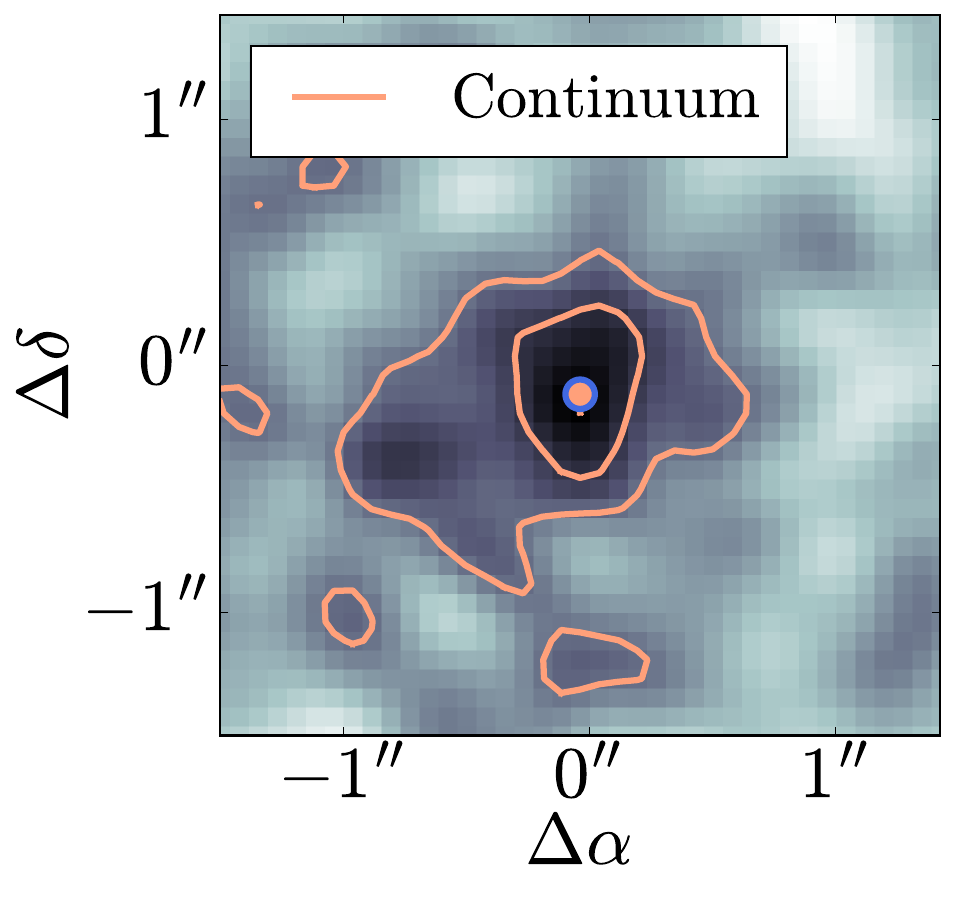} \\
\includegraphics[trim=0 11.2cm 0 0,clip,width=0.30\textwidth]{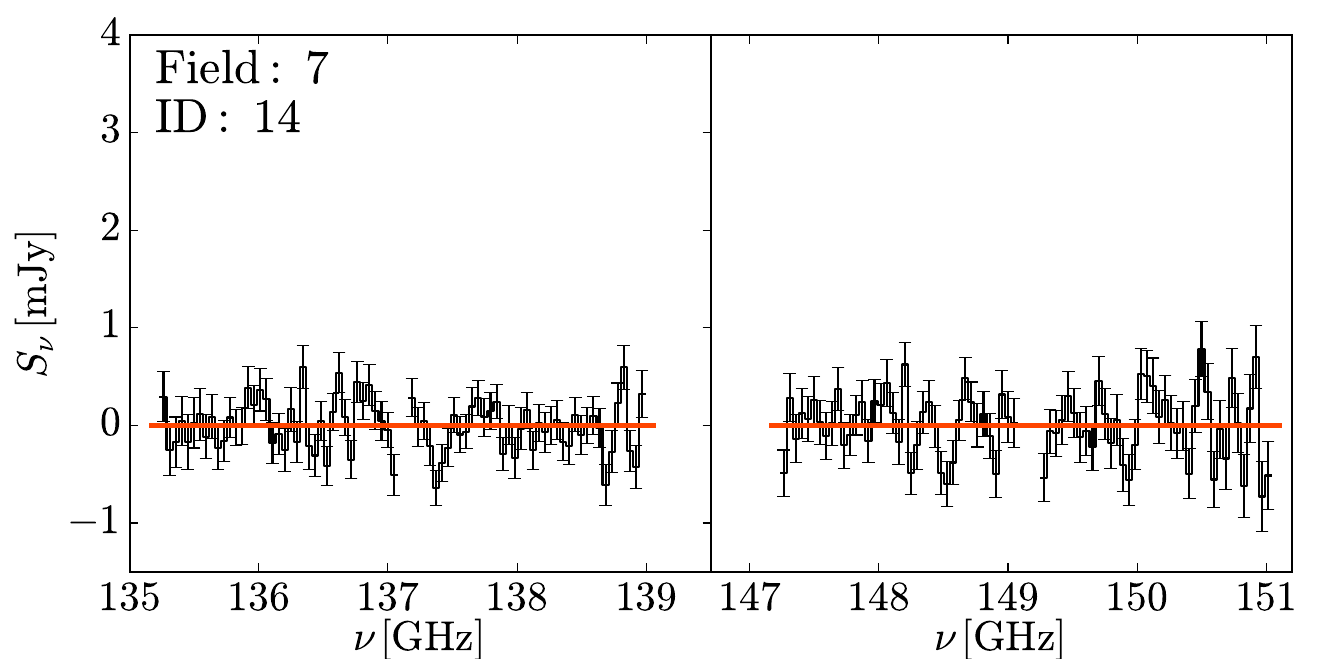}}}
\fbox{
\parbox{0.31\textwidth}{
\centering
\includegraphics[trim=0 0 16.5cm 0,clip,width=0.15\textwidth]{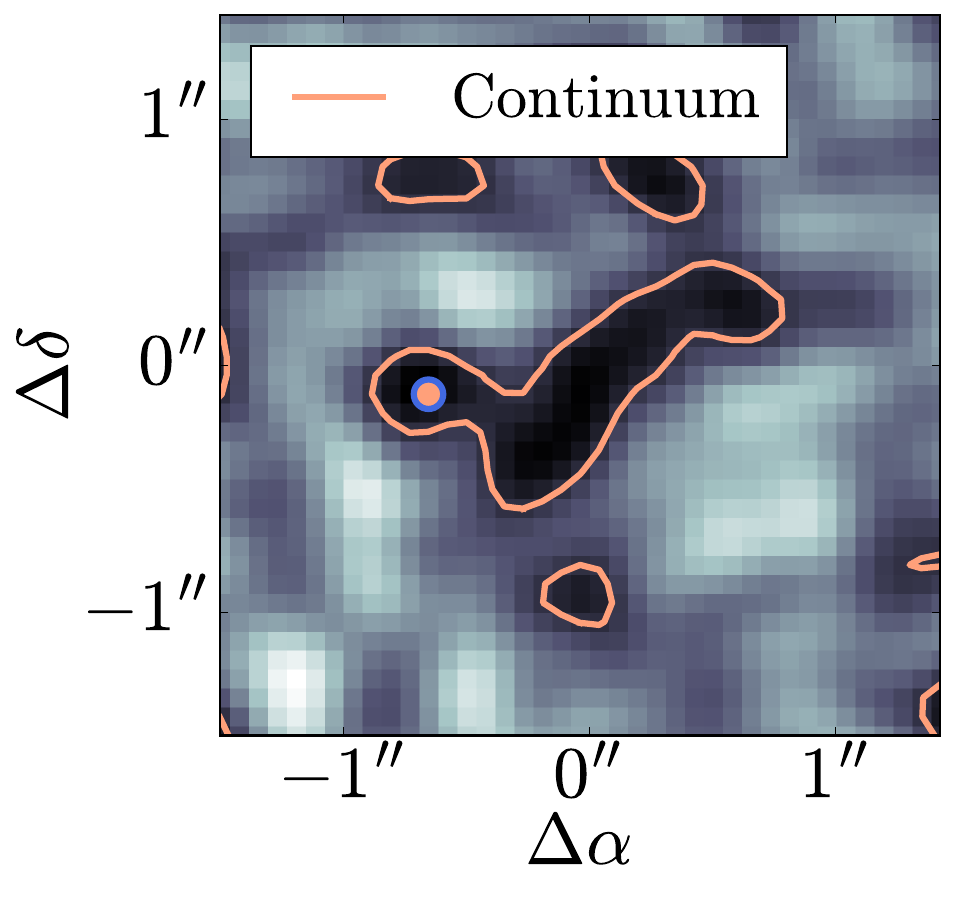} \\
\includegraphics[trim=0 11.2cm 0 0,clip,width=0.30\textwidth]{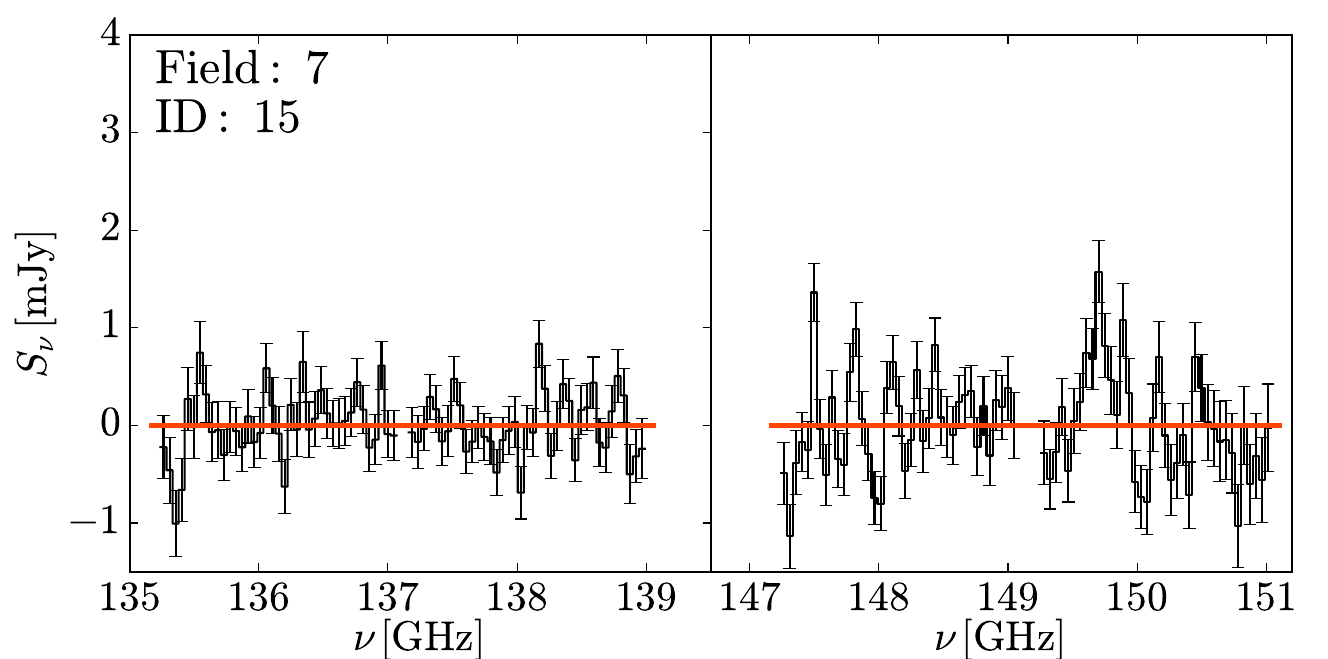}}}
\fbox{
\parbox{0.31\textwidth}{
\centering
\includegraphics[trim=0 0 16.5cm 0,clip,width=0.15\textwidth]{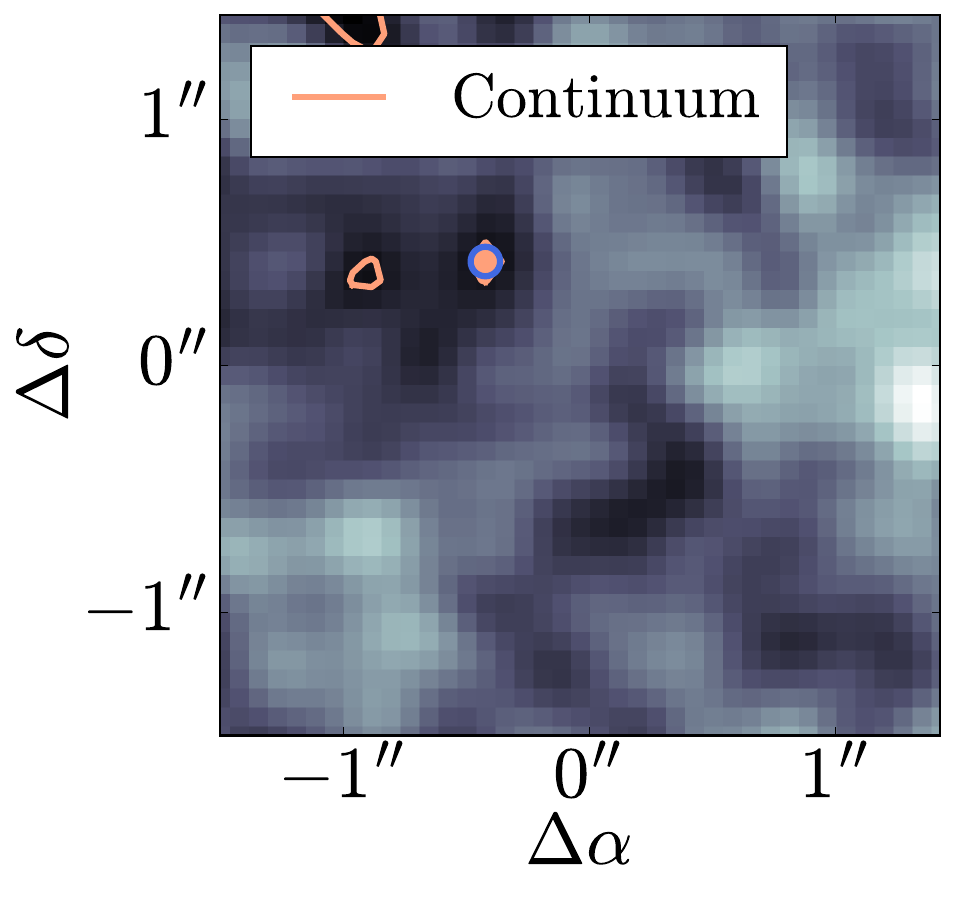} \\
\includegraphics[trim=0 11.2cm 0 0,clip,width=0.30\textwidth]{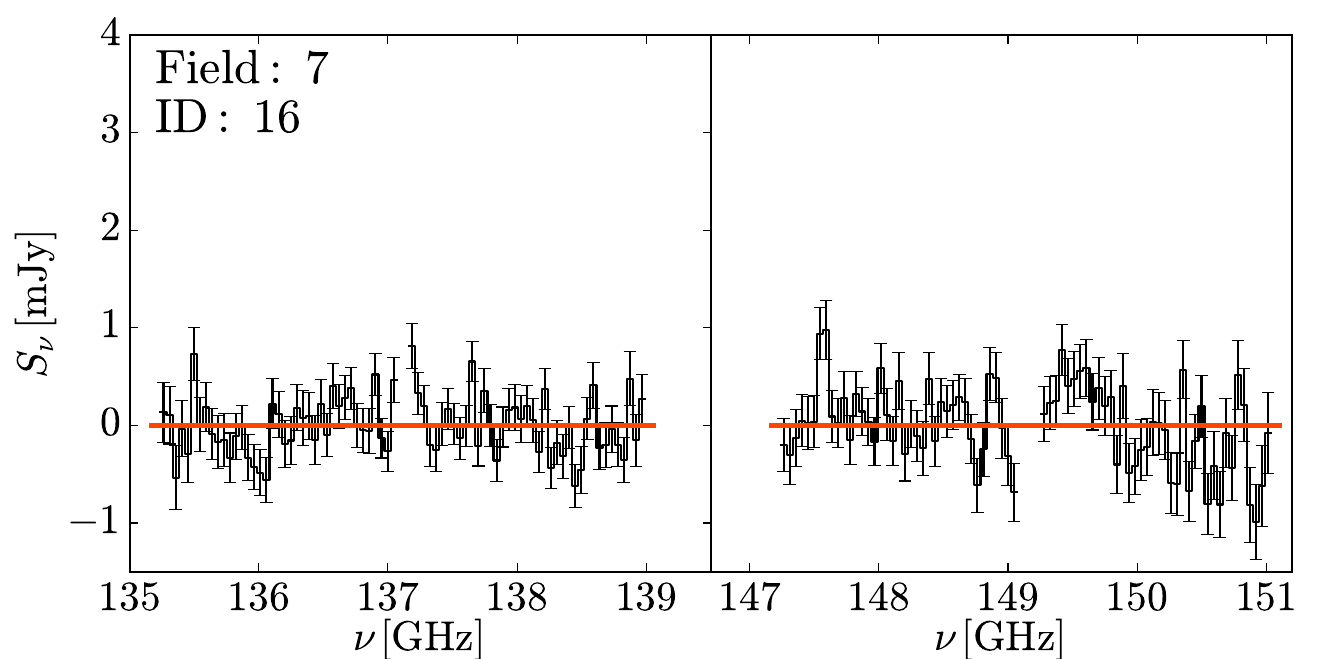}}}
\fbox{
\parbox{0.31\textwidth}{
\centering
\includegraphics[trim=0 0 16.5cm 0,clip,width=0.15\textwidth]{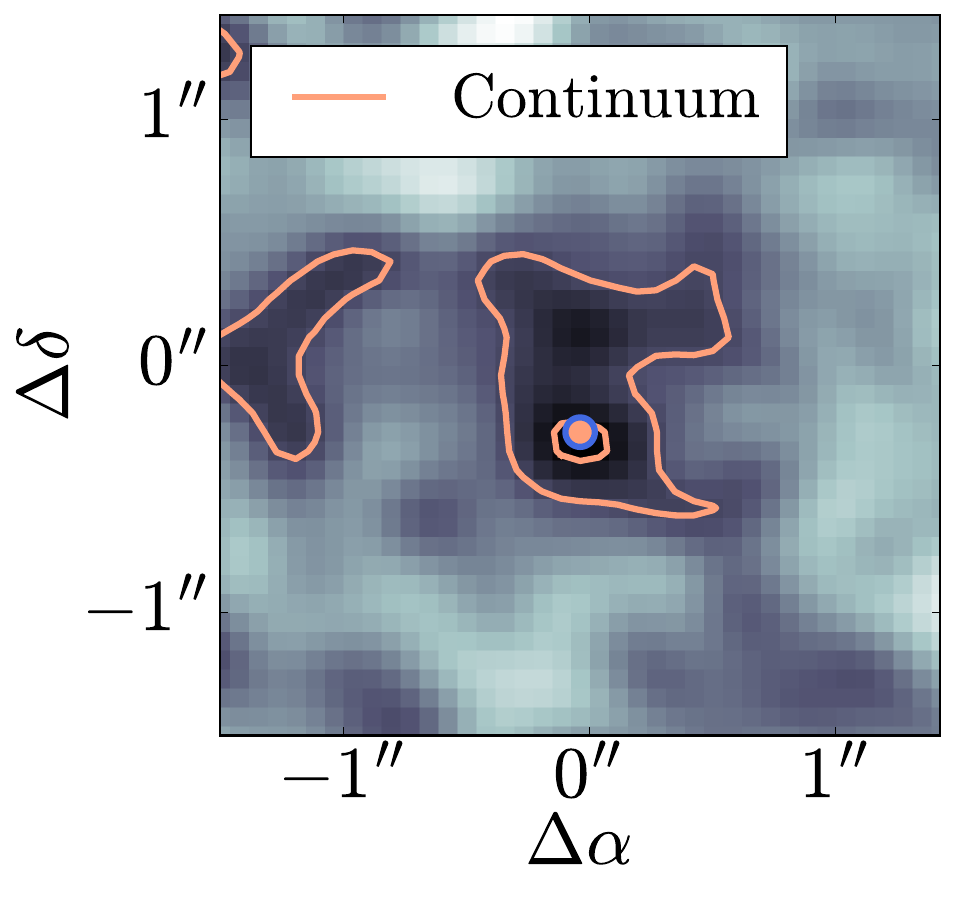} \\
\includegraphics[trim=0 11.2cm 0 0,clip,width=0.30\textwidth]{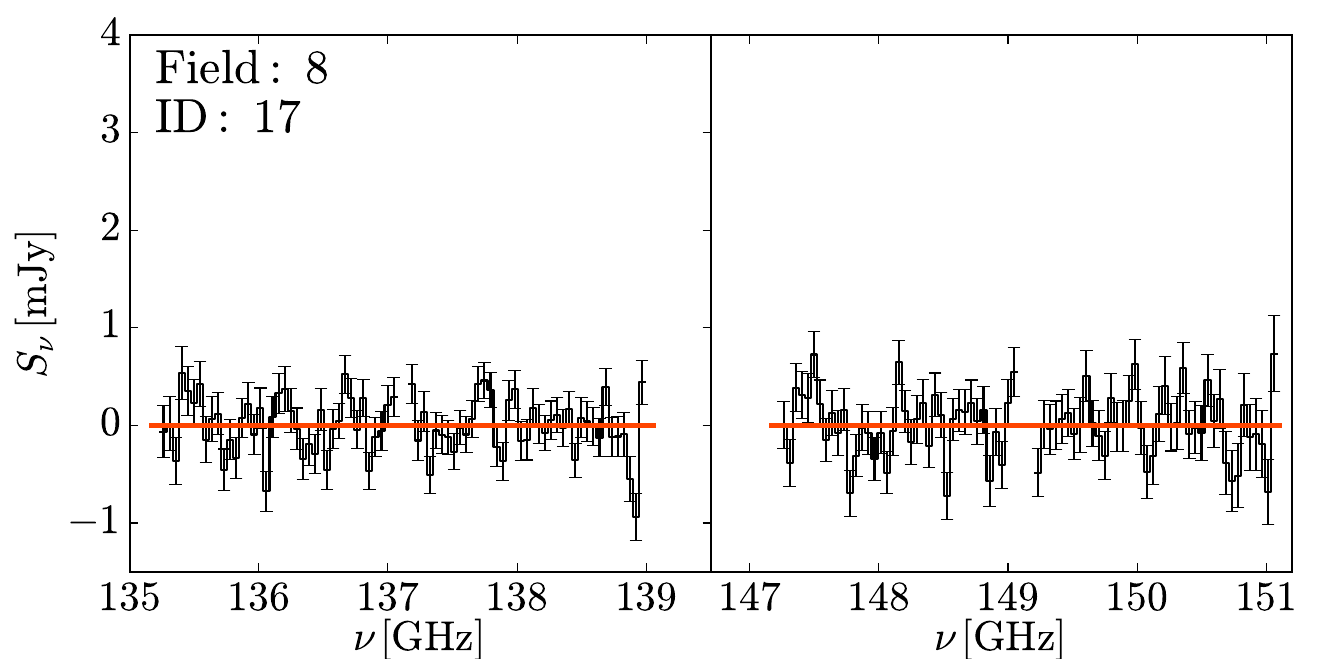}}}
\fbox{
\parbox{0.31\textwidth}{
\centering
\includegraphics[trim=0 0 16.5cm 0,clip,width=0.15\textwidth]{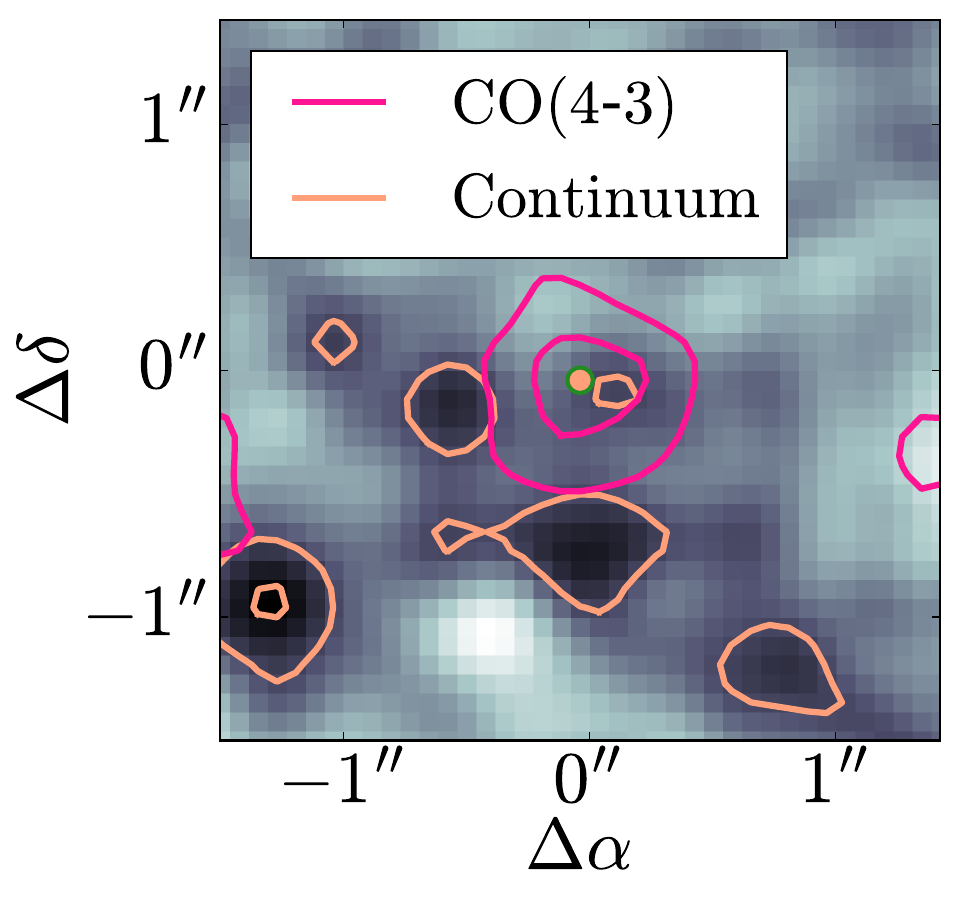}
\includegraphics[trim=0 0 16.5cm 0,clip,width=0.15\textwidth]{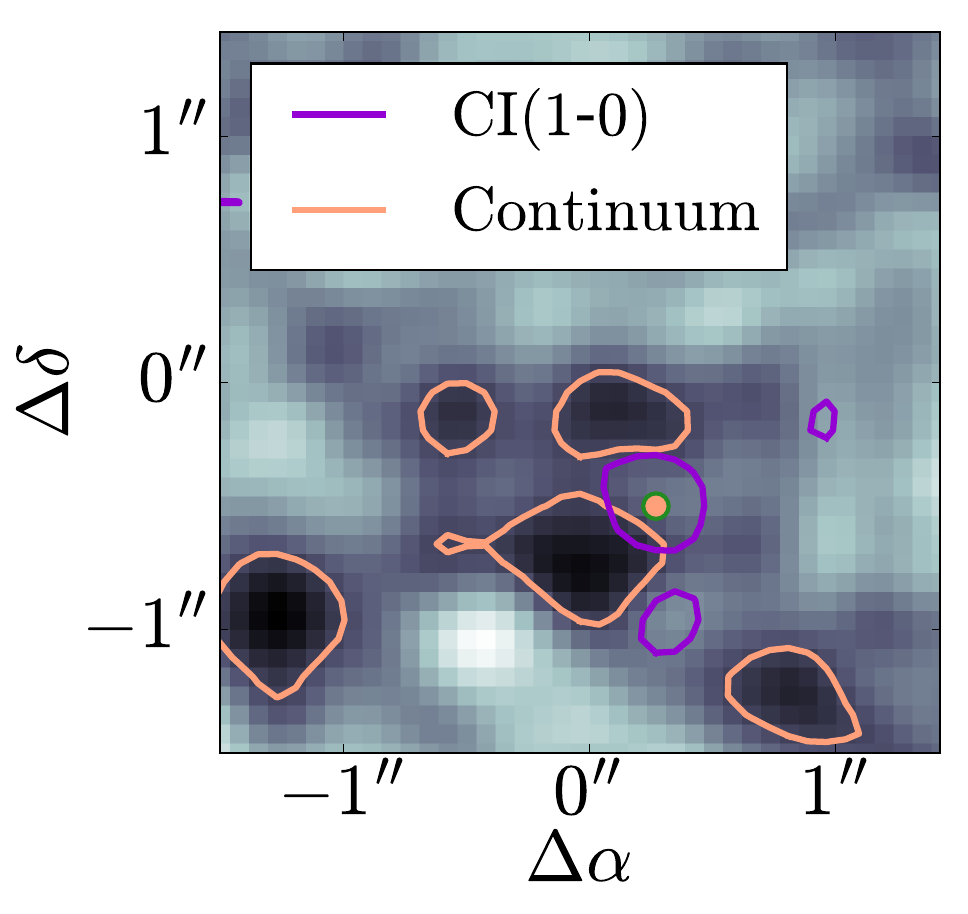} \\
\includegraphics[trim=0 11.2cm 0 0,clip,width=0.30\textwidth]{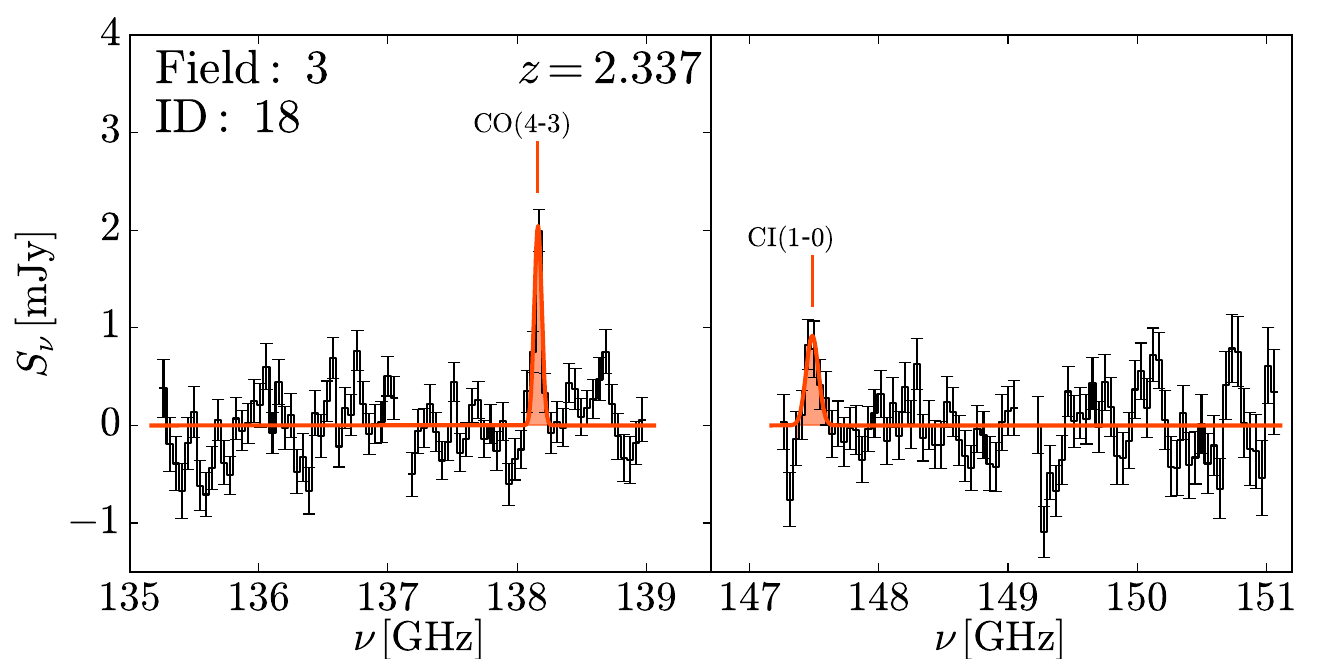}}}
\fbox{
\parbox{0.31\textwidth}{
\centering
\includegraphics[trim=0 0 16.5cm 0,clip,width=0.15\textwidth]{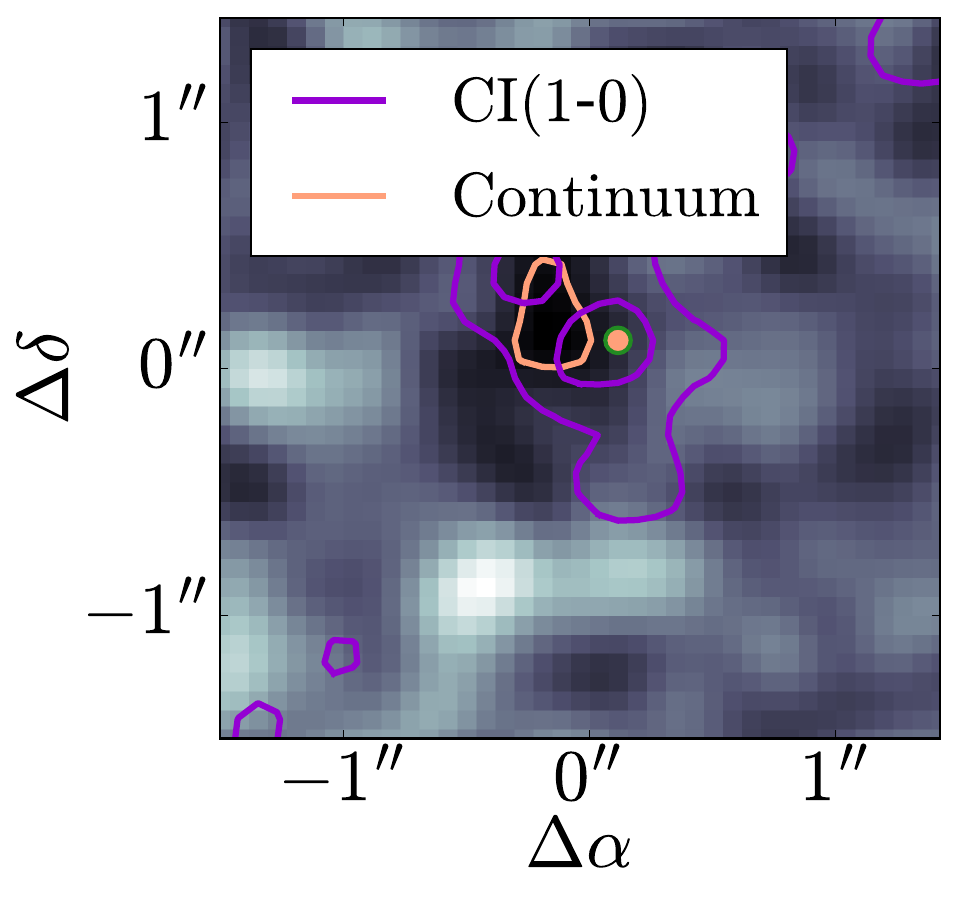} \\
\includegraphics[trim=0 11.2cm 0 0,clip,width=0.30\textwidth]{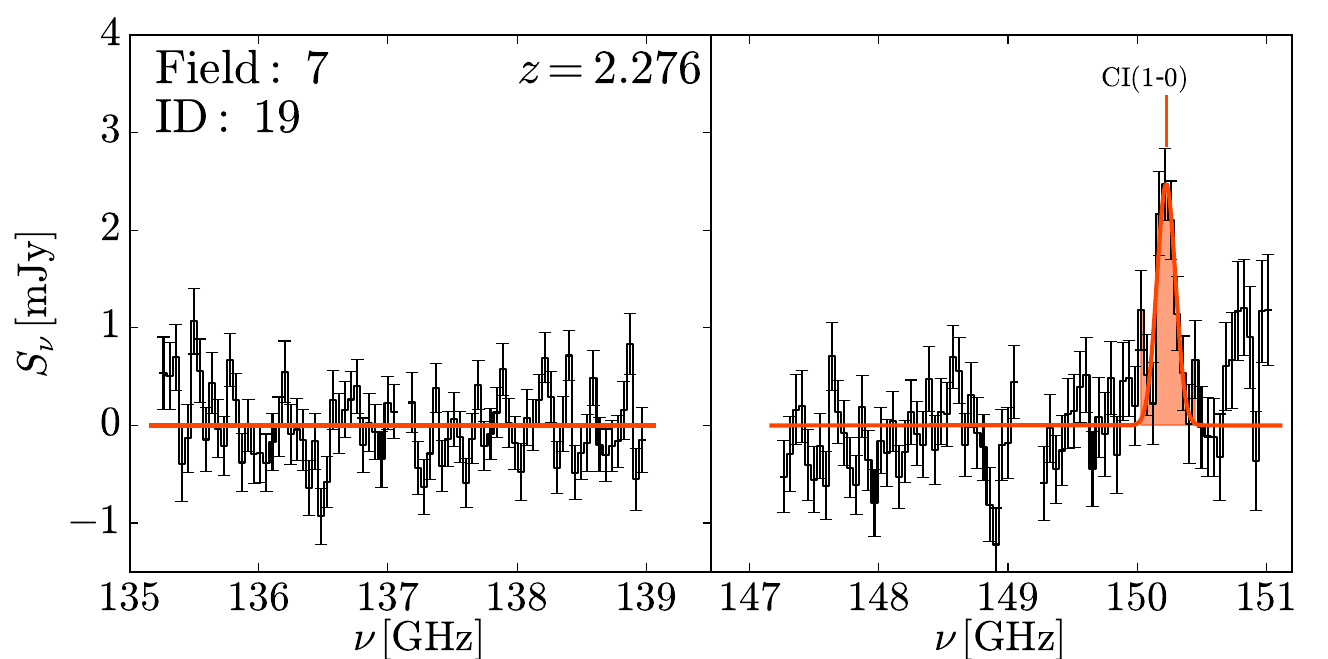}}}
\caption{{Continued.}}
\end{figure*}

\begin{figure*}[h!]
\setcounter{figure}{0}
\fbox{
\parbox{0.31\textwidth}{
\centering
\includegraphics[trim=0 0 16.5cm 0,clip,width=0.15\textwidth]{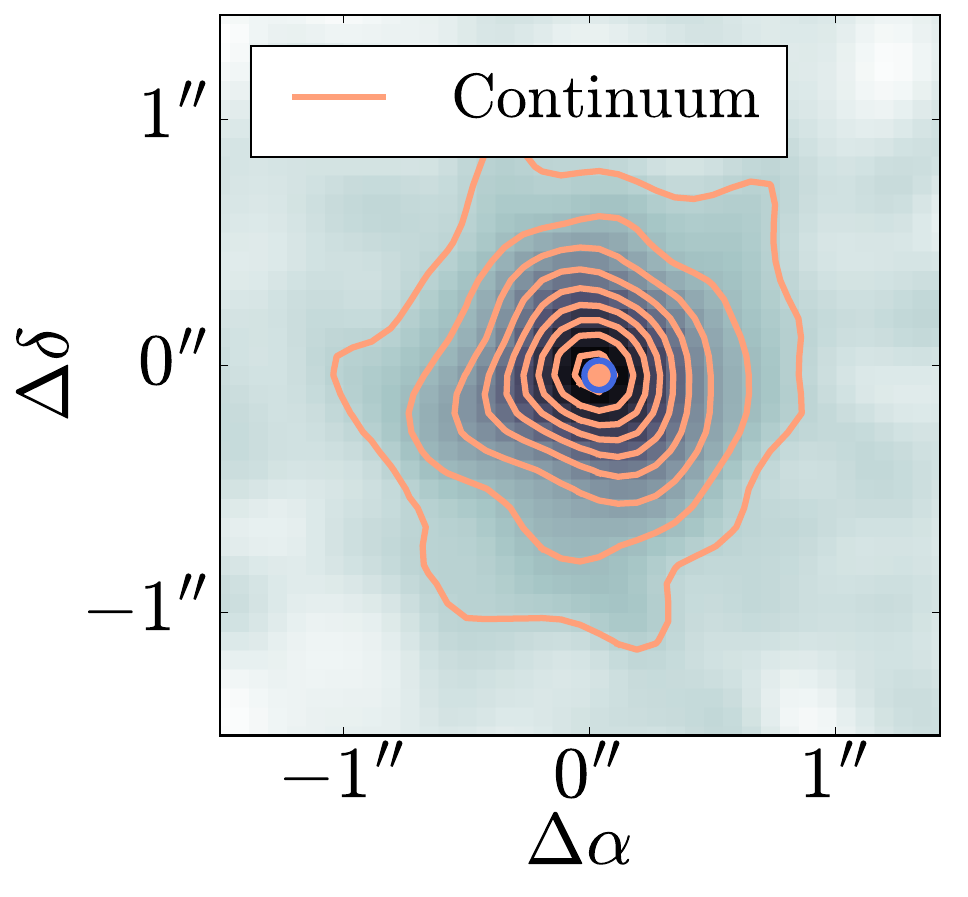} \\
\includegraphics[trim=0 10.1cm 0 0,clip,width=0.30\textwidth]{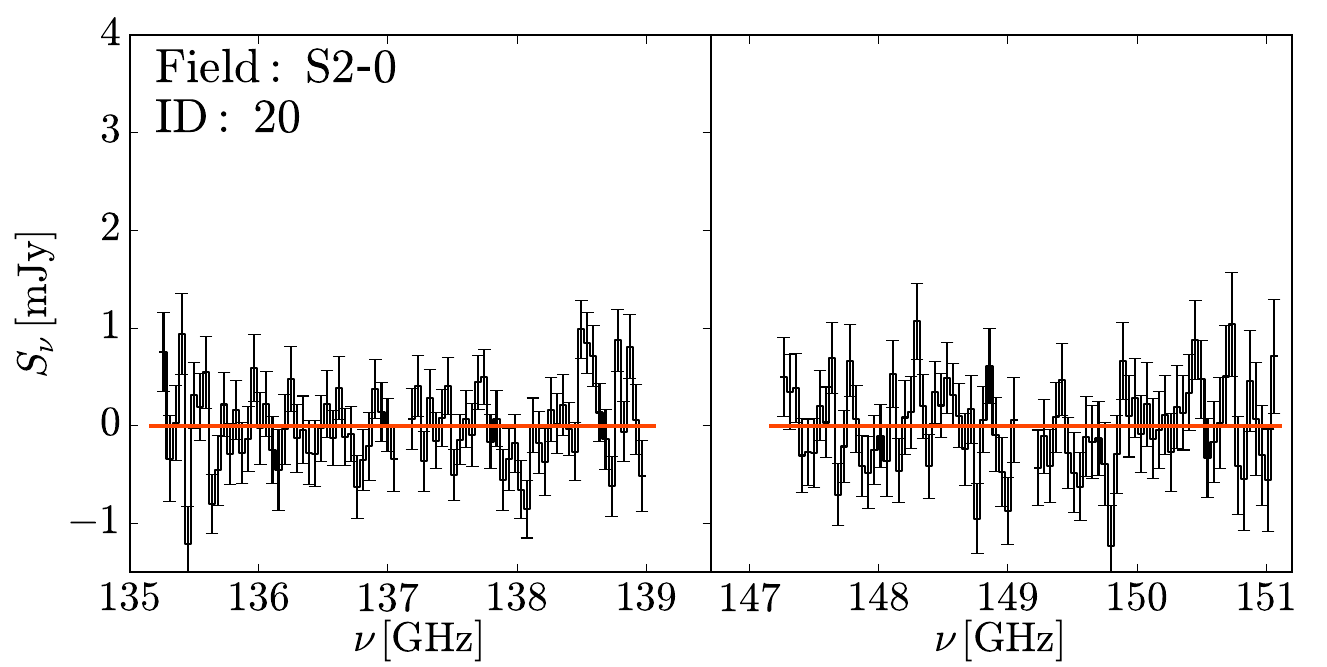}}}
\fbox{
\parbox{0.31\textwidth}{
\centering
\includegraphics[trim=0 0 16.5cm 0,clip,width=0.15\textwidth]{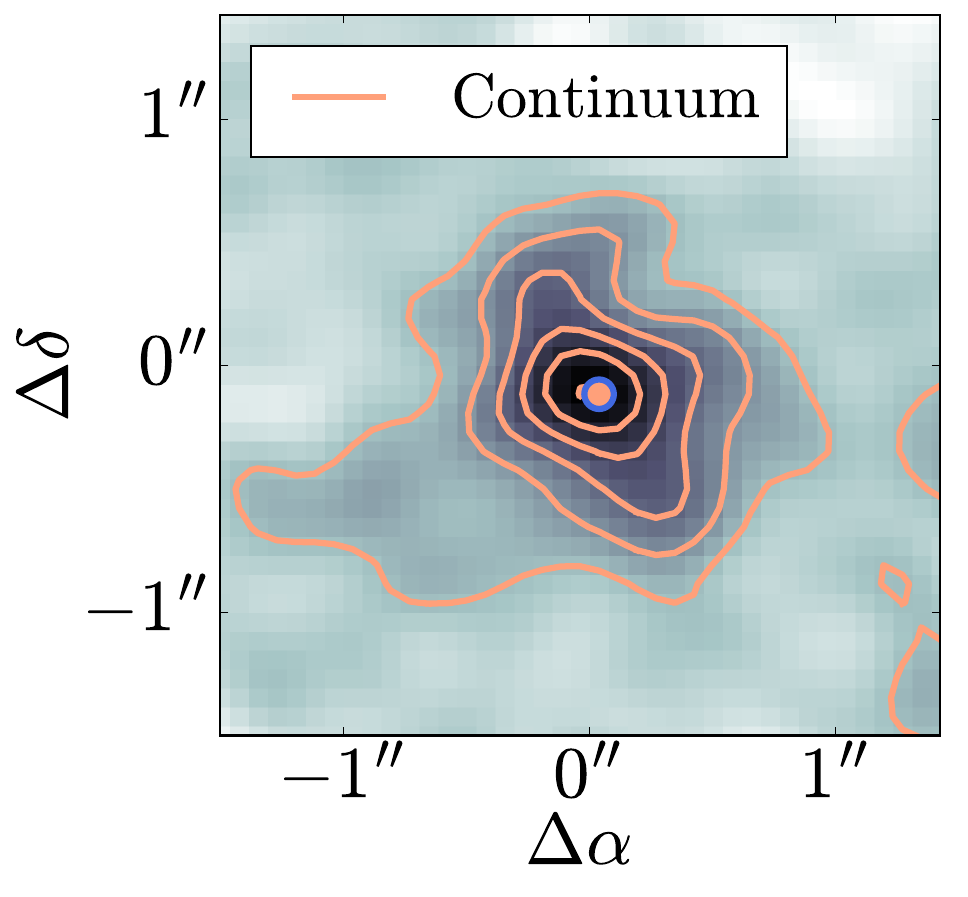} \\
\includegraphics[trim=0 10.1cm 0 0,clip,width=0.30\textwidth]{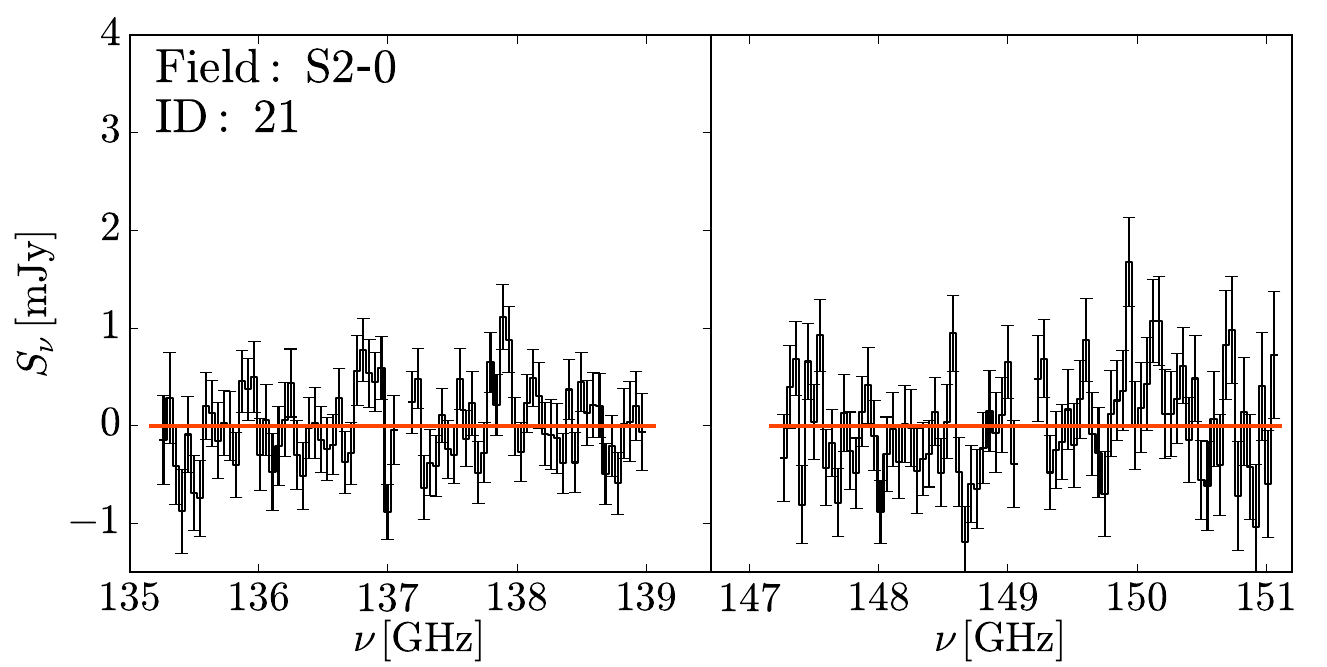}}}
\fbox{
\parbox{0.31\textwidth}{
\centering
\includegraphics[trim=0 0 16.5cm 0,clip,width=0.15\textwidth]{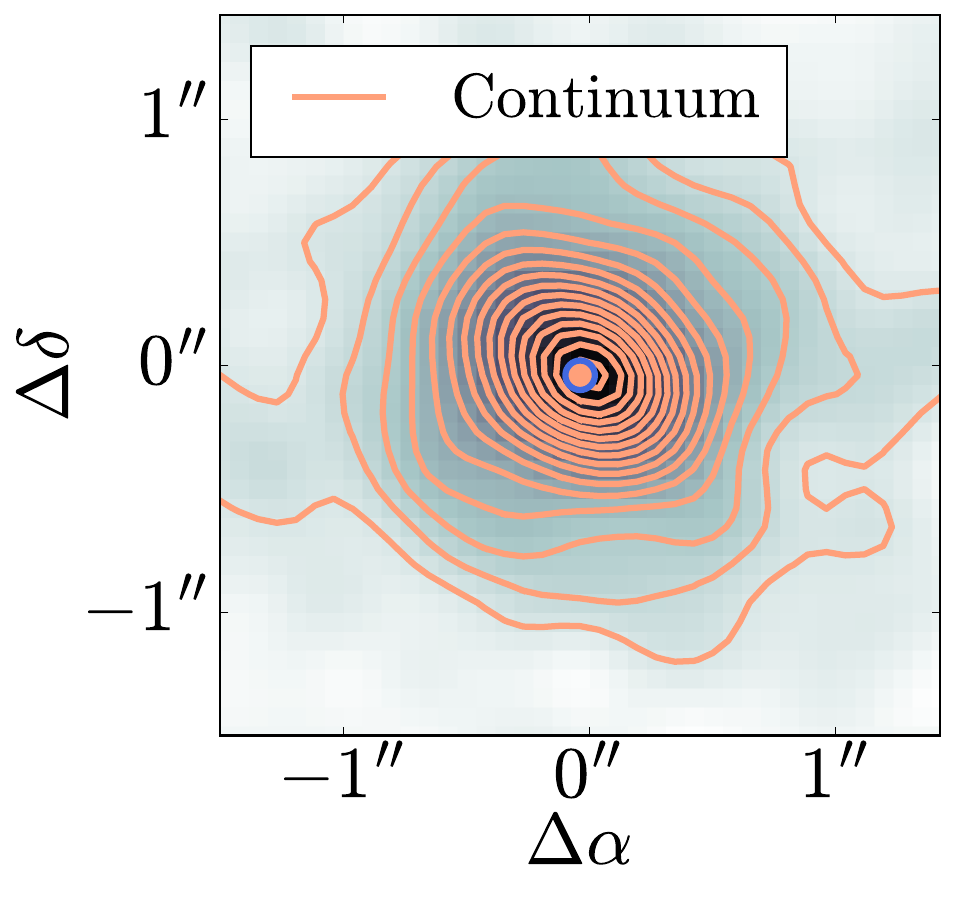} \\
\includegraphics[trim=0 10.1cm 0 0,clip,width=0.30\textwidth]{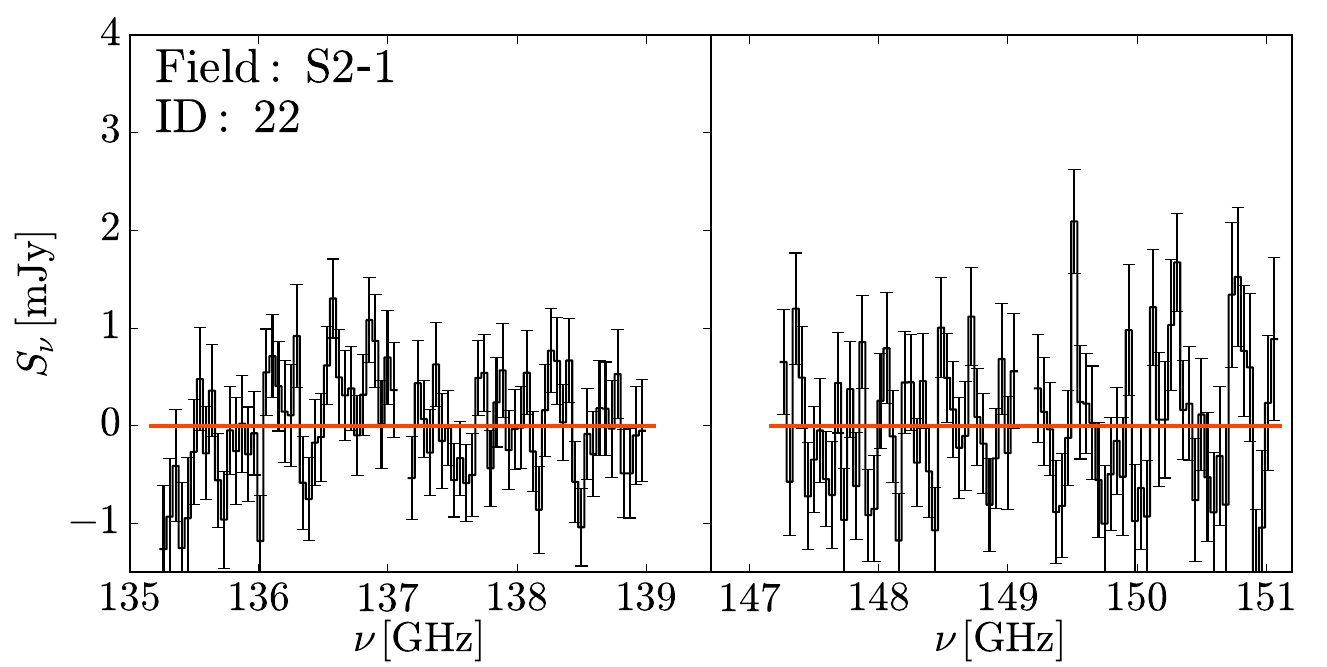}}}
\fbox{
\parbox{0.31\textwidth}{
\centering
\includegraphics[trim=0 0 16.5cm 0,clip,width=0.15\textwidth]{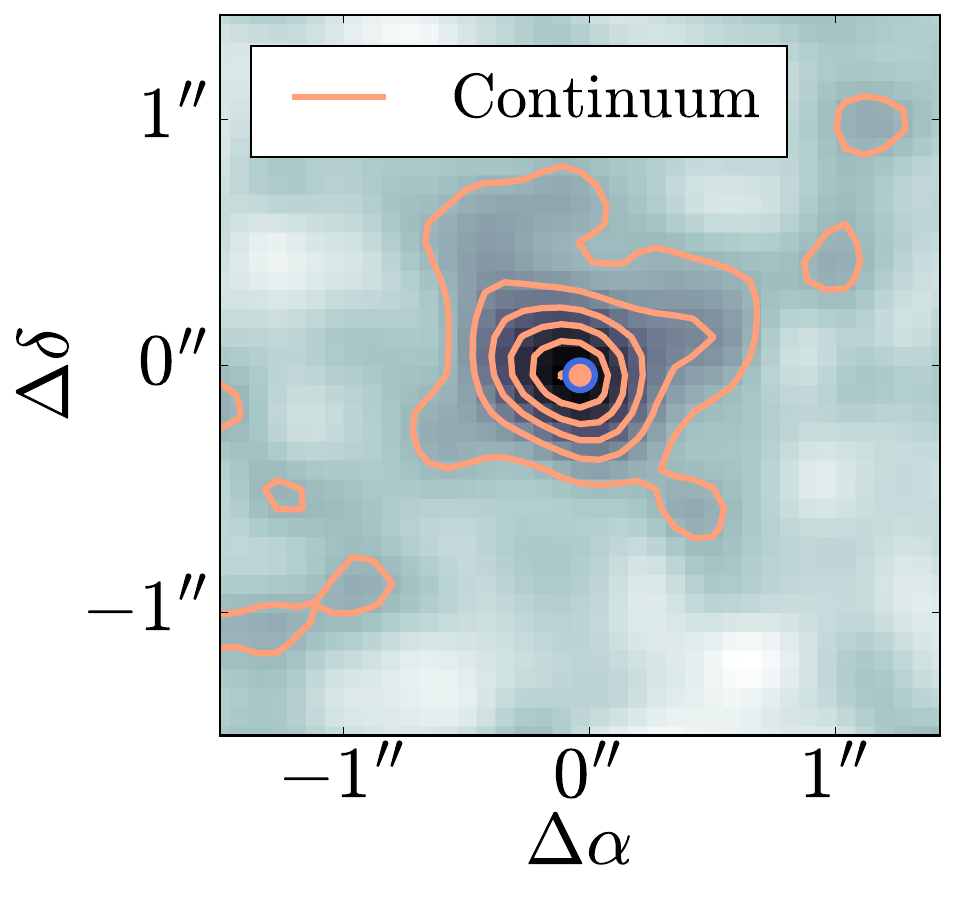} \\
\includegraphics[trim=0 10.1cm 0 0,clip,width=0.30\textwidth]{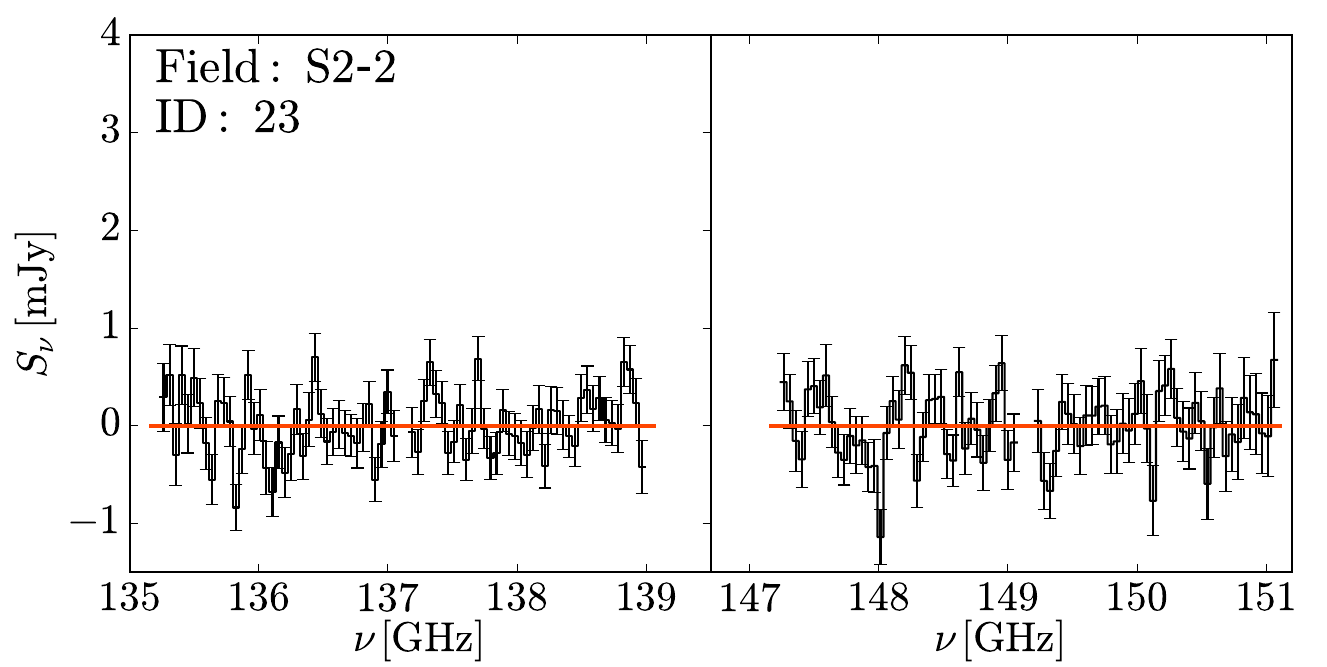}}}
\fbox{
\parbox{0.31\textwidth}{
\centering
\includegraphics[trim=0 0 16.5cm 0,clip,width=0.15\textwidth]{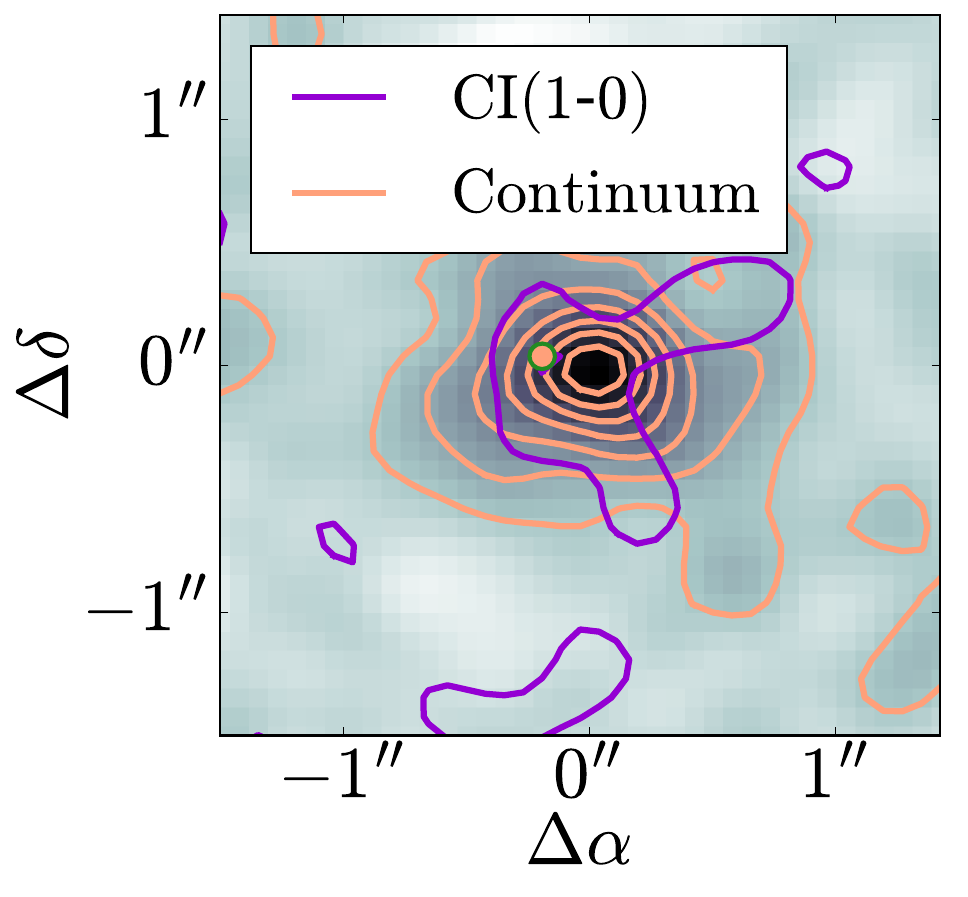} \\
\includegraphics[trim=0 10.1cm 0 0,clip,width=0.30\textwidth]{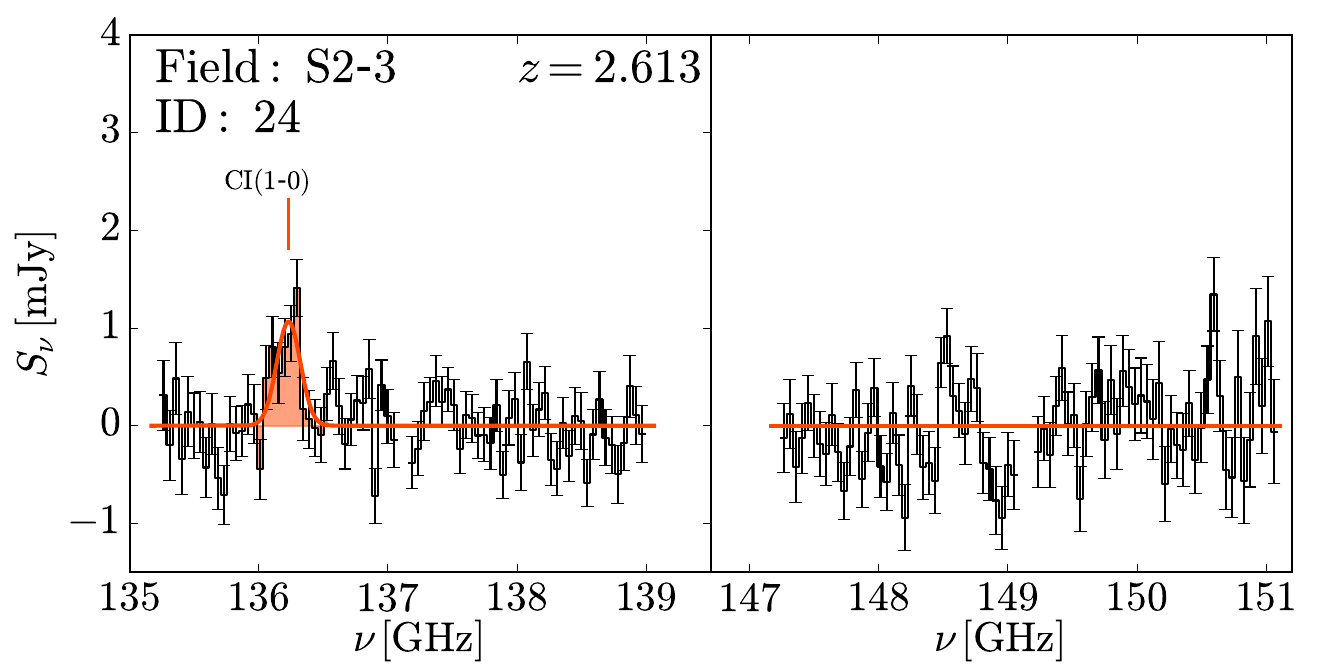}}}
\fbox{
\parbox{0.31\textwidth}{
\centering
\includegraphics[trim=0 0 16.5cm 0,clip,width=0.15\textwidth]{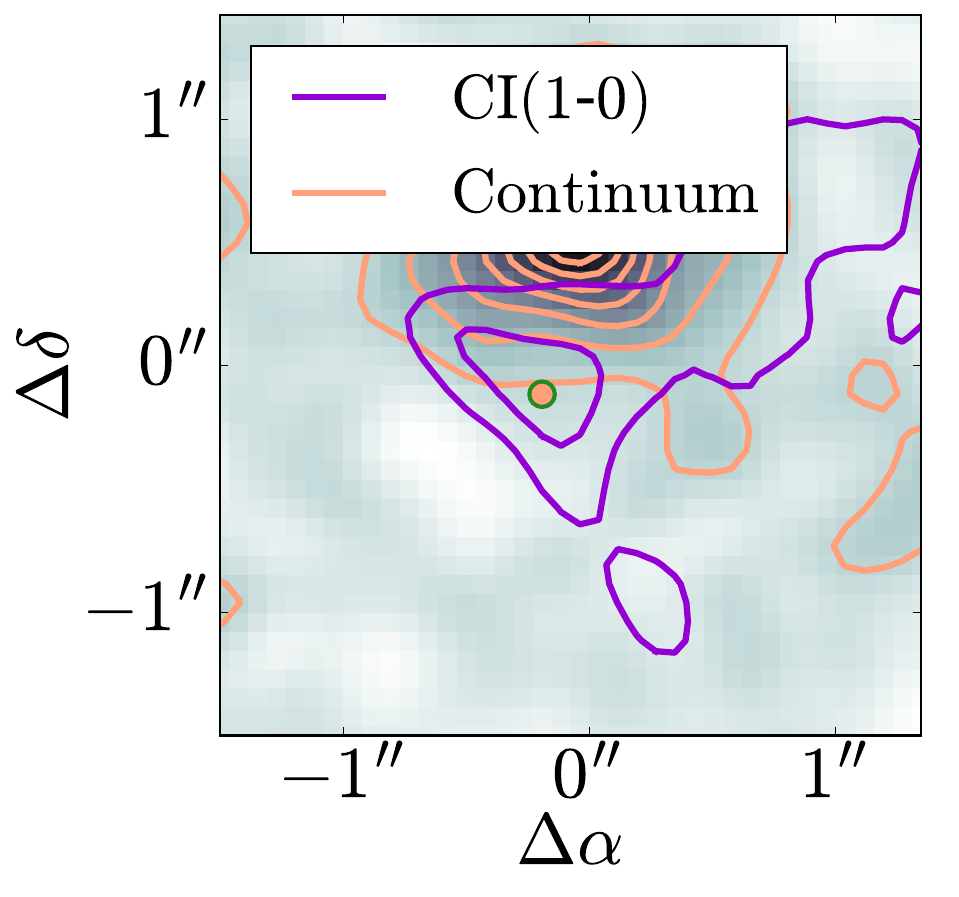} \\
\includegraphics[trim=0 10.1cm 0 0,clip,width=0.30\textwidth]{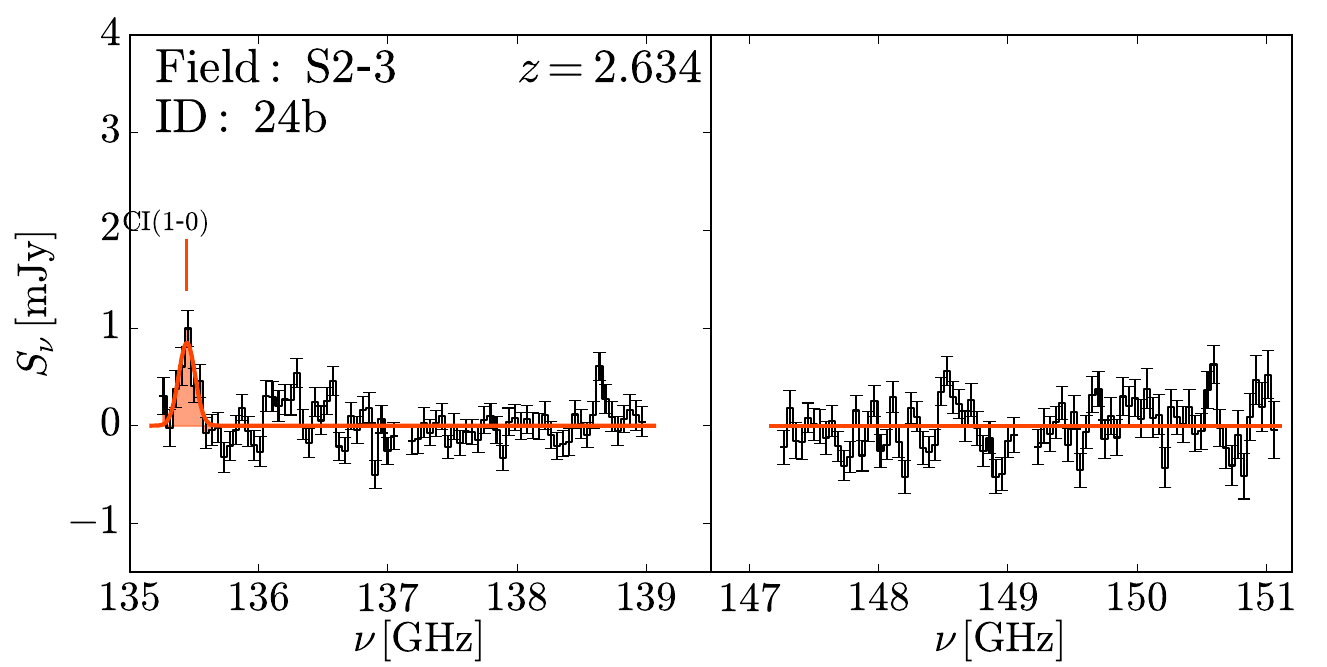}}}
\caption{{Continued.}}
\end{figure*}

\twocolumn
\begin{table*}[h!]
\section{NIR photometric data}
\label{appendix1}

Here we provide the NIR photometry used to fit SEDs and derive physical properties of the galaxies in G073. The $J$ and $K_{\rm s}$-band data are from WIRcam on the CFHT, and the 3.6\,$\mu$m and 4.5\,$\mu$m data are from {\it Spitzer}-IRAC.
\newline

\centering
\caption{NIR and MIR flux densities of the Band~4+6 continuum galaxies found in G073. Note that upper limits correspond to 5$\sigma$.\label{tab:ir_fluxes}}
\begin{tabular}{lccccc}
\hline\hline
\noalign{\vspace{2pt}}
Field & ID & $S_J$ & $S_{K_{\rm s}}$ & $S_\mathrm{3.6}$ & $S_\mathrm{4.5}$ \\
 & & [$\mu$Jy] & [$\mu$Jy] & [$\mu$Jy] & [$\mu$Jy] \\ 
\hline
\noalign{\vskip 1pt}
 1       &     0  &    24.98$\pm$0.44  &     77.15$\pm$0.92  &    137.41$\pm$0.82 &    114.32$\pm$1.08  \\
 1       &     1  &     $<$1.10        &      $<$4.37        &     14.37$\pm$0.83 &     20.30$\pm$1.11  \\
 \hline
 \noalign{\vskip 1pt}
 2    &     2  &     $<$1.10        &      $<$4.37        &      5.47$\pm$0.92 &      7.86$\pm$1.27  \\
 2    &     3  &     8.61$\pm$0.46  &     18.28$\pm$1.47  &     45.30$\pm$0.90 &     57.37$\pm$1.27  \\
 2    &     4  &     $<$1.10        &      $<$4.37        &      6.53$\pm$0.88 &      8.03$\pm$1.02  \\
 \hline
 \noalign{\vskip 1pt}
 3       &     5  &     6.21$\pm$0.44  &     18.19$\pm$0.92  &     39.96$\pm$0.83 &     49.79$\pm$1.11  \\
 3       &     6  &     2.06$\pm$0.44  &      4.60$\pm$0.93  &     16.00$\pm$0.84 &     22.13$\pm$1.12  \\
 3       &     7  &     $<$1.10        &      $<$4.37        &      5.72$\pm$0.85 &      6.24$\pm$1.11  \\
 3    & 18 &     2.09$\pm$0.45  &      8.82$\pm$0.93  &     17.18$\pm$0.84 &     19.61$\pm$1.11  \\
 \hline
 \noalign{\vskip 1pt}
 4    &     8  &     8.40$\pm$0.44  &     20.31$\pm$0.92  &     49.19$\pm$0.82 &     63.65$\pm$1.10  \\
 4    &     9  &     2.78$\pm$0.44  &      6.66$\pm$0.92  &     20.23$\pm$0.83 &     23.34$\pm$1.12  \\
 \hline
 \noalign{\vskip 1pt}
 5       &     10  &    16.31$\pm$0.44  &     34.07$\pm$0.92  &     58.24$\pm$2.85 &     52.33$\pm$3.03  \\
 \hline
 \noalign{\vskip 1pt}
 6    &     11  &     $<$1.10        &      6.35$\pm$0.96  &     14.22$\pm$0.88 &     16.78$\pm$1.17  \\
 6    &     12  &     2.25$\pm$0.46  &     10.30$\pm$0.93  &     23.28$\pm$0.87 &     32.22$\pm$1.16  \\
 \hline
 \noalign{\vskip 1pt}
 7    &     13  &     $<$1.10        &      4.68$\pm$1.10  &     10.40$\pm$1.67 &     14.61$\pm$1.92  \\
 7       &     14  &     $<$1.10        &      $<$4.37        &      2.95$\pm$1.68 &      3.40$\pm$1.93  \\
 7       &     15  &     $<$1.10        &      $<$4.37        &      $<$5.03       &      $<$5.78        \\
 7       &     16  &     6.27$\pm$0.54  &     13.69$\pm$1.01  &     20.54$\pm$1.67 &     13.60$\pm$1.92  \\
 7       &     19  &     3.90$\pm$0.30  &      8.20$\pm$0.54  &     17.19$\pm$1.67 &     16.89$\pm$2.13  \\
 \hline
 \noalign{\vskip 1pt}
 8       &     17  &     5.48$\pm$0.50  &     19.13$\pm$1.00  &     40.68$\pm$0.87 &     40.01$\pm$1.28  \\
 \hline
 \noalign{\vskip 1pt}
 S2-0  & 20 &     $<$1.10        &      $<$4.37        &      $<$5.03       &      $<$5.78        \\
 S2-0  & 21 &     $<$1.10        &      $<$4.37        &      5.08$\pm$0.83 &      8.35$\pm$1.11  \\
 \hline
 \noalign{\vskip 1pt}
 S2-1  & 22 &     $<$1.10        &      $<$4.37        &      3.76$\pm$0.91 &      4.67$\pm$1.31  \\
 \hline
 \noalign{\vskip 1pt}
 S2-2  & 23 &     1.44$\pm$0.45  &      4.28$\pm$0.92  &     14.52$\pm$0.84 &     17.54$\pm$1.11  \\
 \hline
 \noalign{\vskip 1pt}
 S2-3   & 24 &     $<$1.10        &      $<$4.37        &      2.08$\pm$1.08 &   3.00$\pm$1.74  \\
 S2-3  & 25 &     $<$1.10        &      $<$4.37        &      2.08$\pm$1.08 &   3.00$\pm$1.74  \\
\hline
\end{tabular}
\end{table*}

\end{appendix}

\end{document}